\documentclass[aps,prd,twocolumn,amsmath,amssymb,superscriptaddress,floatfix]{revtex4-1}
\usepackage{graphicx}
\usepackage{dcolumn}
\usepackage{bm}
\usepackage{color}
\usepackage{multirow}
\usepackage{xspace}
\usepackage{url}
\usepackage{lineno}
\usepackage[english]{babel}
\usepackage{natbib}

\newcommand{\micron}{$\mu \mathrm{m}$\xspace}

\begin{document}


\title{EMPHATIC: A proposed experiment to measure hadron scattering and production cross sections for improved neutrino flux predictions}

\newcommand{\FNAL}{Fermi National Accelerator Laboratory, Batavia, 
Illinois 60510, USA}
\newcommand{\TRIUMF}{TRIUMF, Vancouver, BC  V6T 2A3, Canada}
\newcommand{\KEK}{Institute of
Particle and Nuclear Study (IPNS), High Energy Accelerator Research
Organization (KEK), Tsukuba, 305-0801, Japan}
\newcommand{\IPMU}{IPMU, Kashiwa, Chiba 277-8583, Japan}
\newcommand{\Nagoya}{Nagoya University, Nagoya, Aichi 464-8601, Japan}
\newcommand{\Kyoto}{Kyoto University, Yoshidahonmachi, Kyoto, Kyoto 606-8501, Japan}
\newcommand{\Winnipeg}{University of Winnipeg, Winnipeg, MB R3B 2E9, Canada}
\newcommand{\Chiba}{Department of Physics, Chiba University, Chiba, Chiba 263-8522, Japan}
\newcommand{\RCNP}{Research Center for Nuclear Physics (RCNP), Ibaraki,
567-0047, Japan}
\newcommand{\Osaka}{Department of Physics, Osaka University, Toyonaka 560-0043, Japan}
\newcommand{\RIKEN}{RIKEN, Wako, 351-0198, Japan}
\newcommand{\Sinica}{Institute of Physics, Academia Sinica, Taipei 11529, Taiwan}
\newcommand{\Tohoku}{Department of Physics, Tohoku University, Sendai 980-8578, Japan}
\newcommand{\ELPH}{Research Center for Electron Photon Science (ELPH), Tohoku University, Sendai, 982-0826, Japan}
\newcommand{\OsakaCity}{Department of Physics, Osaka City University, Osaka 558-8585, Japan}
\newcommand{\Kobe}{Kobe University, Kobe, Hyogo 657-8501, Japan}
\newcommand{\Regina}{Department of Physics, University of Regina, SK, S4S 0A2, Canada}
\newcommand{\York}{York University, Department of Physics and Astronomy, Toronto, Ontario, Canada}
\newcommand{\Wichita}{Department of Mathematics, Statistics, and Physics, Wichita State University, Wichita, Kansas 67206, USA}
\newcommand{\Boston}{Department of Physics, Boston University, Boston, Massachusetts, USA}
\newcommand{\WilliamMary}{Department of Physics, College of William \& Mary, Williamsburg, Virginia 23187, USA}
\newcommand{\Cincinnati}{Department of Physics, University of Cincinnati, Cincinnati, Ohio 45221, USA}
\newcommand{\SDSMT}{South Dakota School of Mines and Technology, Rapid City, South Dakota 57701, USA}
\newcommand{\UTA}{Department of Physics, University of Texas at Austin, Austin, TX 78712, USA}

\affiliation{\Boston}
\affiliation{\Chiba}
\affiliation{\Cincinnati}
\affiliation{\FNAL}
\affiliation{\KEK}
\affiliation{\Sinica}
\affiliation{\IPMU}
\affiliation{\Kobe}
\affiliation{\Kyoto}
\affiliation{\Nagoya}
\affiliation{\OsakaCity}
\affiliation{\Osaka}
\affiliation{\RCNP}
\affiliation{\Regina}
\affiliation{\Tohoku}
\affiliation{\ELPH}
\affiliation{\RIKEN}
\affiliation{\SDSMT}
\affiliation{\UTA}
\affiliation{\TRIUMF}
\affiliation{\Wichita}
\affiliation{\WilliamMary}
\affiliation{\Winnipeg}
\affiliation{\York}

\author{T.~Akaishi}
\affiliation{\Osaka}

\author{L.~Aliaga-Soplin}
\affiliation{\FNAL}

\author{H.~Asano}
\affiliation{\RIKEN}

\author{A.~Aurisano}
\affiliation{\Cincinnati}

\author{M.~Barbi}
\affiliation{\Regina}

\author{L.~Bellantoni}
\affiliation{\FNAL}

\author{S.~Bhadra}
\affiliation{\York}

\author{W-C.~Chang} 
\affiliation{\York}

\author{L.~Fields}
\affiliation{\FNAL}

\author{A.~Fiorentini} 
\affiliation{\SDSMT}

\author{M.~Friend}
\affiliation{\KEK}

\author{T.~Fukuda}
\affiliation{\Nagoya}

\author{D.~Harris}
\affiliation{\FNAL}

\author{M.~Hartz}
\affiliation{\IPMU}
\affiliation{\TRIUMF}

\author{R.~Honda} 
\affiliation{\Tohoku}

\author{T.~Ishikawa} 
\affiliation{\ELPH}

\author{B.~Jamieson}
\affiliation{\Winnipeg}

\author{E.~Kearns}
\affiliation{\Boston}

\author{N.~Kolev}
\affiliation{\Regina}

\author{M.~Komatsu} 
\affiliation{\Nagoya}

\author{Y.~Komatsu} 
\affiliation{\KEK}

\author{A.~Konaka}
\affiliation{\TRIUMF}

\author{M.~Kordosky}
\affiliation{\WilliamMary}

\author{K.~Lang}
\affiliation{\UTA}

\author{P.~Lebrun}
\affiliation{\FNAL}

\author{T.~Lindner}
\affiliation{\Winnipeg}
\affiliation{\TRIUMF}

\author{Y.~Ma}
\affiliation{\RIKEN}

\author{D.~A.~Martinez Caicedo}
\affiliation{\SDSMT}

\author{M.~Muether}
\affiliation{\Wichita}

\author{N.~Naganawa} 
\affiliation{\Nagoya}

\author{M.~Naruki} 
\affiliation{\Kyoto}

\author{E.~Niner} 
\affiliation{\FNAL}

\author{H.~Noumi}
\affiliation{\RCNP}

\author{K.~Ozawa} 
\affiliation{\KEK}

\author{J.~Paley}
\thanks{Co-spokesperson: Jonathan M. Paley (jpaley@fnal.gov)}
\affiliation{\FNAL}

\author{M.~Pavin}
\affiliation{\TRIUMF}

\author{P.~de Perio}
\affiliation{\TRIUMF}

\author{M.~Proga}
\affiliation{\UTA}

\author{F.~Sakuma}
\affiliation{\RIKEN}

\author{G.~Santucci} 
\affiliation{\York}

\author{T.~Sawada} 
\affiliation{\OsakaCity}

\author{O.~Sato} 
\affiliation{\Nagoya}

\author{T.~Sekiguchi}
\thanks{Co-spokesperson: Tetsuro Sekiguchi (tetsuro.sekiguchi@kek.jp)}
\affiliation{\KEK}

\author{K.~Shirotori}
\affiliation{\RCNP}

\author{A.~Suzuki} 
\affiliation{\Kobe}

\author{M.~Tabata}
\affiliation{\Chiba}

\author{T.~Takahashi}
\affiliation{\RCNP}

\author{N.~Tomida}
\affiliation{\RCNP}

\author{R.~Wendell}
\affiliation{\Kyoto}

\author{T.~Yamaga}
\affiliation{\RIKEN}

\collaboration{The EMPHATIC Collaboration}
\noaffiliation

\date{\today}

\begin{abstract}
Hadron scattering and production uncertainties are a limiting systematic on accelerator and atmospheric neutrino flux predictions.  New hadron  measurements are necessary for neutrino flux predictions with well-understood and reduced uncertainties.  We propose a new compact experiment to measure hadron scattering and production cross sections at beam energies that are inaccessible to currently operating experiments.  These measurements can reduce the current 10\% neutrino flux uncertainties by an approximate factor of two.
\end{abstract}

\maketitle


\section{\label{sec:Intro}Introduction}
The discovery of neutrino oscillations triggered a new era of dedicated neutrino oscillation experiments.  The global effort to determine the 3-flavor oscillation parameters is entering an unprecedented level of precision.  Maximizing the scientific output of the current (T2K\cite{T2KLatest} and NOvA\cite{NOvALatest}) and future (Hyper-Kamiokande\cite{H2KProposal} and DUNE\cite{DUNEIDR}) long-baseline neutrino oscillation experiments requires improved modeling and reduced systematic uncertainties of neutrino flux.  Dedicated $\nu-A$ scattering experiments, BSM physics searches and any other measurement that relies on a single detector exposed to a beam of neutrinos also greatly benefit from improved modeling of neutrino flux predictions.  

We propose a new compact experiment capable of high-rate data collection to measure hadron production cross sections that are particularly relevant to neutrino flux predictions.  The compact design is enabled by the use of precision tracking detectors.  Particle identification is accomplished using Cherenkov detectors and fast timing electronics.  Although the goal of the experiment is to collect data to reduce neutrino flux prediction uncertainties, the data will benefit the general HEP and scientific community that relies on modeling hadron interactions.

This Proposal proceeds as follows: in Section \ref{sec:FluxImpact} we motivate the need for improved flux predictions within the context of the global neutrino program. 
Section \ref{sec:FluxPredictions} describes in more detail how new precise hadron production data will impact neutrino flux predictions and backgrounds to other rare processes.
Section \ref{sec:Experiment} describes the concept and design of the proposed experiment, including details of the subsystems under consideration.
Details of proof-of-principle measurements from data collected in 2018, which have yielded results that will be submitted for publication soon, are provided in Section \ref{sec:POPRun}. 
In Section \ref{sec:ProposedMeasurements} we list measurements with specific targets and beam momenta that will improve predictions, and demonstrate how reduced hadron scattering and production uncertainties will impact the Fermilab neutrino beam flux uncertainties.
In Section \ref{sec:RunPlan} we present a staged run plan taking into account the physics drivers and subsystem readiness. 

\section{\label{sec:FluxImpact} Impact of Flux Uncertainties on Measurements by Current and Future Neutrino Experiments}
The relatively large neutrino flux uncertainties are mitigated in accelerator-based neutrino oscillation experiments using a two-detector scheme.  In this approach, a near detector is used to characterize the event rate as a function of neutrino flavor and energy at a location where the effect of neutrino oscillation on the spectrum is negligible.  That information is used to improve predictions for the expected spectra at a far detector typically located at or near the location where the oscillation is maximized.  This technique is extremely effective, and in the case where functionally identical detectors are used at both the near and far locations, such as the MINOS and NOvA experiments, flux uncertainties become a negligible systematic uncertainty.

However, it is important to stress that having a robust \emph{a priori} prediction of the absolute neutrino flux with reduced uncertainties is extremely beneficial to the 3-flavor oscillation measurements for DUNE and T2HK.  These improvements reduce the risk to the 3-flavor measurement program, in particular to that of $\delta_\mathrm{CP}$, that our current hadron production models are not far outside of our current estimates.  Furthermore, these improvements reduce the possibility that any observations of discrepancies between near detector data and simulation will be incorrectly ascribed to flux mis-modeling, rather than the modeling of neutrino-nucleus scattering or detector response.  

The current and future neutrino oscillation experiments and non-oscillation experiments such as MINERvA have very rich physics programs using data from a single detector.  For detectors located near the neutrino production source, the intense beam of neutrinos enables high-statistics measurements of neutrino cross sections and searches for signals of more exotic phenomena such as short-baseline sterile neutrino searches, neutrino magnetic moments, non-standard interactions (NSI) in neutrino production and interactions, and dark matter production in the neutrino beamline.  All of these measurements and searches are ultimately systematically limited by the $\sim10\%$ flux uncertainty.  The neutrino cross section measurements from the MINERvA, NOvA, T2K and LAr SBN experiments are critical for the future DUNE and T2HK long-baseline neutrino experiments.  Therefore improvements in neutrino flux predictions for the current experimental program will benefit the long-term program.  

The DUNE near detector offers exciting opportunities to search for signs of Beyond the Standard Model (BSM) physics (see, eg, \cite{PONDD}).  However, any search for BSM physics will be limited by the flux uncertainty.  As an example, the MINERvA collaboration recently reported on a data-driven constraint on the on-axis medium-energy NuMI flux using neutrino-electron elastic scattering with a systematic uncertainty below 3\%\cite{MINERvA_nu_e}.  We expect that the DUNE experiment will have even smaller uncertainties for the same measurement.  Since the theoretical uncertainties on neutrino-electron elastic scattering are also at the level of 1-2\%, this measurement can alternatively be used to constrain new physics, for example NSI or neutrino magnetic moments using GeV energy neutrinos.  Any reduction to the current 10\% flux uncertainties will directly translate into improved limits on new physics.

\section{\label{sec:FluxPredictions}Impact of Hadron Production Uncertainties on Predictions of Neutrino Fluxes and Backgrounds to Other Rare Processes}
The primary motivation for this experiment is to reduce accelerator and atmospheric neutrino flux uncertainties.  However, we note that new hadron production cross section data benefit other fields of high energy particle and nuclear physics, in particular collider and fixed-target experiments and searches for rare phenomena in accelerator-produced beams of muons and mesons.

\subsection{Accelerator Neutrino Flux Predictions}

\begin{figure*}[htpb]
\begin{center}
\includegraphics[width=0.48\textwidth]{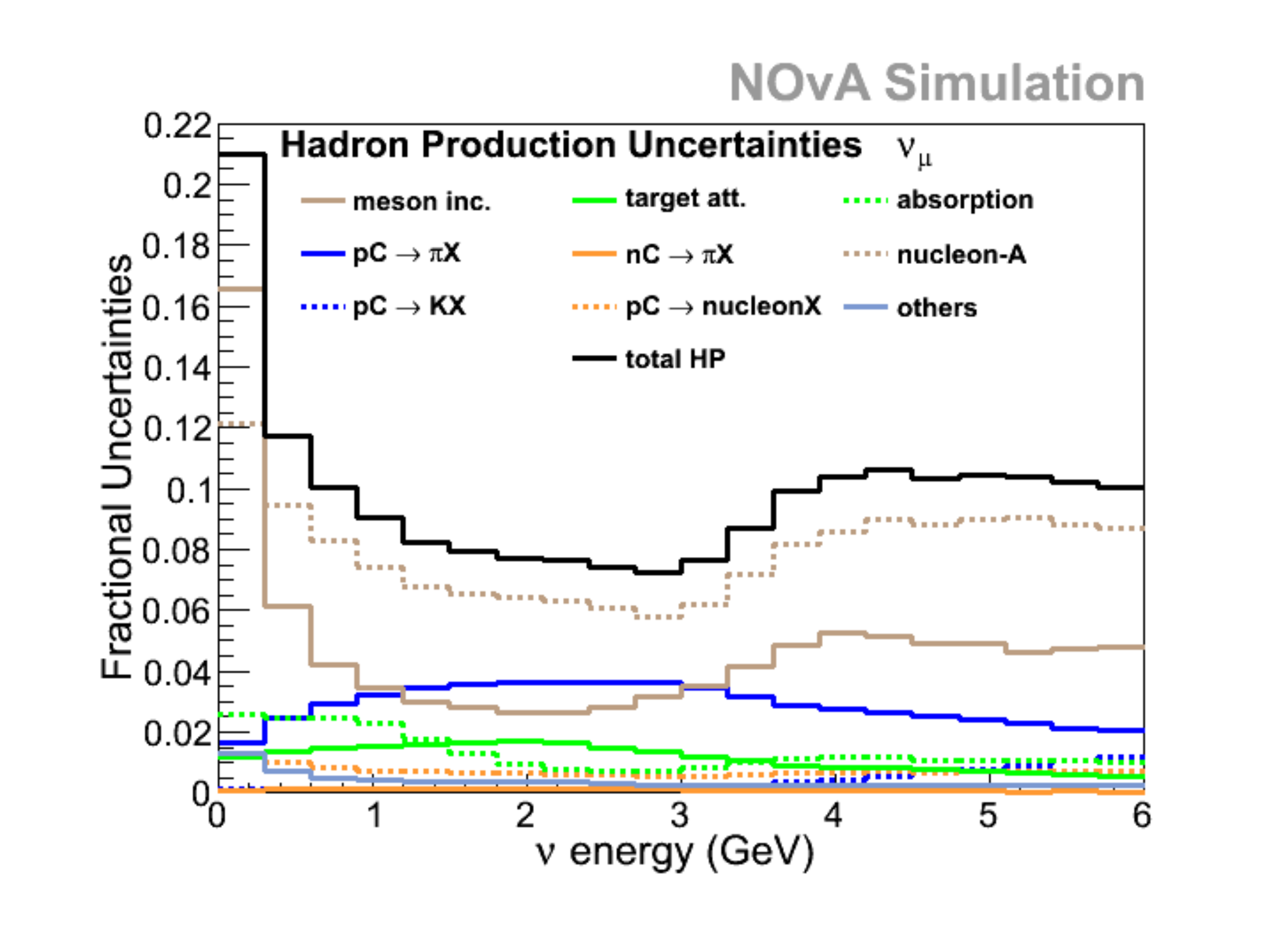}
\includegraphics[width=0.48\textwidth]{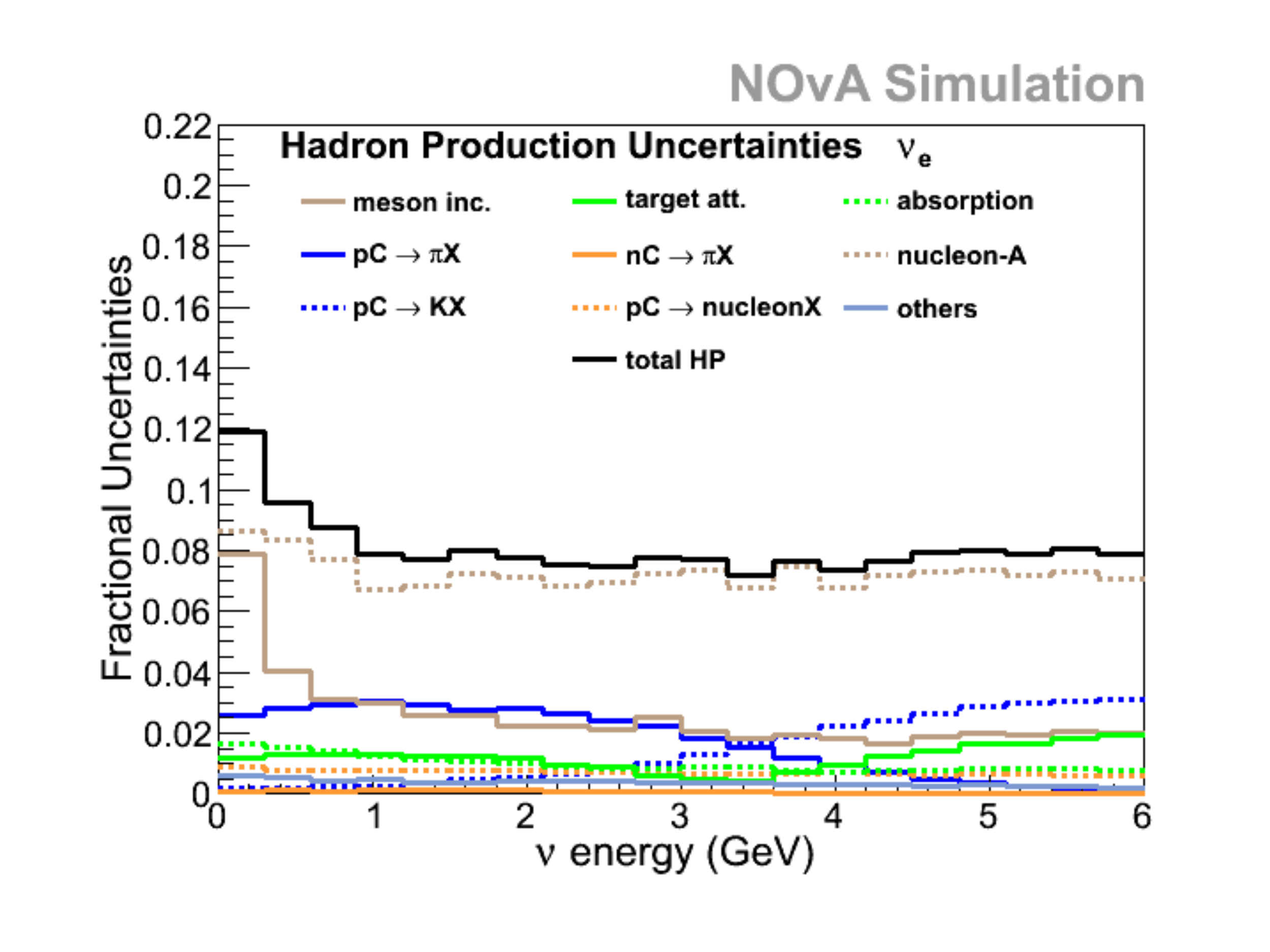}
\caption{Fractional uncertainties of the $\nu_\mu$ (left) and $\nu_e$ (right) flux as a function of neutrino energy in the 14.6 mrad off-axis NOvA near detector due to hadron production uncertainties \cite{NOvAFluxCurrent}.}
\label{fig:NOvAFlux}
\end{center} 
\end{figure*}

\begin{figure*}[htpb]
\begin{center}
\includegraphics[width=0.48\textwidth]{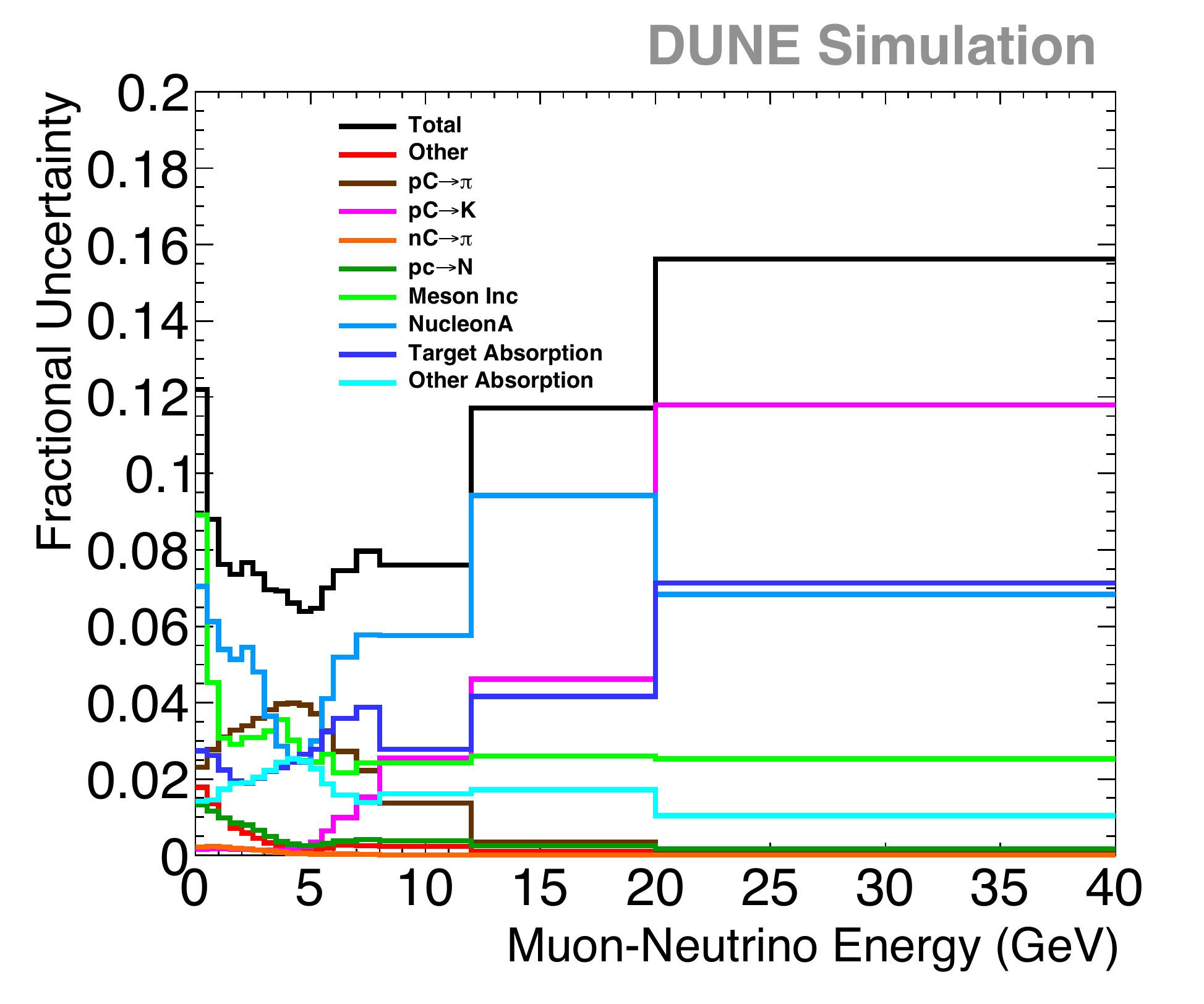}
\includegraphics[width=0.48\textwidth]{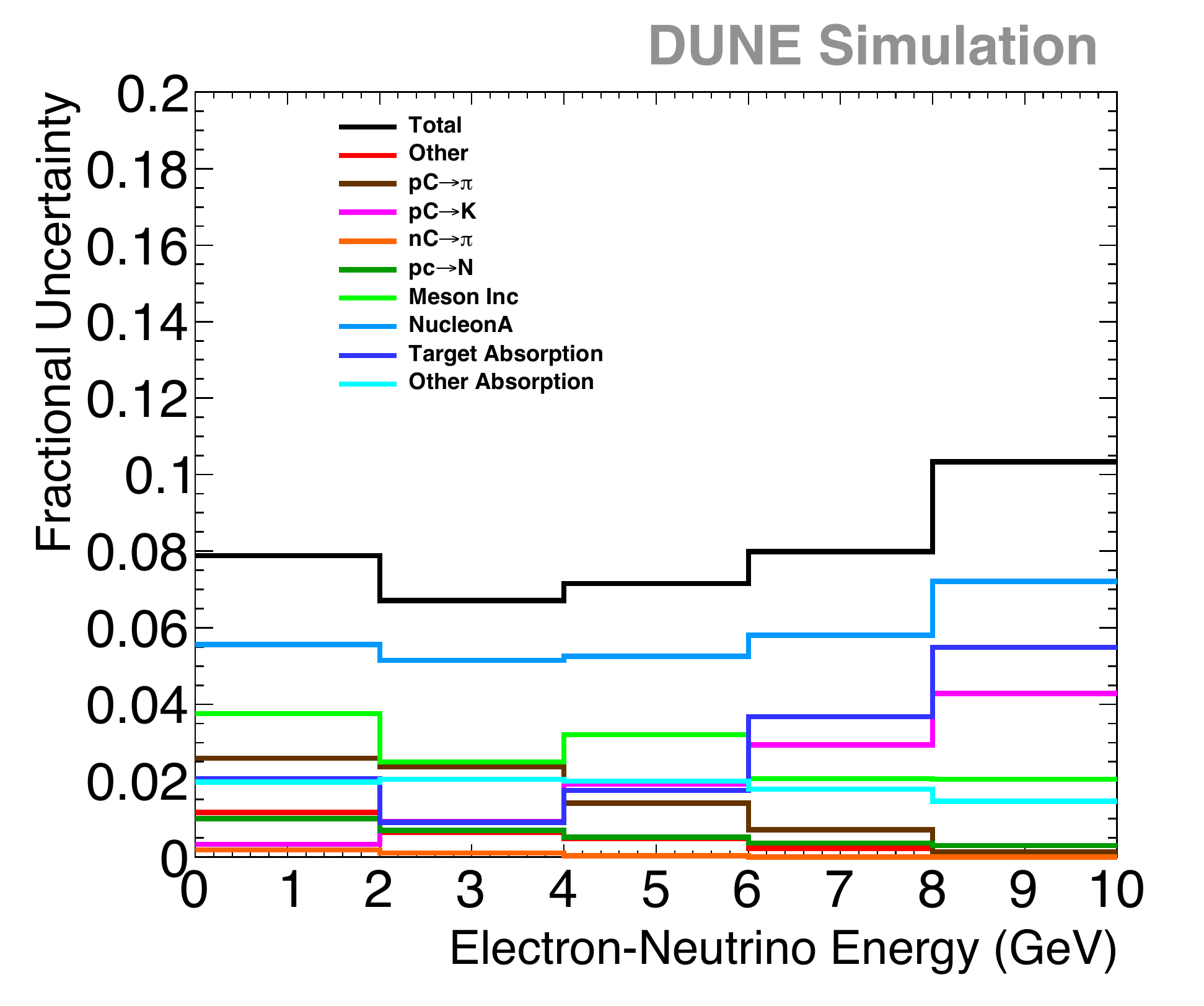} \\
\includegraphics[width=0.48\textwidth]{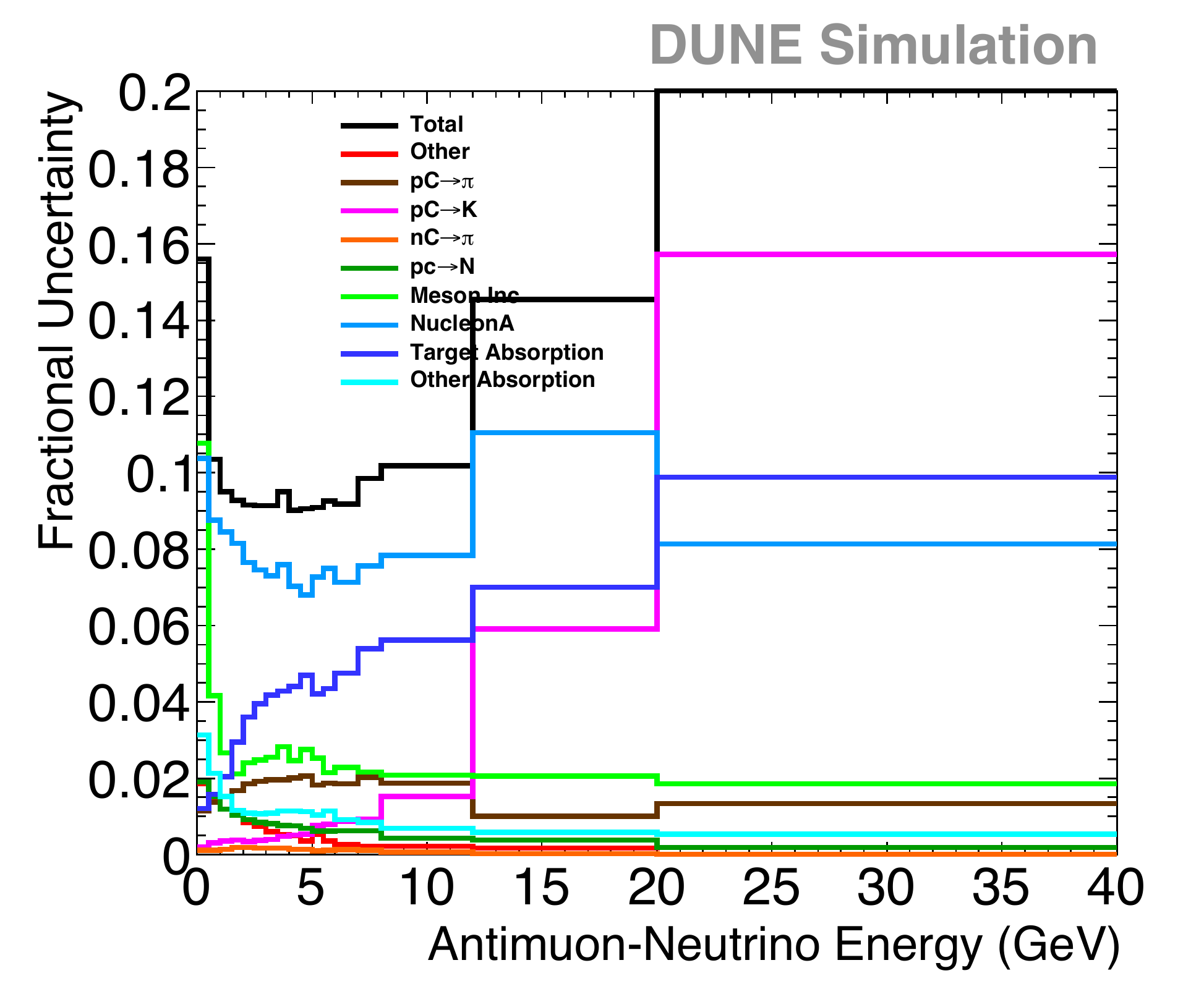}
\includegraphics[width=0.48\textwidth]{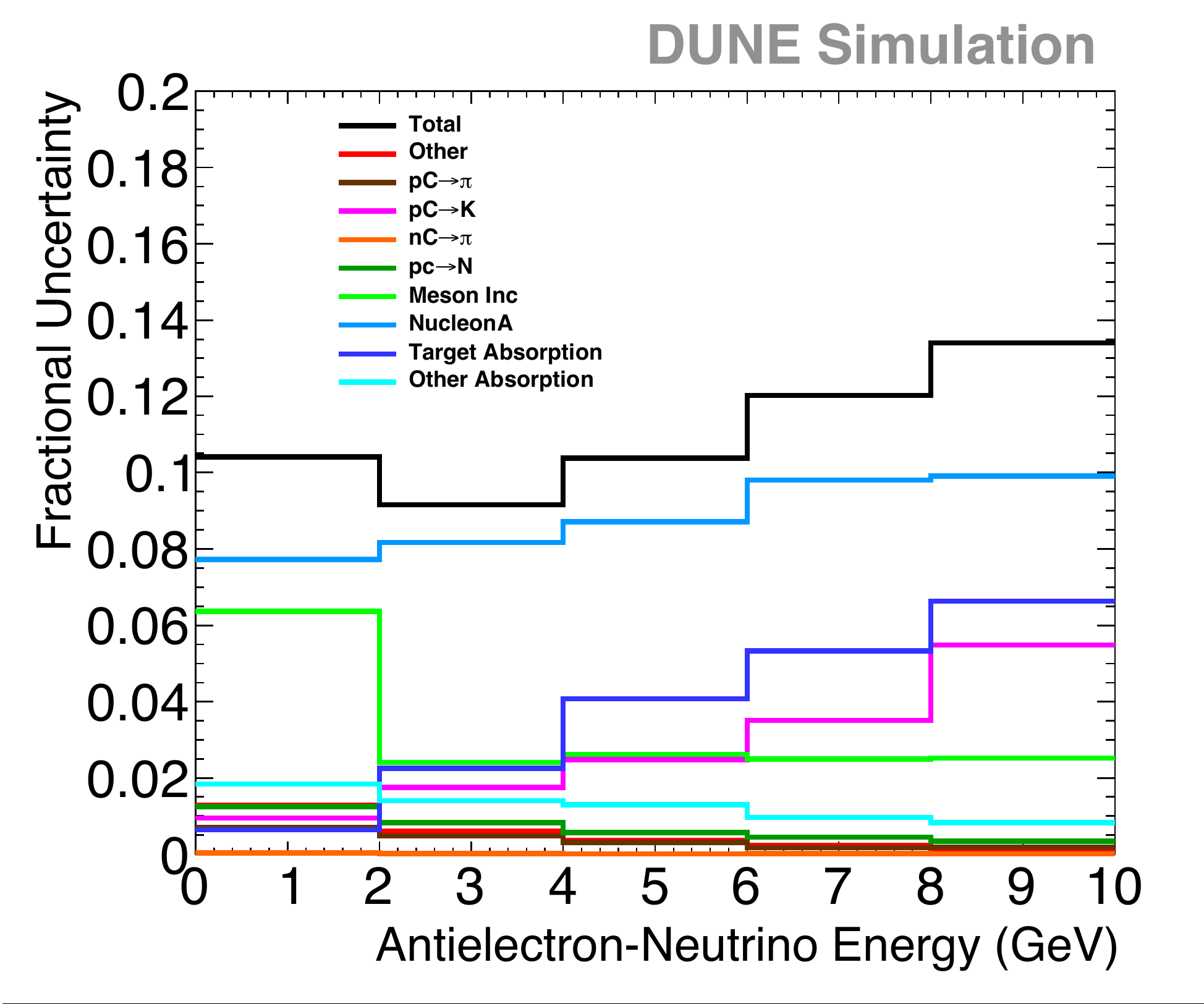} 
\caption{Fractional uncertainties of the $\nu_\mu$ (top left), $\nu_e$ (top right), $\bar{\nu}_\nu$ (bottom left) and $\bar{\nu}_e$ (bottom right) fluxes as a function of neutrino energy in the on-axis LBNF neutrino (forward-horn-current) beam due to hadron production uncertainties \cite{DUNEFluxCurrent}.}
\label{fig:DUNEFlux}
\end{center} 
\end{figure*}

Conventional accelerator-based neutrino beams for oscillation experiments are produced by colliding 10-100 GeV-scale-energy protons on thick ($>1$ interaction length) carbon or beryllium targets.  A series of pulsed toroidal magnets (horns) are used to focus positive [negative] hadrons into a decay volume, where pions and kaons produced in the target and horns decay, producing a predominantly $\nu_\mu$ [$\bar{\nu}_\mu$] beam.  

The energy spectrum of the neutrino beam is rather broad, and the individual neutrino energy is unknown prior to a $\nu-A$ interaction.  Accelerator-based neutrino experiments validate their neutrino energy measurement by comparisons of their observed energy distribution with predicted energy spectra based on Monte Carlo simulations including the production, transport and decay of hadrons produced in the target and surrounding materials.  These simulations are constrained and improved using external data and any in-situ measurements.  Examples of external data are measurements by dedicated hadron production experiments and test-stand measurements of the toroidal focusing magnetic fields.  Examples of in-situ measurements are the position and angle of the primary proton beam on the target, and measurements of the surviving hadron and muon spatial distribution at the end of the decay volume.

T2K and the planned Hyper-K experiment measure neutrino oscillations with the J-PARC neutrino beam created from the collision of 30~GeV protons on a graphite target.  Fermilab currently operates two neutrino beams, NuMI and BNB.  The NuMI beam is created with collisions of 120~GeV on a graphite target, and the BNB beam is created with collisions of 8~GeV protons on a beryllium target.  The planned Fermilab-based DUNE experiment will use a new neutrino beam created with collisions of 60-120 GeV protons on a graphite target.  In all cases, the produced hadrons can undergo further interactions in the target, the horn material and the decay volume walls.  

The dominant systematic uncertainty on the predicted neutrino flux for all accelerator-based experiments arises from uncertainties in hadron production cross sections.  Figures \ref{fig:NOvAFlux} and \ref{fig:DUNEFlux} show the fractional systematic uncertainty in the flux as a function of neutrino energy for the NuMI beam at the off-axis NOvA near detector and the LBNF on-axis beam at the DUNE far detector, respectively.  The uncertainties are drawn separately for the different kinds of hadronic interactions that produce neutrinos seen by each experiment.  In both cases, the hadron production has been corrected using external hadron production data, primarily from the NA49~\cite{NA49} and NA61/SHINE~\cite{NA61} experiments via the prescription described in \cite{PPFX}.  Although the largest contribution to the flux is from pions produced in the primary interaction $p+C$, this pion production mode has been measured at the level of a few percent by the NA49 and NA61/SHINE experiments \cite{NA49}.  Instead, the dominant uncertainties in the predicted fluxes are from uncertainties of quasi-elastic scattering (deflections off of a single nucleon) and particle production in secondary and tertiary interactions in the target for which no or very limited hadron production data exist.  NA61/SHINE at CERN plans to measure more hadron production cross sections at energies above 15 GeV.  However, the beamline for that experiment is not designed for momenta below 15 GeV, and although TPCs have been added to cover the forward-tracking region, it is unclear yet if these will enable precision quasi-elastic scattering measurements.  For the NuMI and LBNF flux, 40\% uncertainties in cross sections are assumed for these secondary and tertiary interactions in the target \cite{PPFX}.  

\begin{figure*}[hbt]
\begin{center}
\includegraphics[width=0.48\textwidth]{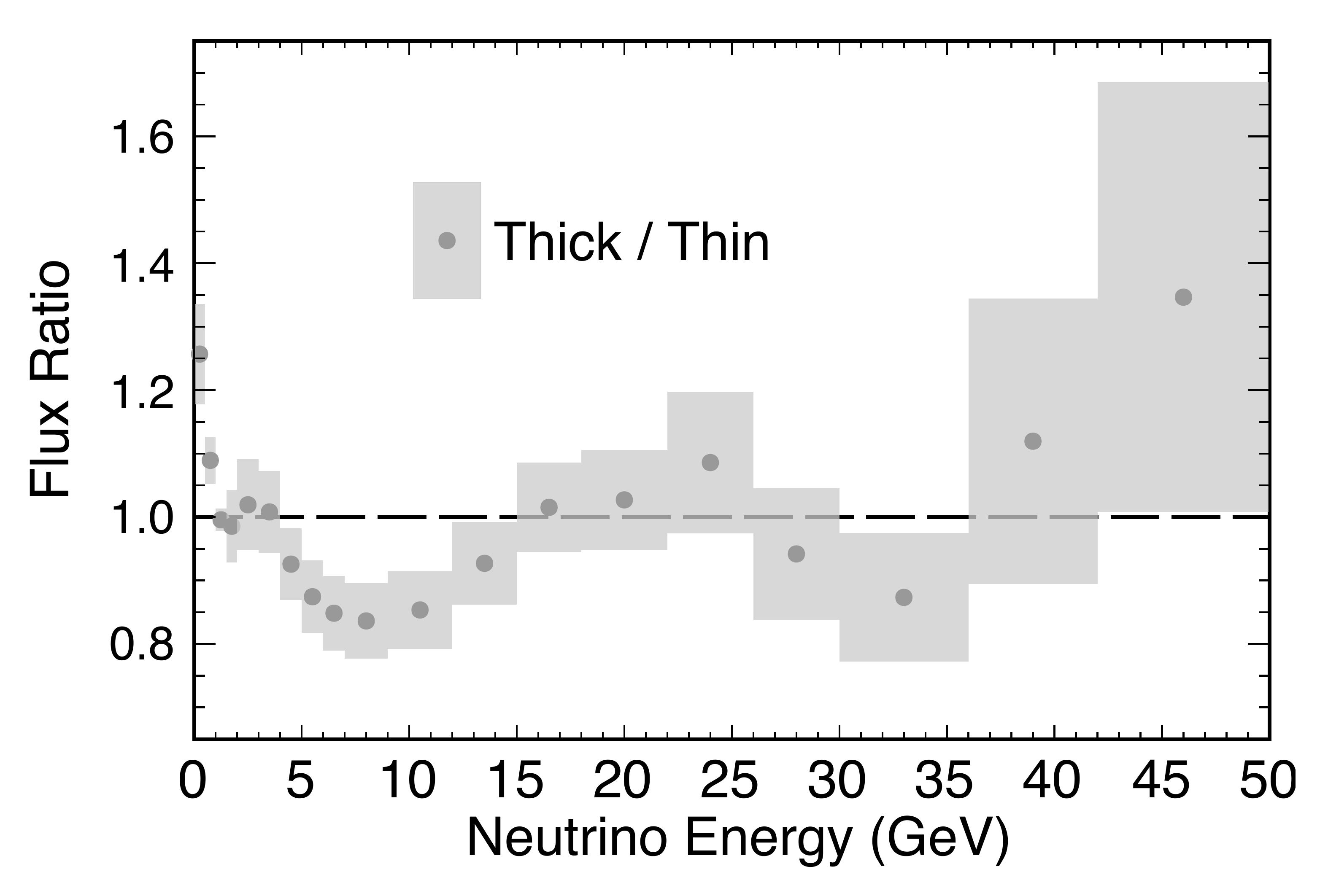}
\includegraphics[width=0.48\textwidth]{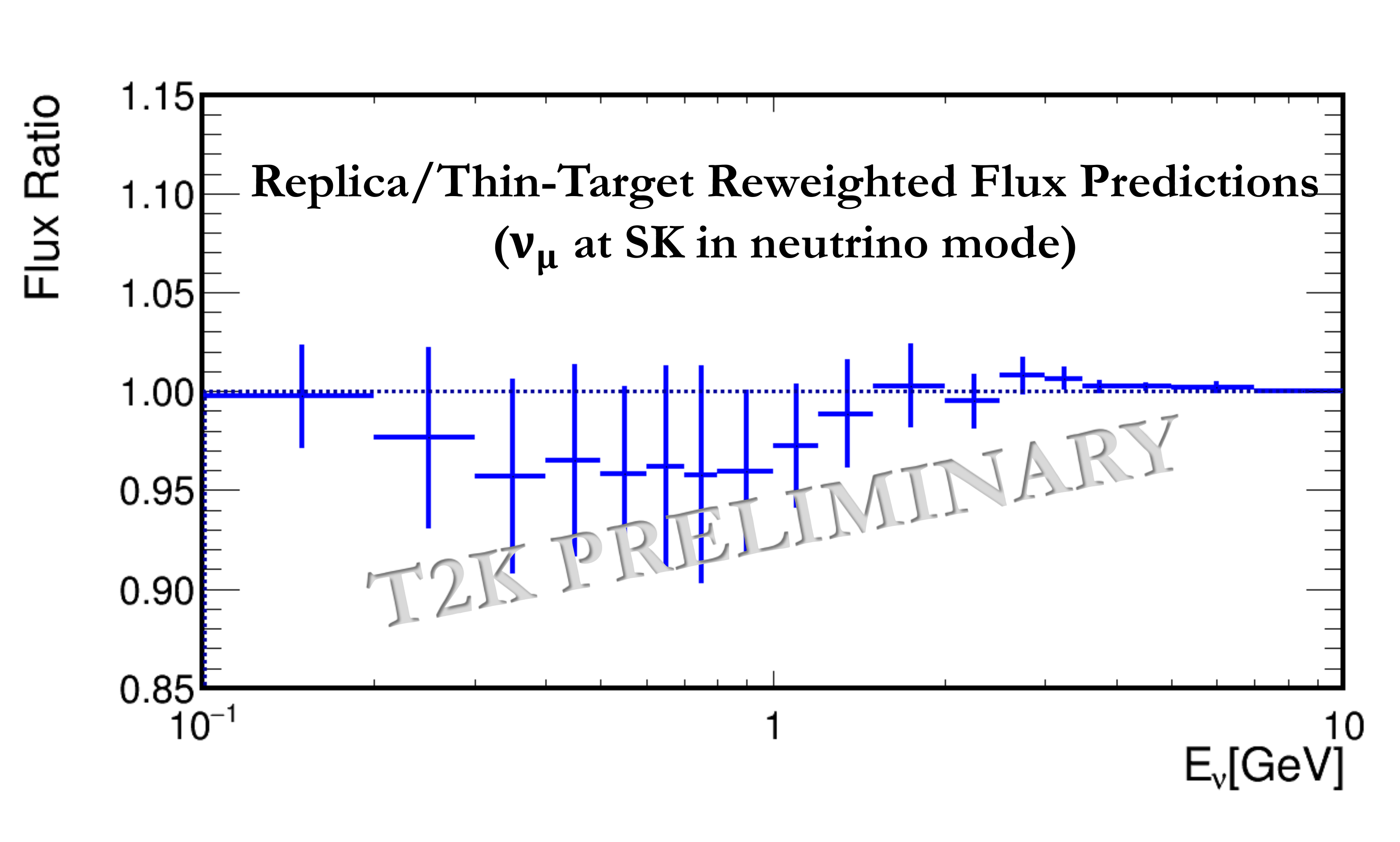}
\caption{Comparisons of flux predictions from replica target measurements to thin-target measurements for the NuMI (left) on-axis beam \cite{PPFX} and the T2K off-axis beam \cite{T2K-2009NA61}}.
\label{fig:ThinThickComp}
\end{center} 
\end{figure*}

NA61/SHINE has published pion and kaon production measurements using a T2K replica target \cite{Abgrall:2016jif}, has recently collected data on a replica NuMI medium energy target, and has proposed future similar measurements for the LBNF target.  Using such precise replica target measurements to constrain flux predictions is preferable to using thin-target measurements, since the secondary interactions inside of the target do not have to be modeled.  However, the timescale for when such measurements will be completed is unknown.  The hadron production yield measurements for the NuMI low energy production target by the MIPP experiment and the T2K production target by the NA61/SHINE experiment took 7-8 years to complete.  

Furthermore, both MINERvA and T2K experiments have reported a ~5-10\% discrepancy between flux predictions based on these replica target measurements and predictions based on thin-target measurements, as shown in Fig.~\ref{fig:ThinThickComp}.  In the case of the MINERvA's observation, the replica (``thick'') data are from the MIPP experiment \cite{MIPP}, and the replica data for T2K were collected in the NA61/SHINE experiment \cite{T2K-2009NA61}.  The thin target data used in the predictions for the MINERvA and T2K fluxes were highly correlated.  Since these thin target measurements are used to tune hadron production models used in simulations throughout the HEP field, it is important to make independent measurements.

\begin{table}[tb]
\begin{center}
\caption{Fraction of simulated hadronic interactions in the T2K flux that are tuned by replica or thin target data~\cite{T2KFluxTuning}.}
\begin{tabular}{l|*{4}{>{\raggedleft\arraybackslash}p{3em}}}
\hline
 & \multicolumn{4}{c}{Fraction of Hadronic Interactions}\\ \hline
Horn Mode & $\nu_{\mu}$ & $\bar{\nu}_{\mu}$ & $\nu_{e}$ & $\bar{\nu}_{e}$  \\ \hline
Neutrino Mode & 0.97 & 0.87  & 0.91 & 0.77 \\
Antineutrino Mode & 0.87 & 0.96 & 0.77 & 0.92 \\ \hline
\end{tabular}
\label{tab:t2kfluxtuned}
\end{center}
\end{table}

Although replica target measurements can be used to significantly reduce the uncertainty of the main component of the neutrino beam flux, other components of the beam are not well constrained from these measurements.  Table~\ref{tab:t2kfluxtuned} shows the fraction of simulated hadronic interactions in the T2K target that are tuned to measurements by NA61/SHINE~\cite{Abgrall:2015hmv,Abgrall:2016jif}, including the replica target measurements, and HARP experiment thin target measurements~\cite{Apollonio:2009bu}.  For the $\nu_\mu$ and $\bar{\nu}_\mu$ flux from focused pions and kaons, about 96\% of interactions are tuned by hadron production data.  Measurements of the remaining 4\% will ensure significant mis-modeling in the unmeasured region does not bias the flux prediction.  However, the intrinsic $\bar{\nu}_\mu$ component of the beam born from defocused pions and kaons are important background sources for searches of CP violation in all long-baseline neutrino oscillation experiments.

Measurements of the intrinsic $\nu_e$ and $\bar{\nu}_e$ interactions in near detectors will also be used to measure differences between the electron- and the muon-neutrino interaction cross section~\cite{E61Proposal}.  A precise prior calculation of the $\nu_e$ and $\bar{\nu}_e$  fluxes is needed.  The fractions of hadronic interactions in the $\nu_e (\bar{\nu}_e)$ T2K fluxes that are not tuned by data range from 9\% to 23\% of the total.  Most of the modeled interactions that are not tuned by data involve the scattering and production of a kaon. To make precise relative measurements of neutrino interaction cross sections, these kaon scattering processes should be constrained by measurements.

\begin{figure*}[htbp]
\begin{center}
\includegraphics[width=0.98\textwidth]{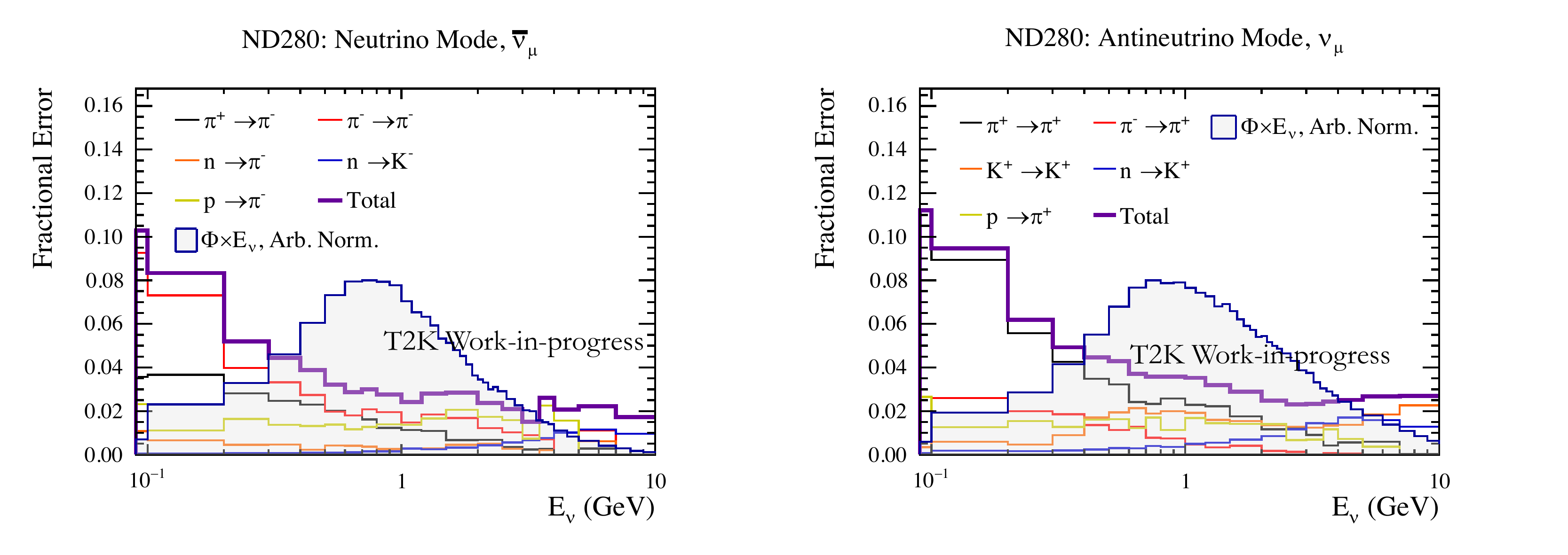}
\caption{Fractional uncertainties of the wrong-sign flux (i.e. $\bar{\nu}_\mu$ in neutrino-mode and $\nu_\mu$ in anti-neutrino mode), in the T2K near detector, located 2.5 degrees off axis from the beam center, as a function of neutrino energy, due to untuned hadronic interactions~\cite{T2KFluxTuning}.}
\label{fig:T2KUntunedInteractionErrors}
\end{center} 
\end{figure*}

In the prediction of the $\nu_\mu$ background in the $\bar{\nu}_\mu$-dominant beam, 13\% of hadron interactions are not tuned by measurements, and most of these are interactions of pions in the beam line material. The fractional uncertainty on this ``wrong-sign'' flux calculation due to these unconstrained interactions is shown in Figure \ref{fig:T2KUntunedInteractionErrors} for T2K to be 3\% around the $\bar{\nu}_\mu$ flux peak and 4\% in the low energy flux tail in antineutrino mode. In antineutrino mode operation of the beam, this wrong-sign background is enhanced by the larger neutrino interaction cross section compared to the antineutrino interaction cross section (see Figure \ref{fig:T2KUntunedInteractionErrors}). The wrong-sign muon neutrinos can undergo oscillations to an electron neutrino that is indistinguishable in a water Cherenkov detector from an electron antineutrino from muon antineutrino oscillations.  An incorrect modeling of this background will bias a measurement of CP asymmetry. 

The Booster neutrino beam line provided neutrino and antineutrino beams for the MiniBooNE experiment and is providing beams for the ongoing short baseline LAr detector program.  The calculation of the MiniBooNE neutrino flux~\cite{AguilarArevalo:2008yp} relies on pion production data from
the HARP~\cite{Catanesi:2007ab} and Brookhaven E910~\cite{Chemakin:2007aa} experiments.  The HARP experiment does not provide kaon production 
data, which is relevant for the $\nu_e$ flux calculation, so data from a  number of experiments with beam momenta ranging from 9.5~GeV/c to 24.0~GeV/c~\cite{Aleshin:1977bb,Eichten:1972nw,Dekkers:1965zz,Abbott:1991en,Marmer:1971gd,Vorontsov:1988pu,Allaby:1970jt} are fitted with
an empirical parameterized model, which is used to model kaon production for the flux calculation.  The largest contribution to the 
uncertainty on the total $\nu_{\mu}$ flux is
14.7\% and arises due the uncertainty on the modeling of pion production.  The $\nu_e$ flux uncertainty has a dominant contribution of 11.5\%
from the uncertainty on the modeling of $K^{+}$ production.

\begin{figure}
\begin{center}
\includegraphics[width=0.40\textwidth]{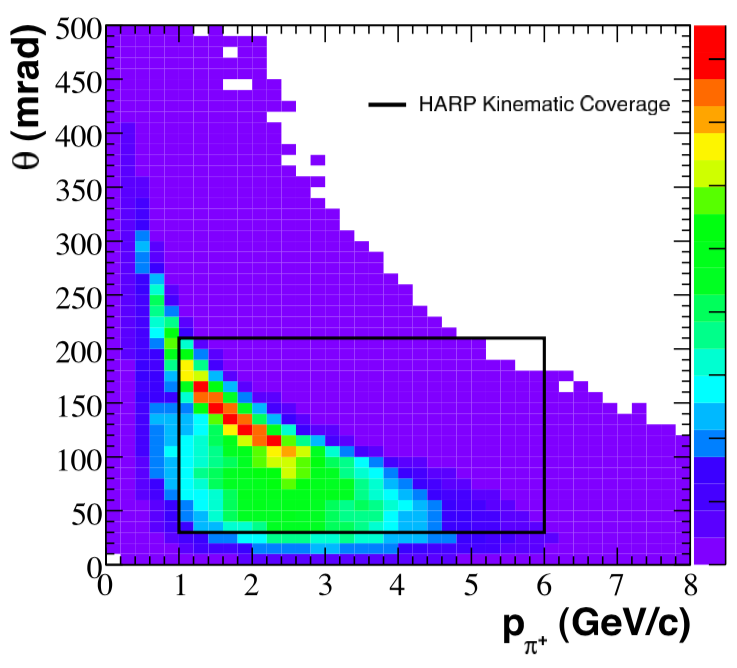}
\caption{Figure taken from~\cite{AguilarArevalo:2008yp} shows the parent pion phase space for the $\nu_{\mu}$ flux prediction at the MiniBooNE
detector.  The black box contains the region covered by HARP measurements~\cite{Catanesi:2007ab}}.
\label{fig:harp_booster_flux}
\end{center} 
\end{figure}

While  HARP pion production data were collected with a Be target at the
Booster proton beam momentum of 8.9~GeV/c, the data does not cover the full phase space needed for the Booster neutrino flux, as shown in 
Fig.~\ref{fig:harp_booster_flux}.  The MiniBooNE flux calculation uses an empirical parameterized model fitted to the HARP and E910 data 
in order to extend the covered phase space.  However, it is necessary to increase the data uncertainties by 35\% to achieve an acceptable 
goodness of fit.  The 
EMPHATIC measurements will expand on the HARP phase space to cover the pion momentum range from 0.4~GeV/c to $>$8~GeV/c and the pion scattering 
angle from 0~mrad to 350~mrad.  This will cover the full phase space for the Booster neutrino flux, and eliminate the need for a fit to
multiple data sets with error inflation.  Hence, we expect that the uncertainty from pion production modeling can be reduced by at least 35\%.

\begin{figure}
\begin{center}
\includegraphics[width=0.48\textwidth]{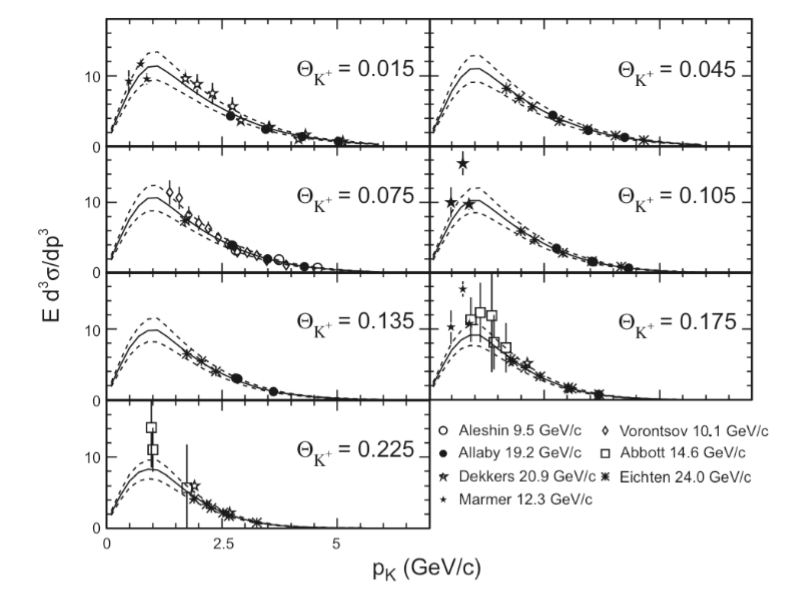}
\caption{Figure taken from~\cite{AguilarArevalo:2008yp} shows the fit to kaon production data 
sets~\cite{Aleshin:1977bb,Eichten:1972nw,Dekkers:1965zz,Abbott:1991en,Marmer:1971gd,Vorontsov:1988pu,Allaby:1970jt}}.
\label{fig:miniboone_kaon_fit}
\end{center} 
\end{figure}

The fit to kaon production data, shown in Fig.~\ref{fig:miniboone_kaon_fit}, is used to model the kaon production in the
Booster neutrino flux calculation.  To achieve a consistent fit to the data, one potential data set is excluded due to inconsistencies and
another has a 500\% normalization uncertainty applied.  An additional scaling of all data errors by 50\% is applied to achieve an acceptable 
goodness of fit.  As shown in Fig.~\ref{fig:miniboone_kaon_fit}, the data at low momentum ($<$1.5~GeV/c) is sparse and generally lies
above the fitted empirical model.  The EMPHATIC experiment will measure charged kaon production over the full phase space relevant
for the Booster neutrino flux, including the sparsely covered low momentum region.  We expected that the uncertainty from kaon production 
modeling can be reduced by at least 50\%.  The EMPHATIC data will also remove the reliance on Feynman scaling assumptions that are made
in order to apply data from different beam energies to this calculation.

\subsection{Atmospheric Flux Predictions}

Atmospheric neutrinos are the byproduct of primary cosmic ray interactions with nuclei in the atmosphere. 
Though the basic production mechanism is the same as for accelerator neutrinos, there are significant differences in the details. 
The primary cosmic ray flux is dominated by protons but has contributions from heavier nuclei, such as helium, and spans many decades in energy in contrast to the well-controlled beam energy at an accelerator. 
Since air is composed of oxygen and nitrogen, the interaction targets differ from accelerators.  The production of primary kaons is important to the $\bar{\nu}_e$ flux. The ratio of $\pi/K$ production is estimated at 10\%, and currently has a pull of nearly that size in the Super-K atmospheric neutrino fit\cite{SKLatestAtm}.
 
Atmospheric neutrino production is a shower process, and therefore compared to accelerator beam production, the secondary interactions of hadrons down to low energies is a crucial element.
The Earth itself acts as an absorber for muons that have enough energy to reach the surface, meaning that the atmospheric neutrino flux below O(10)~GeV has a significant $\nu_{e}+\bar{\nu}_{e}$ component. 
Perhaps most significantly, it is not possible to make \textit{in-situ} measurements of the primary cosmic ray or secondary meson spectrum and as a result atmospheric neutrino flux calculations are even more dependent upon external data sets than accelerator neutrino calculations. 
Indeed, calculations rely on dedicated cosmic ray experiments to measure the flux and particle content of the primary and secondary meson fluxes.  Hadron production measurements are used to tune MC simulations of its subsequent interactions in the air. 

In practice this means that uncertainties in atmospheric neutrino fluxes are directly associated with uncertainties in those external data sets.  Extrapolations are often used to account for gaps in the existing data. 
While the primary cosmic ray spectrum has been measured to O(1)\% by AMS02 and BESS, the atmospheric neutrino flux below 20~GeV suffers from large uncertainties stemming from deficiencies in the exiting hadron interaction and cosmic ray muon data.  Although it is possible to tune the hadronic interaction generators used in flux calculations using muon data, as is done in the Honda model, those generators are themselves built upon hadron production data.
Since the atmospheric neutrino flux below 2 GeV is sensitive to CP violation and that between 2 and 10 GeV is sensitive to the  mass hierarchy, improved hadron production data is of central interest for oscillation measurements with this source.  

A complication arises from the fact that the hadronic phase space for producing a neutrino of given energy and direction is large. 
Indeed the CP-sensitive flux has sizeable contributions from the $1 \sim 20$~ GeV/c mesons produced at O(100) mrad relative to the primary projectile. 
Existing hadron production data is particularly lacking in this region and as a result 30\% uncertainties accompany the mesons used to compute the low energy atmospheric neutrino flux.  
Accordingly the atmospheric neutrino flux carries absolute uncertainties of between 10\% and 25\% in this energy range and similarly the ratio of electron neutrino to antineutrinos is known only to 5\%. 
Both of these can be improved with measurements of $<10$ GeV/$c$ $\pi^{\pm}$ produced in the interactions of protons below 20 GeV.

\begin{figure*}[htpb]
\begin{center}
\includegraphics[width=0.48\textwidth]{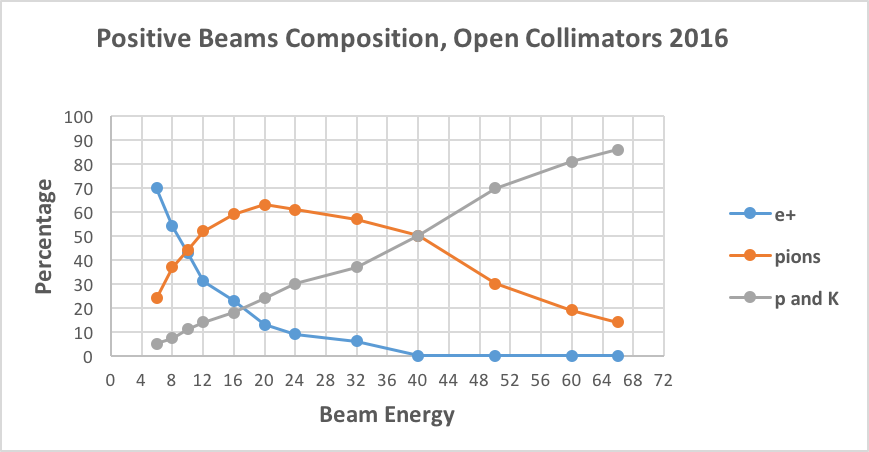}
\includegraphics[width=0.48\textwidth]{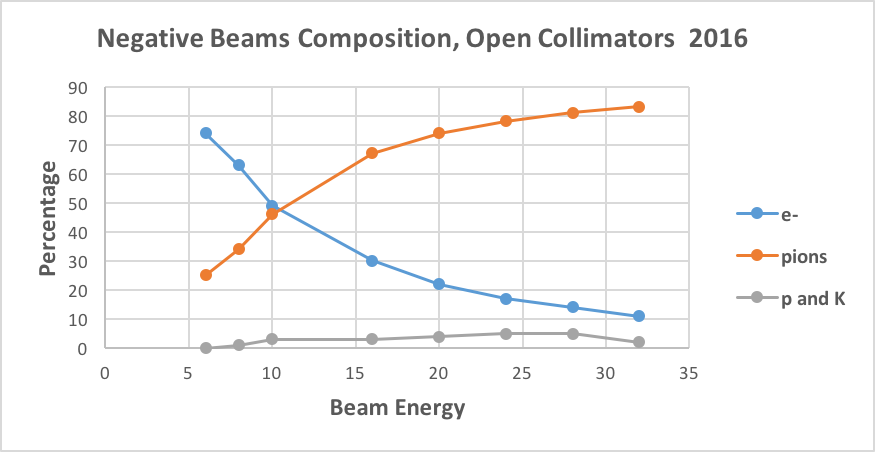}
\caption{FTBF beam particle fractions as a function of beam momentum for positive (left) and negative (right) beams \cite{FTBF}.}
\label{fig:FTBFBeam}
\end{center} 
\end{figure*}

\subsection{Predictions of Backgrounds for Rare Processes}
The suppression of backgrounds is a key driver in the design of experiments that are searching for rare processes.  The specifications of the detectors in such experiments are often determined by simulation studies of known background processes.  Uncertainties in hadron production translate directly into uncertainties in the rates of the background processes.  Conservative uncertainties are typically assumed in the design of the experiment in order to ensure success of the experiment.  Reliable estimates and reduced uncertainties of these rates can significantly reduce the cost of an experiment and build confidence in the design sensitivities.

A case in point is the J-PARC E50 experiment~\cite{Noumi}, which was proposed to investigate charmed baryons at the J-PARC high-momentum beam line using an intense secondary hadron beam. 
The experiment aims to use the missing mass technique to systematically measure the excitation energy, production rates and decay products of charmed baryons via the $\pi^{-}\;{+}\;p \rightarrow Y^{*+}_c\;{+}\;D^{*-}$ reaction at 20 GeV/$c$.

The decay chain of the $D^{*-}$ meson, $D^{*-} \rightarrow \bar{D}^0\;{+}\;\pi^{-}$ (branching ratio of 67.7\%) 
and $\bar{D}^0 \rightarrow K^{+}\;{+}\;\pi^{-}$ (branching ratio of 3.88\%), 
is detected by using a spectrometer~\cite{CBS} for reconstructing the invariant mass of the $D^{*-}$ and $\bar{D}^0$ mesons.
One of the experimental challenges is to measure the production events of charmed hadrons 
in the multitude of background events having the same final states of ($K^+$, $\pi^-$, $\pi^-$) as the decay chain of the $D^{*-}$ meson. 
While the cross section of charmed baryon production is expected to be order of 1 nb~\cite{Hosaka}, 
the background of the ($K^+$, $\pi^-$, $\pi^-$) state is estimated to be $\sim$2.4 mb. 
The main contribution of the background is the multi-meson production including the strangeness.
Strangeness production background processes were simulated by the JAM code~\cite{JAM}, 
while the PYTHIA code~\cite{PYTHIA} was also used for the comparison.
The most effective method of background reduction is via $D^*$ tagging.
Both the mass region of the $\bar{D}^0$ mass and the Q-value ($Q=M(K^+\pi^-\pi^-)-M(K^+\pi^-)-M_{\pi^-}$) corresponded to the $\bar{D}^{*-}$ decay event are selected.
By using the $D^*$ tagging, 
the background reduction of $\sim2\times10^6$ in the total invariant mass region from the $K^+$ and $\pi^-$ reconstruction can be achieved from the JAM simulation result. 
Around the mass region of $\bar{D}^0$ (1.84$-$1.87 GeV/$c^2$), the background reduction of $\sim$500 was achieved.

The cross section of the background and its reduction were only estimated from simulations and calculations by using the hadronic reaction models.
However, there are factors of 2$-$3 differences by comparing between different hadronic reaction models. 
The background level depends on the cross section and the charged track multiplicity which causes combinatorial background.
The background rates, and the approach to reducing the number of selected backgrounds described above, can be measured and tested in an experiment such as the one described in this paper.

\section{EMPHATIC: Experiment to Measure the Production of Hadrons At a Testbeam In Chicagoland \label{sec:Experiment}}
The EMPHATIC Collaboration proposes a series of precise measurements of hadron scattering and production measurements using a compact, table-top-size spectrometer located at the Fermilab Test Beam Facility (FTBF).  The measurements will be made on a broad range of nuclear targets and beam momenta that are particularly relevant for GeV-scale neutrino production.
In this section we describe the beam, beam particle id, spectrometer and secondary particle id systems of the proposed experiment.  The EMPHATIC detectors are based on well established technologies and either already exist, or are readily constructed.

\subsection{Beam and Beam Particle Id }

The FTBF provides beams of particles between 2-120 GeV/c.  Details of the beam at the FTBF can be found online\cite{FTBF}, but we briefly summarize the pertinent features here.  The beam is delivered over $\sim4$ seconds, once per minute.  The 120 GeV/c beam is extracted directly from the Main Injector, and is therefore a pure proton beam.  Secondary beams of pions, kaons, protons and electrons are available at momenta as low as 2 GeV/c. The intensity, spot size and momentum spread of the beam are all tunable, with highest particle rates over 100 kHz, typical spot sizes of 1-2 cm$^2$, and $\Delta p/p \sim$ 2-3\%.   

The FTBF provides particle identification via gas Cherenkov detectors.  Figure \ref{fig:FTBFBeam} shows typical beam composition determined from data collected from the Cherenkov detectors filled with nitrogen, with the pressure adjusted to optimize particle separation.  Pion identification is achievable at 5 GeV, but kaon-proton separation is not achievable below approximately 18 GeV.  

\subsubsection{Beam Aerogel Cherenkov Detector }
\begin{figure}[htbp]
\begin{center}
\includegraphics[width=0.48\textwidth]{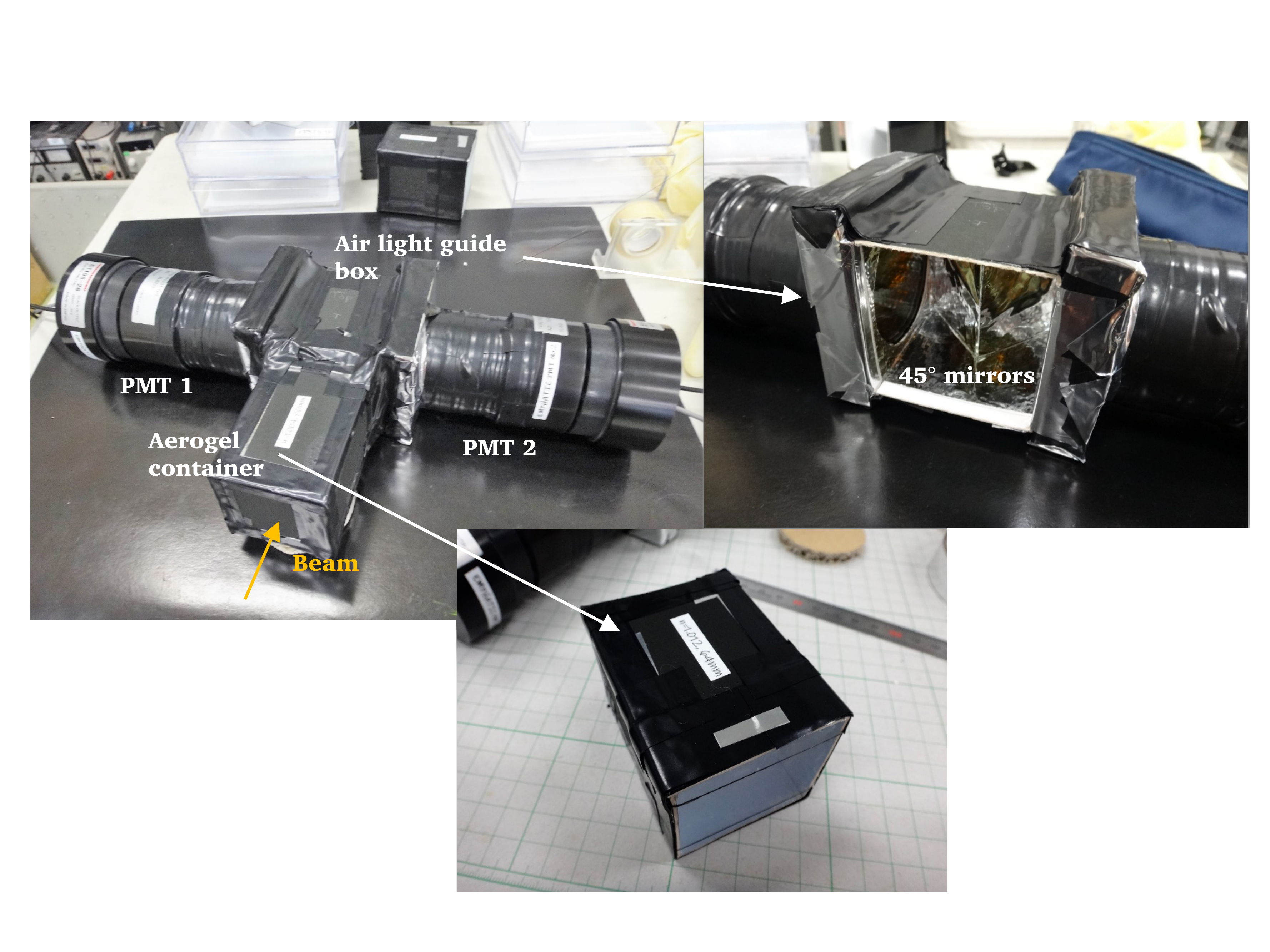}
\caption{Photo of a Beam Aerogel Ckov detector prototype.}
\label{fig:BACkov}
\end{center} 
\end{figure}

To extend the  range of beam particle identification to lower momenta, EMPHATIC will employ a beam aerogel Cherenkov (BACkov) detector between the gas Cherenkov detectors and the EMPHATIC spectrometer.  The BACkov is an array of threshold-type Cherenkov counters that use silica aerogels as the radiators. The array will consist of counters with different Cherenkov thresholds (i.e., aerogels with different refractive indices, $n$) to distinguish protons from kaons and pions between 4--12 GeV/$c$. Aerogels with $n$ = 1.026, 1.007, and 1.003 will be employed for the proton/meson separation at 4, 8 and 12 GeV/$c$, respectively. Each of the counters are equipped with 1--3  tubes (PMTs), depending on the expected number of Cherenkov photons emitted from aerogel radiators with different indices. A prototype of a BACkov counter is shown in Fig.~\ref{fig:BACkov}.  The fringe magnetic field at the location of the PMTs is expected to be low (eg, $<$ 25 Gauss), so no additional shielding will be required to operate the PMTs.  However, PMTs with high quantum efficiency are needed in particular for efficiently detect photons emitted from ultralow index (i.e., low density) aerogels. The lateral size of a single-layer aerogel block will be 5 cm $\times $ 5 cm to cover beams with a diameter of approximately 3 cm. Aerogel blocks with a thickness of 2 cm are stacked so that the total thickness is 6--16 cm or more (the number of the aerogel layer will be optimized to maximize the photon detection efficiency). Cherenkov photons are directly transported to PMT(s) using reflector plates lined with aluminized mylar sheets. 

The operation of the $n$ = 1.003 ultralow index aerogel counter has in the past been a great challenge, since the scattering length of such low-index aerogels was $\sim 1$ cm, resulting in very low photon yield.  Much higher transparency hydrophobic silica aerogel blocks have been produced at Chiba University with new fabrication techniques, enabling the practical use of such aerogels for applications in experiments \cite{Aerogel1,Aerogel2}.  Test beam measurements at the ELPH Laboratory, Tohoku University have demonstrated an average photoelectron yield of $\sim 5$ for a 16-cm thick aerogel detector of $n$ = 1.003, resulting in a $\sim 99\%$ detector efficiency for $\beta \sim 1$ particles.  

The BACkov counters will be installed on the upstream side of the target (and SSD array) in the beam line. In combination with the upstream gas Cherenkov counters ($n \sim $ 1.001) at the FTBF, we will be able to separate kaons from pions as well as to separate protons from kaons at 4--12 GeV/$c$.

\subsection{Spectrometer Overview}

\begin{figure*}[htpb]
\begin{center}
\includegraphics[width=\textwidth]{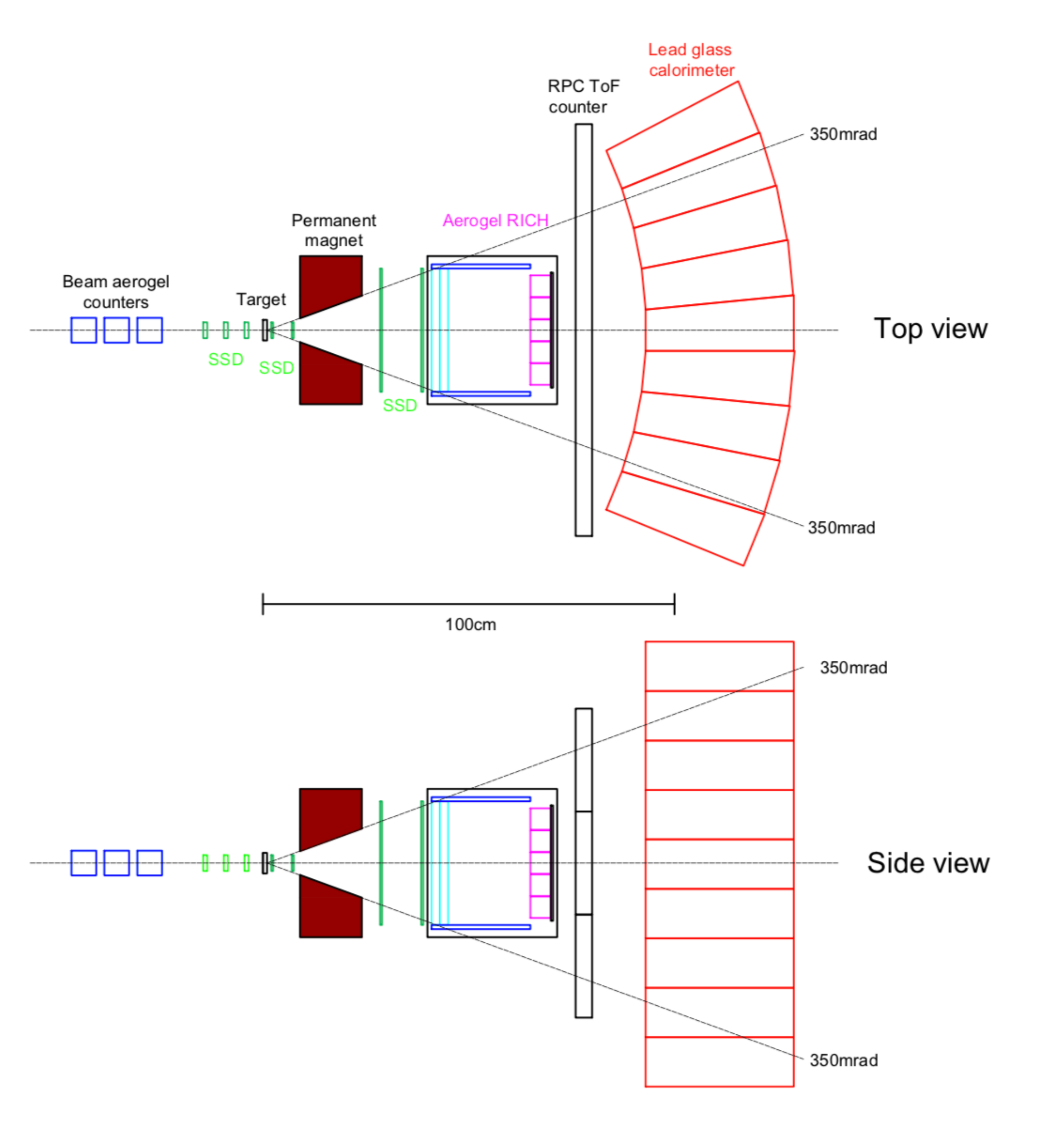}
\caption{Schematic of the configuration for the EMPHATIC spectrometer.}
\label{fig:spectrometer}
\end{center} 
\end{figure*}

Figure \ref{fig:spectrometer} is a schematic of the EMPHATIC spectrometer.  The system is less than 2 m in length, which is an order of magnitude smaller than other hadron production experiments, such as HARP, NA49/NA61, and MIPP experiments. The compact system makes it easier to cover large solid angle with small detector components, which helps in reducing systematic uncertainties by reducing the dead materials, in addition to the reduced detector cost. The EMPHATIC spectrometer takes advantage of precise tracking using silicon strip detectors (SSDs) and a high field permanent magnet.  The angular acceptance of the spectrometer is approximately 400 mrad, which covers the region of interest for accelerator-based neutrino beams.  
Particle identification will be done by a compact aerogel ring imaging Cherenkov (ARICH) detector, a time-of-flight (ToF) wall, and a lead glass calorimeter array. 
The EMPHATIC ARICH, based on the Belle II forward end-cap detector, detects Cherenkov photons with multi-anode PMTs and is read out with FPGA-based TDCs.  The ARICH will be capable of $K/\pi$ separation up to 7--8 GeV/$c$ with a multi-track capability. 
Particle identification below 2 GeV/c is covered by the resistive plate chamber (RPC)-based time-of-flight system developed for J-PARC E50 experiment with a timing resolution of 70 psec. The lead-glass calorimeter identifies electrons as well as punch through muons, both of which provide monochromatic energy deposit. Lead glass can also catch gammas and neutrons which are tagged by no hit in the RPC and separated by their arrival time.

\subsubsection{Silicon Strip Detectors}
\label{sec:SiStripDets}

\begin{figure}[htpb]
\begin{center}
\includegraphics[width=0.48\textwidth]{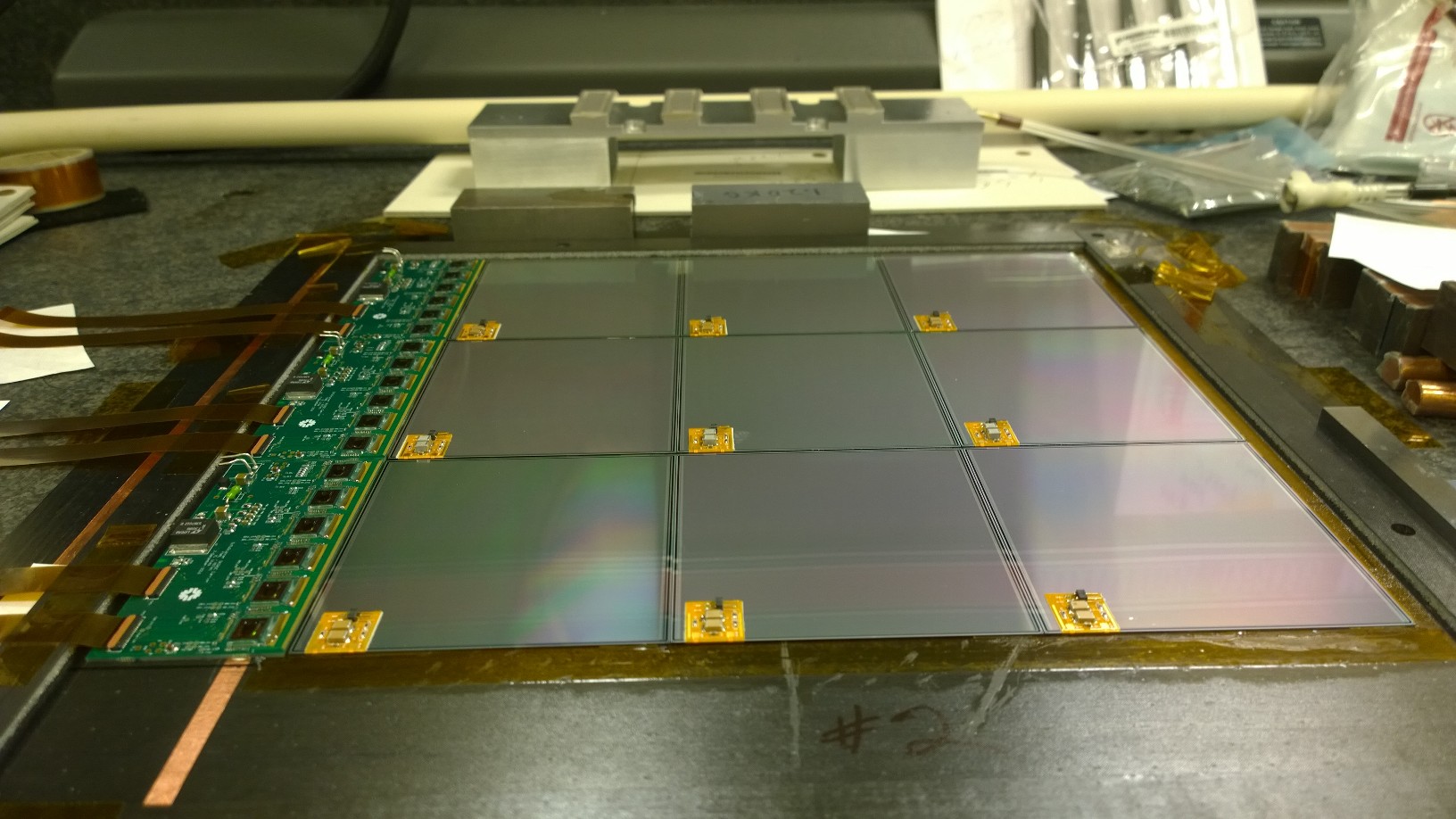}
\caption{30 $\times$ 30 cm$^2$ active area 122 \micron pitch silicon strip detectors, to be used for downstream tracking.}
\label{fig:LASSD}
\end{center} 
\end{figure}

Real-time digitized trajectories of charged particles, both upstream and downstream of the target, will be recorded using silicon strip detectors (SSDs, shown in Fig.~\ref{fig:spectrometer}).  Upstream of the target, SSDs with 25 cm$^2$ active area available at the FTBF can be used.  These detectors are 300 \micron-thick, have a readout track pitch of 60 \micron with 30 \micron pitch intermediate strips.  A test run in January 2018 demonstrated a resolution of better than $10$ \micron position resolution for these detectors.  Further downstream, large-area $30 \times 30$ cm$^2$ SSDs with 122 \micron pitch will be used (see Fig.~\ref{fig:LASSD}).  The CMS Collaboration has demonstrated a resolution of better than $25$ \micron for these detectors \cite{CMSTracking}.

The SiDet facility at Fermilab has already produced prototypes of the $30 \times 30$ cm$^2$ SSDs.  These detectors are constructed of 9 $10 \times 10$ cm$^2$ silicon wafers, read out using the SKIROC2\cite{SKIROC2} front end chip, with a stand-alone data acquisition system.  Due to the density and size of these chips, every other strip is read out by a chip, and the prototype has only been tested with single-sided readout.  Furthermore, the SKIROC2 chips were designed for readout of calorimetry readout and are not best suited for readout of silicon strip detectors.  Alternative VATAGP7 or VATAGP8 chip sets, which were designed for just this type of application, are also being investigated.  These chips have a matching density and size, so that all strips on one side and are expected to be easier to integrate into the mounting board and read out.  In either case, whether one uses the SKIROC2 or VATAGP7/8, some minor engineering of up to a few months of effort will enable the full readout of the detector.

\subsubsection{Emulsion Detectors }

Further details of interaction features can be studied using nuclear emulsion detectors.  These detectors record charged particle trajectories with sub-micron resolution.  The precision resolution of emulsion detectors reduces the ambiguity of whether hadrons interacted in target or detector material to a negligible systematic uncertainty.  Because of the resolution capability, emulsion detectors have $\sim 4\pi$ acceptance.  

\begin{figure}[htbp]
\begin{center}
\includegraphics[width=0.48\textwidth]{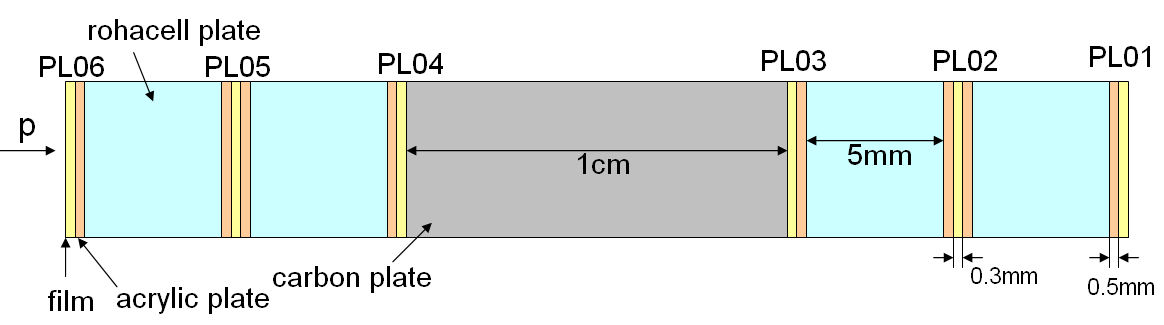}
\caption{Schematic of the emulsion cloud chamber (ECC) that will be used to record fine details of hadron-nucleon final state particle kinematics.}
\label{fig:ECC1}
\end{center} 
\end{figure}

An Emulsion Cloud Chamber (ECC) is constructed as a sandwich structure of thin nuclear emulsion films and the target material. The target material can be material such as carbon, iron, aluminum and water, and ECCs can be constructed with a variety of target thicknesses and shapes.  The ECCs will be positioned in the same location as the target in Fig. \ref{fig:spectrometer}. An automatic scanning system is able to recognize tracks up to $|tan \theta| < $3.0 (where $\theta$ is the track angle with respect to the perpendicular to the emulsion film) \cite{LAtracking1,LAtracking2}. Therefore not only high energy tracks but also nuclear fragments produced in the nuclear evaporation process by an excitation of the target nucleus can be detected with high efficiency \cite{HadronAnaly}.  It is envisioned that some fraction of data collected will include a ECC in the spectrometer configuration.  The ECC will be placed on a motion table that will be moved in between beam spills to avoid overlapping beam trajectories (pileup).

Nuclear emulsion gel, containing 45 or 35$\%$ AgBr crystal by volume ratio is produced at the facility of Nagoya University. Two emulsion layers, each 60-70 $\mu m$ thick, are formed on both surfaces of a 180 or 210 $\mu m$ thick polystyrene plate. As shown in Fig. \ref{fig:ECC1}, three emulsion plates are placed at upstream and downstream of target material to detect incident particles and secondary particles. The angular accuracy of an emulsion film is approximately 2 mrad, but is further improved to 0.1 mrad by installing a rohacell plate (5mm-thickness, density is approximately 0.05 g/cm$^3$) between emulsion films. The ECCs are packed in vacuum to fix the structure and shield the emulsion films from light before beam exposure.


An automatic scanning system \cite{HTS} in Nagoya University is used to measure the energy depositions of charged particles traversing the emulsion films. The alignment between plates is determined from matching energy depositions of cosmic-rays and beam particles. Track segments are separately formed for the plates upstream and downstream of the target, and then connected between upstream and downstream.  Interaction vertices are also reconstructed using the same technique.  Finally all primary and secondary particle trajectories reconstructed in the ECCs are matched to data recorded in the neighboring silicon strip detectors using the pattern matching method, which allows one to extract a time stamp to be used for particle identification.  Studies are underway to determine the maximum beam spill intensity such that pileup in the ECC is a negligible effect.

\subsubsection{Magnet}
\label{sec:Magnet}

A compact spectrometer magnet is required to measure the momentum of secondary particles.  Given the resolution of the SSDs described above, a $B\cdot dl \sim$ 0.2 Tm results in momentum resolutions of better than 7\% below 10 GeV/c, and approximately 10\% at 20 GeV/c.  A cost effective approach to constructing such a magnet is to use a Halbach array\cite{Halbach} of Neodymium permanent magnets. By lining up the orientation of the magnets with the desired field line, one can enhance the magnetic field in the gap region.  

\begin{figure*}[htpb]
\begin{center}
\includegraphics[width=0.95\textwidth]{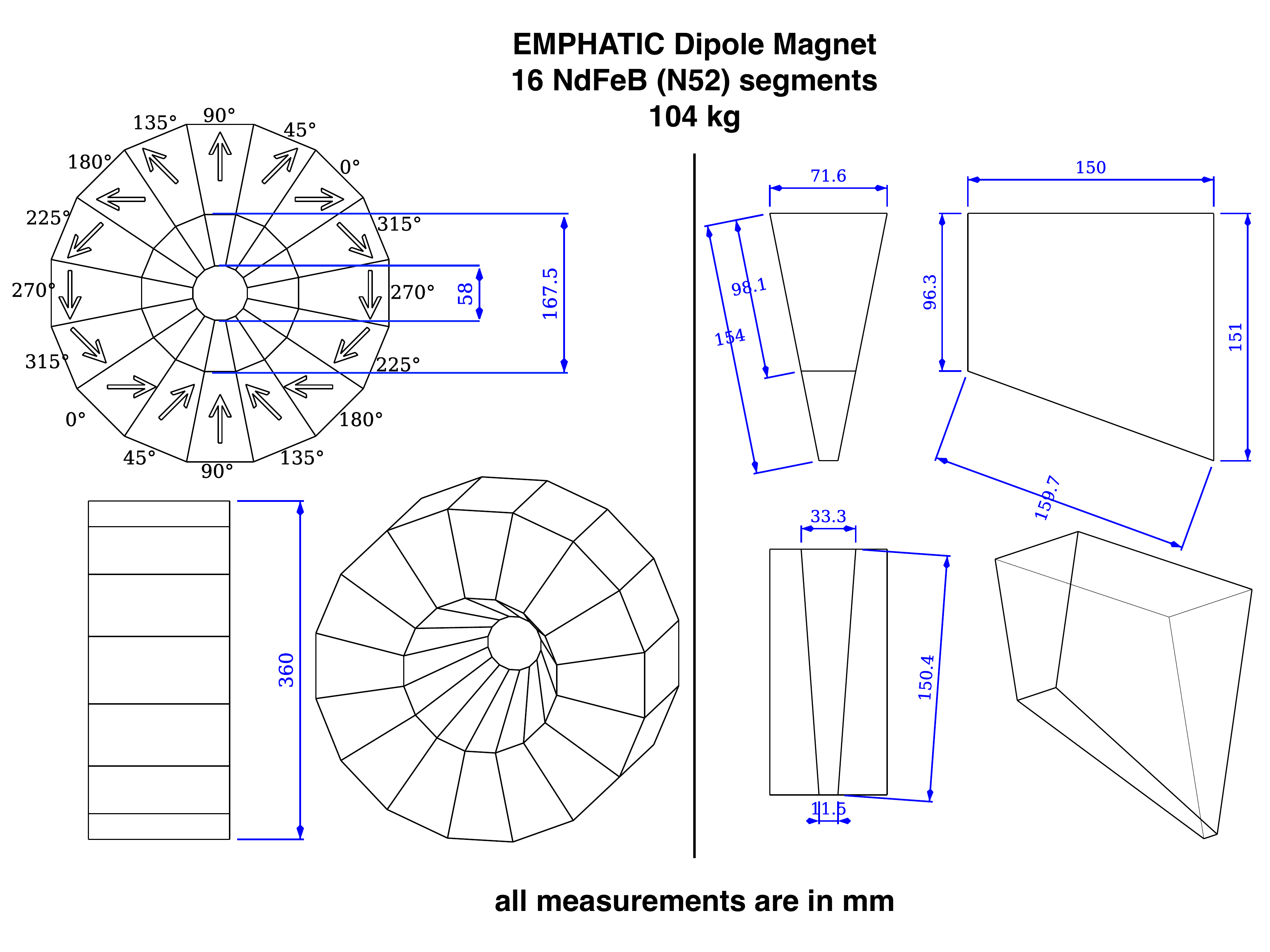}\\
\includegraphics[width=0.95\textwidth]{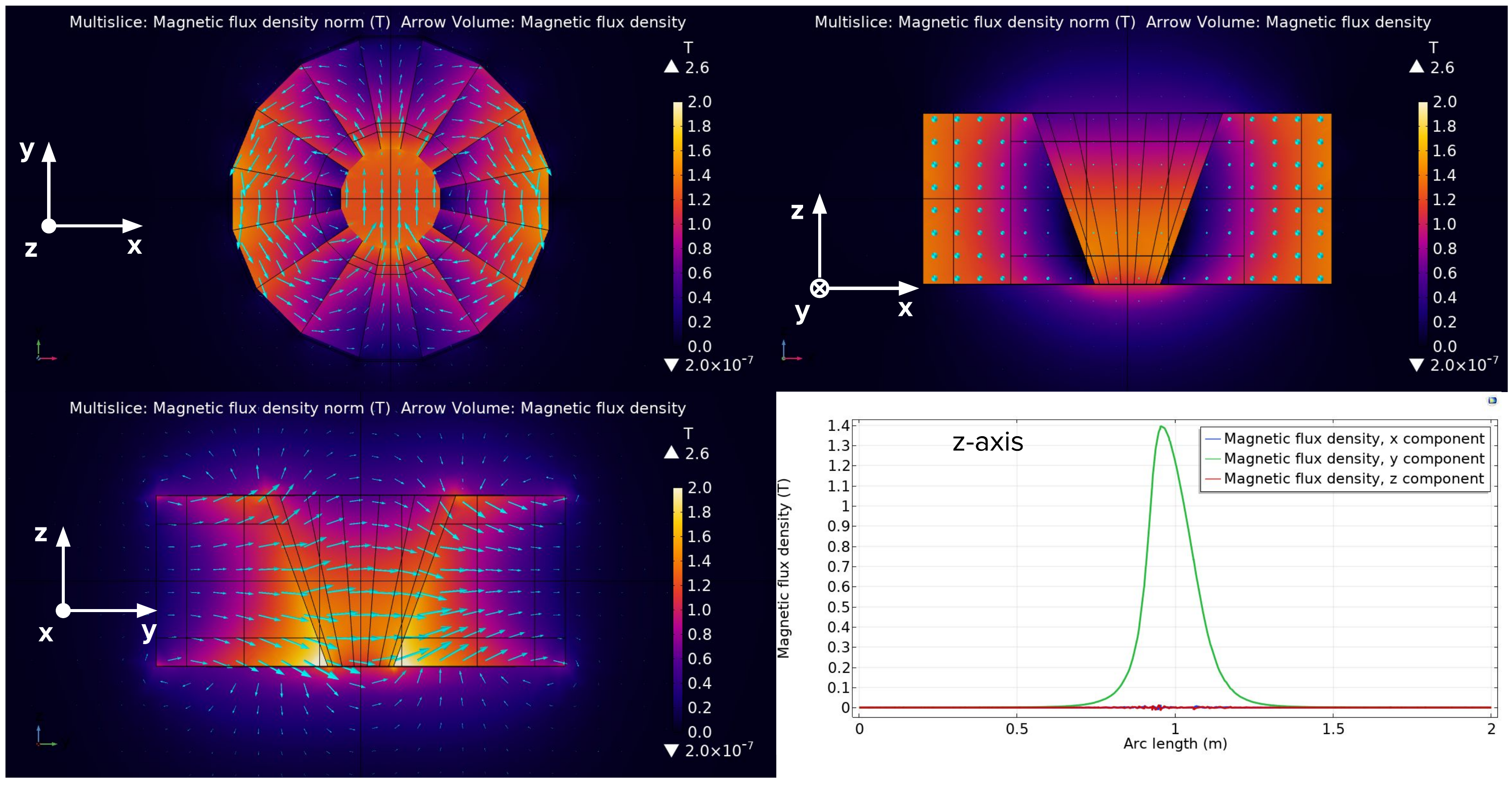}
\caption{Top: Preliminary design of the EMPHATIC permanent spectrometer magnet. Bottom: Magnetic field maps generated by COMSOL Multiphysics 5.4 simulation, using the configuration shown above.}
\label{fig:MagnetDesign}
\end{center} 
\end{figure*}

A preliminary design of the proposed magnet is shown in the top of Fig.~\ref{fig:MagnetDesign}.  A total of 16 segments of NdFeB (N52) are used to form a dipole magnetic with a mass of approximately 104 kg.  The upstream aperture diameter is 5.80 cm, and the downstream aperture diameter is 16.75 cm.  The length of the magnet and aperture is 15 cm.  The bottom of Figure~\ref{fig:MagnetDesign} shows the magnetic field maps, calculated using the COMSOL Multiphysics 5.4 simulation assuming perfectly uniform and identical N52 segments.  At the center of the magnet the field is peaked at approximately 1.4T, and there are very small non-dipole magnetic moments and fringe field at the downstream location of the aerogel RICH PMTs.  The field will be measured throughout the entire volume of the spectrometer post assembly.

\subsubsection{Particle Momentum Resolution Studies}
\begin{figure}[htpb]
\begin{center}
 \includegraphics[width=0.48\textwidth]{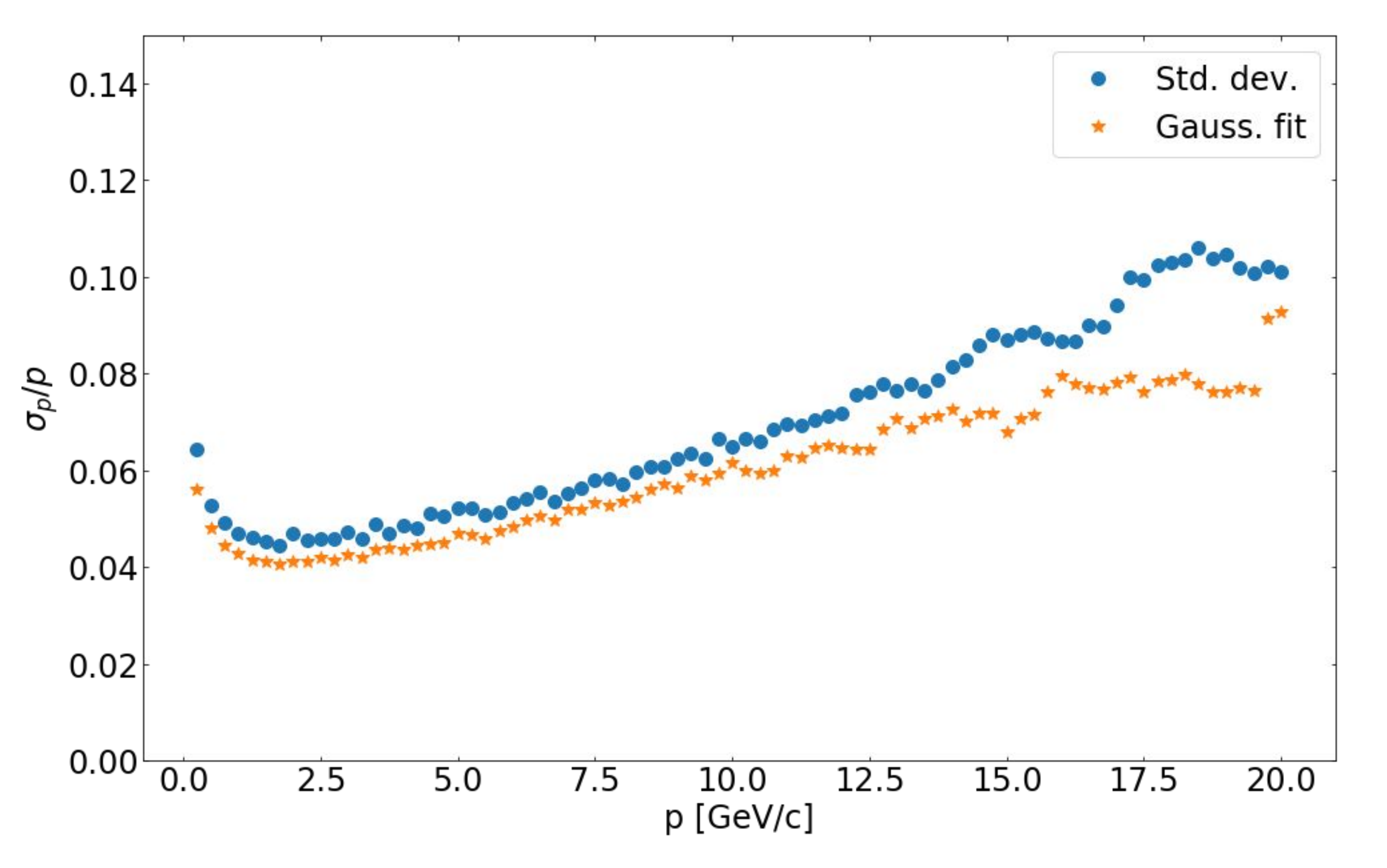}
\end{center}
\caption{Momentum resolution of reconstructed charged particles using the silicon strip detectors and magnetic described in Sections \ref{sec:SiStripDets} and \ref{sec:Magnet}.}
\label{fig:MomRes}
\end{figure}

Using the magnetic field maps shown in the previous section, and the resolution of the silicon strip detectors described in Section \ref{sec:SiStripDets}, we have performed a study to determine the pion momentum resolution using the configuration shown in Fig.~\ref{fig:spectrometer}.  In this study, positively charged pions were tracked through the spectrometer in air using the GEANT4 simulation that incorporated the magnetic field maps generated by the COMSOL simulation described above, and a geometry that includes the silicon strip detector materials.  The pions were uniformly distributed across the downstream face of the target shown in Fig.~\ref{fig:spectrometer}, and aimed exactly along the longitudinal axis of the magnet.  This study is a worst-case scenario, since only forward-going pions were considered; particles passing near the walls of the magnet aperture will experience a stronger magnetic field and will bend even more.  The positions of the pions were recorded at the location of the silicon strip detectors, and smeared according to their known resolution.  The trajectories of the pions were then reconstructed using a Kalman filter algorithm\cite{Kalman1,Kalman2,Kalman3} with the same magnetic field maps used in the simulation.
The fractional resolution determined of the reconstructed pion momentum as a function of the true pion momentum, determined by either the standard deviation or the width from a Gaussian fit to the distribution, is shown in Fig.~\ref{fig:MomRes}.  At very low momentum, the resolution is dominated by multiple scattering.  At higher momenta, the measured deflection angle becomes more quantized due to finite strip pitch, resulting in an apparent oscillatory behavior in the resolution.  Nevertheless, the energy resolution for all pions below 20 GeV is better than 10\% and better than 6\% below 10 GeV.  

\subsection{Particle Identification Downstream of the Target}
\subsubsection{Downstream Aerogel Cherenkov Detector}
\begin{figure}[htpb]
\begin{center}
 \includegraphics[width=0.48\textwidth] {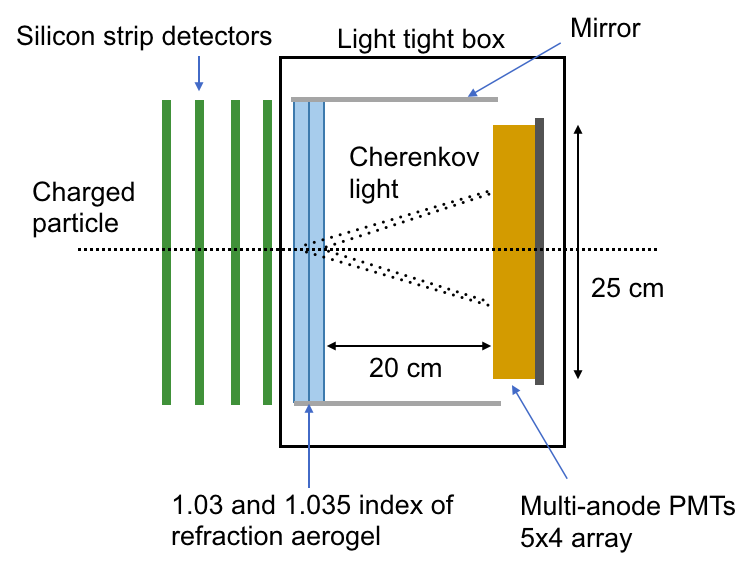}
\end{center}
\caption{Conceptual design of the aerogel RICH detector.}
\label{fig:arich_concept}
\end{figure}

A downstream ring-imaging Cherenkov detector with aerogel radiators (ARICH) will be used for identification of forward particles with momenta above 2~GeV/c.
The detector designed based on the aerogel-based proximity-focusing ring-imaging Cherenkov (ARICH) counter developed for the Belle II endcap particle identification \cite{ARICH1,ARICH2}. Aerogel plates with a total thickness of approximately 4 cm \cite{Aerogel3} are followed by multi-anode PMTs located 20~cm downstream of the aerogel radiators. Unlike the Belle~II experiment, the detector will be operated outside of a magnetic field, so standard multi-anode PMTs (Hamamatsu R12700) will be used for lower cost and better efficiency.  The conceptual design of the ARICH detector is shown in Fig.~\ref{fig:arich_concept}.  This configuration uses a 5x4 array of multi-anode PMTs and mirrors to reflect Cherenkov light outside of the PMT array acceptance onto the PMT array.  The angular acceptance of the detector can also be increased by building a larger array of multi-anode PMTs, budget permitting.

In the ARICH detector, each channel typically detects 1 or 0 photo-electrons, so a measurement of time at the one photo-electron threshold level is sufficient.
We use the GSI TRB3 FPGA-based TDC for the front-end electronics.
The GSI TRB3 is a flexible FPGA based readout board, which has five FPGAs.  One FPGA in the center collects events from the other four FPGAs, 
each of which collect data from an add-on board to the TRB3.  The add-on boards being used for the ARICH readout are 48-channel fast 20-ps resolution TDCs.  This is the same electronics that has been used in the HADES experiment.  The maximum trigger rate of 700 kHz will be more than adequate for measurements in the EMPHATIC experiment. 

\begin{figure}[htpb]
\begin{center}
 \includegraphics[width=0.48\textwidth] {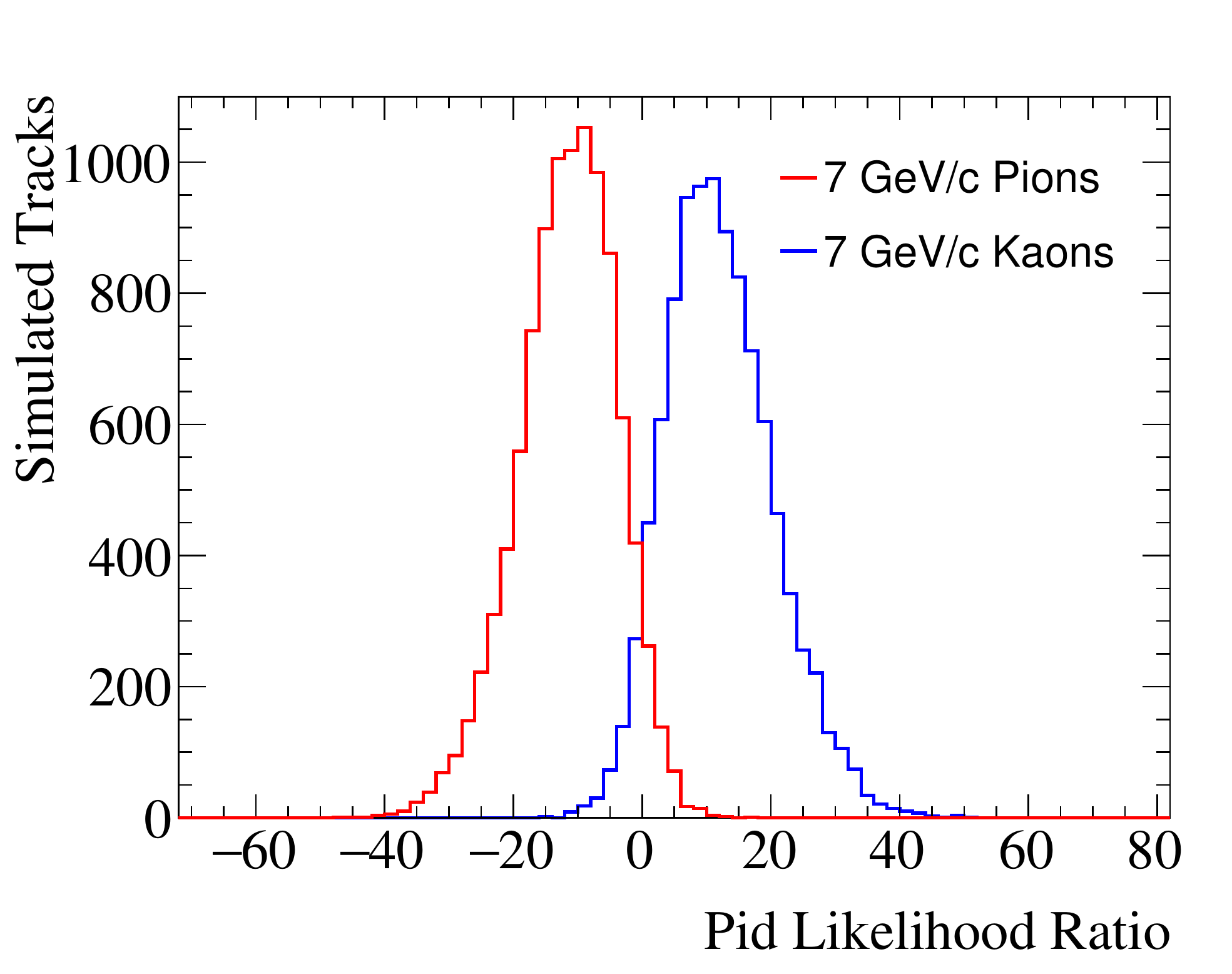}
 \includegraphics[width=0.48\textwidth] {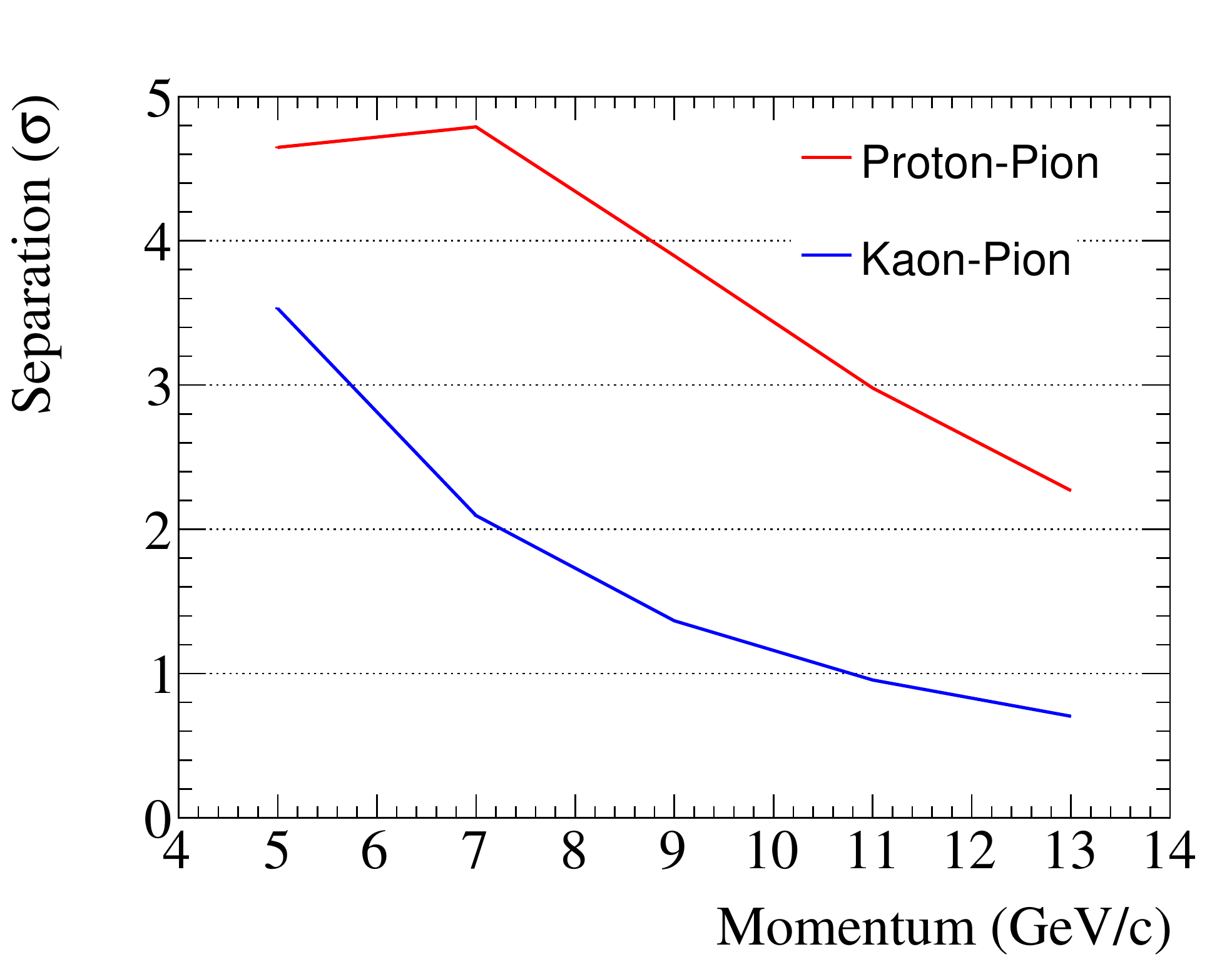}
\end{center}
\caption{Simulated particle identification performance of the ARICH detector.  Top: The likelihood ratio distributions for simulated 7~GeV/c pions and kaons. Bottom: The mean separation of kaon(proton) distributions from pion distributions normalized by the quadrature sum of the kaon(proton) distribution RMS and pion distribution RMS.}
\label{fig:arich_pid}
\end{figure}

Simulation studies of the ARICH have been performed with a realistic GEANT4 simulation that includes the optical properties of the aerogel, the quantum efficiency of the R12700 PMTs, and the expected resolution of the upstream tracker and spectrometer.  For a 9~GeV/c pion, on average 17 Cherenkov photons are detected by the multi-anode PMT array.  Fig.~\ref{fig:arich_pid} illustrates the simulated particle identification performance.  A 2$\sigma$ separation of pions and kaons is possible below 7~GeV/c, while a 2$\sigma$
separation of protons and pions is possible below 13~GeV/c.

\begin{figure}[htpb]
\begin{center}
 \includegraphics[width=0.48\textwidth] {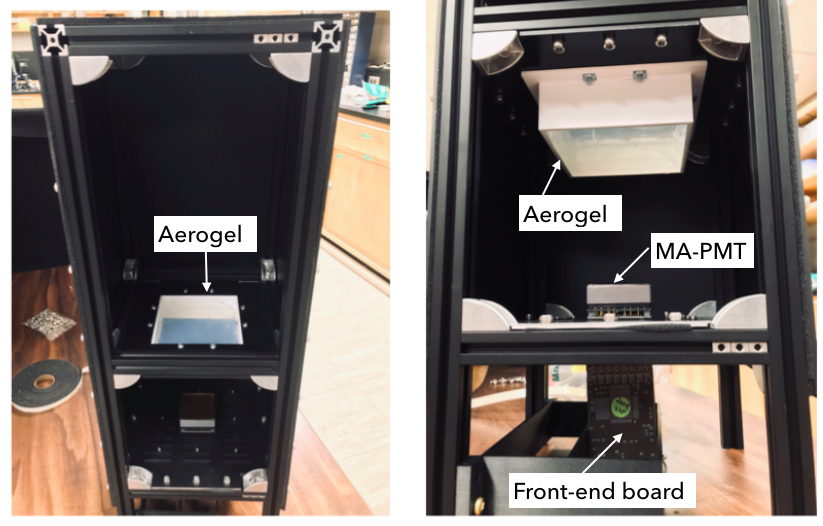}
\end{center}
\caption{Pictures of the ARICH detector prototype.  The dark box houses the aerogel radiators and up to 9 5~cm$\times$5~cm size photosensors.}
\label{fig:arich_prototype}
\end{figure}

 Improvements to the ARICH detector design to extend particle identification to higher momentum are being investigated.  Aerogels with refractive indices below 1.03 are considered to extend the sensitivity to higher momentum.  In the nominal design shown in Fig.~\ref{fig:arich_concept}, two layers of aerogel with increasing refractive index are used to improve the Cherenkov ring focusing.  Further improvements to the focusing can be achieved by using more than two thin layers of aerogel radiators with increasing refractive index. To take advantage of the improved focusing, silicon photomultiplier (SiPM) arrays with 3~mm$\times$3~mm sized channels are considered as a replacement for the multi-anode PMTs, which have 6~mm$\times$6~mm channel size.  The SiPMs also have a higher photon detection efficiency compared to the multi-anode PMTs, which will further improve the resolution, and the cost per photosensitive area is reduced.  The nominal ARICH design and these improvements will be tested in August 2019 with a prototype ARICH detector operating in the TRIUMF M11 beam line, which will provide $\sim$80 MeV electrons and positrons to test the ARICH response to $\beta\approx1$ particles.  Fig.~\ref{fig:arich_prototype} shows the prototype ARICH detector during its construction.

\begin{figure}[htpb]
\begin{center}
\includegraphics[width=0.48\textwidth]{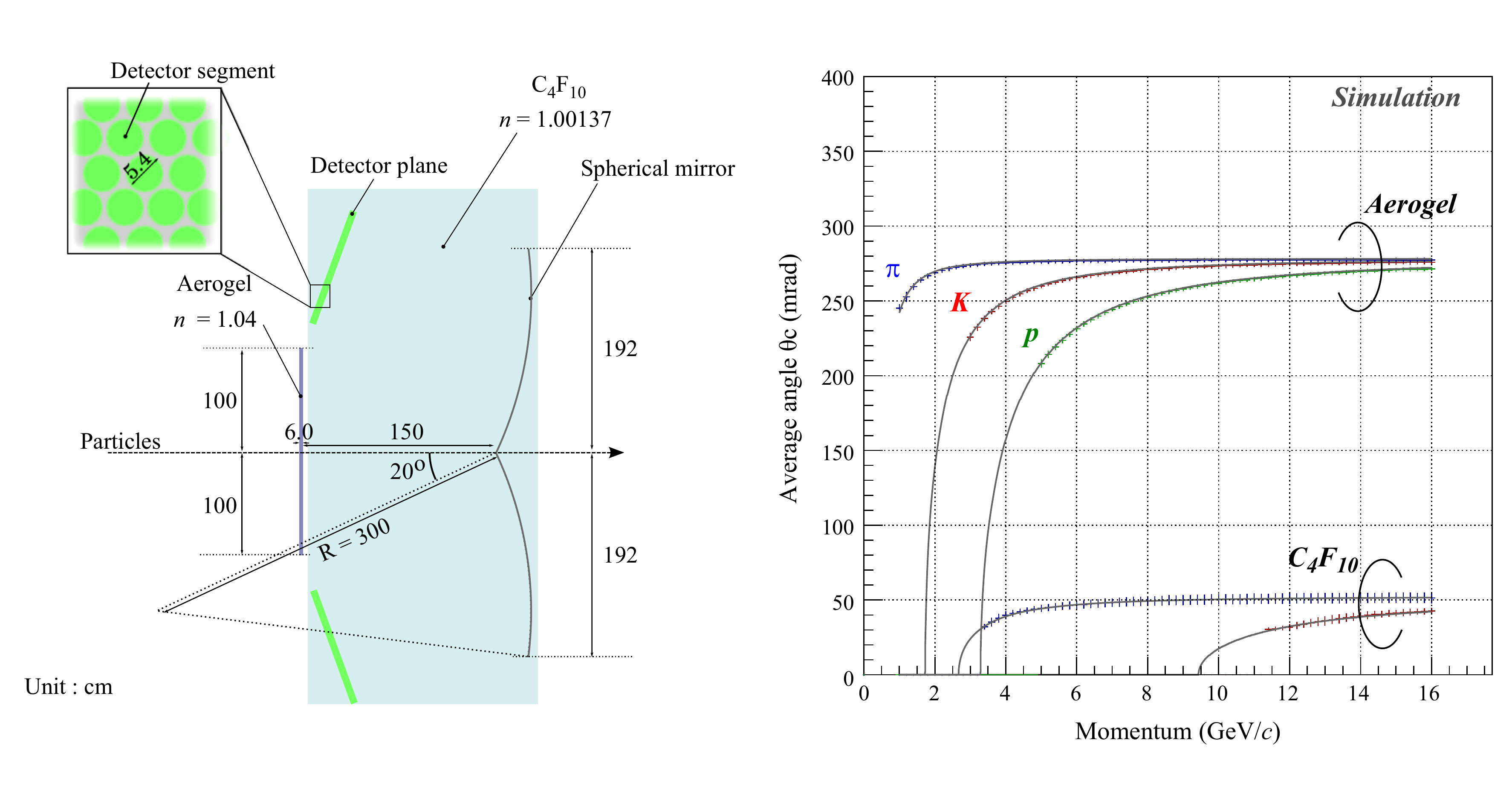}
\caption{Left: Schematic of the proposed E50 RICH.  Right: Cherenkov ring angle size as a function of particle momentum.  Both figures taken from \cite{Yamaga}.}
\label{fig:E50RICH}
\end{center} 
\end{figure}

The J-PARC E50 group is also pursuing a design of the RICH detector that combines aerogel radiators with a C$_4$F$_{10}$ or C$_4$F$_{8}$O gas, as shown in Fig.~\ref{fig:E50RICH} to give better particle identification at high momentum.  This detector is planned as a particle identification upgrade for EMPHATIC.

\subsubsection{ToF Detectors}
The ToF system consists of two separate detectors, one to measure the time of the incident beam particle before it hits the target ($T0$), and the second to measure the time of the secondary particles downstream.  
Both detectors require are a timing resolution of less than 100 ps(rms) and a fast response in order to handle high-counting rate conditions.
The upstream $T0$ detector will be an acrylic Cherenkov radiator attached to a MPPC, and the downstream detector will consist of resistive plate chambers (RPCs).

\begin{figure}[htbp]
\begin{center}
\includegraphics[width=0.48\textwidth]{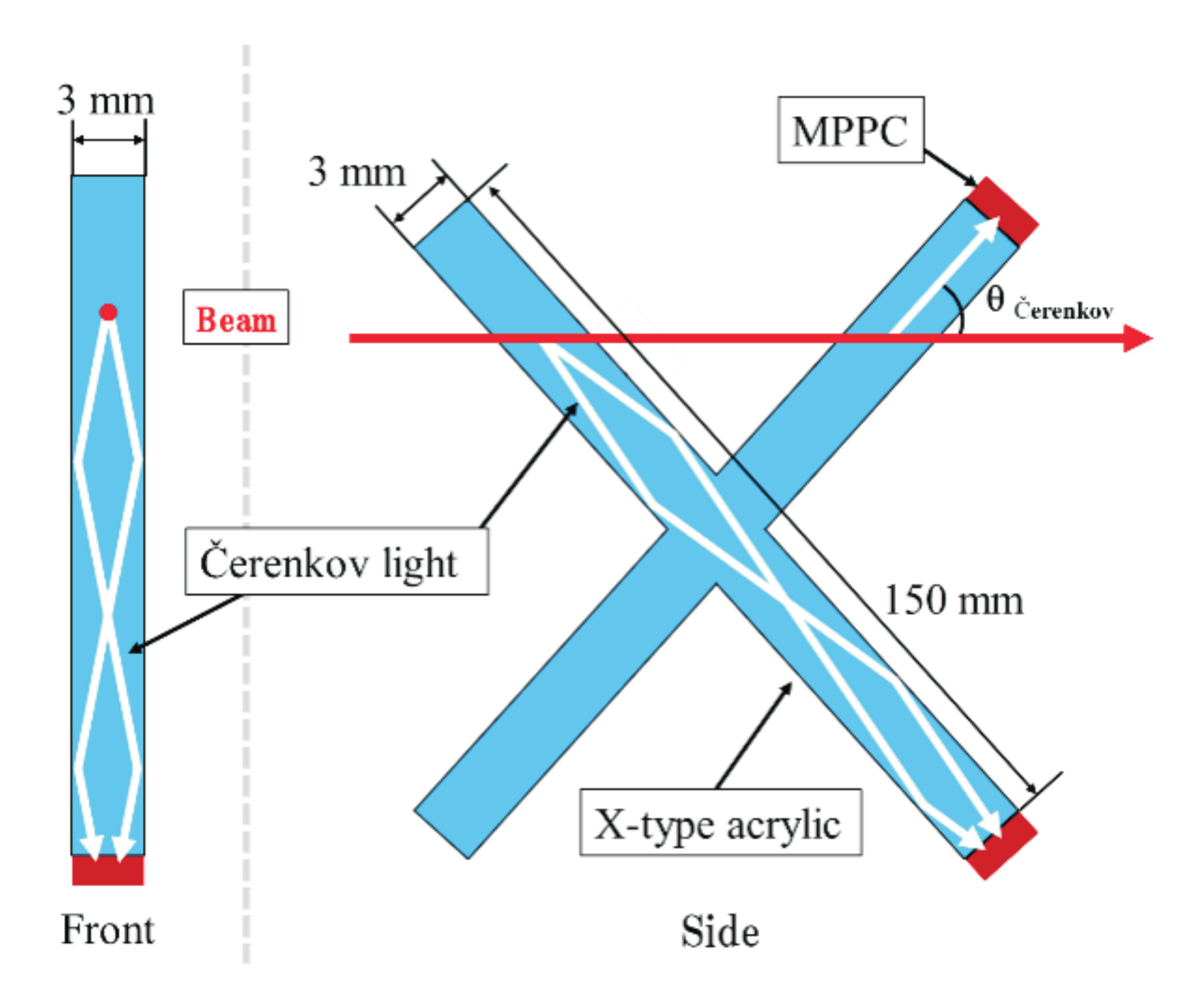}
\caption{"X-type" $\check{\mathrm{C}}$erenkov radiator. }
\label{fig:Xtype}
\end{center}
\end{figure}

The acrylic Cherenkov $T0$ detector consists of an array of x-shaped radiators, shown in Fig.~\ref{fig:Xtype}.  The beam crosses both acrylic bars, and the Cherenkonv light is internally reflected into a  Hamamatsu S13360-3050PE.  This multi-pixel photon counter (MPPC) has good intrinsic time resolution of $\sim$200 ps(FWHM) for the one photoelectron detection~\cite{Hamamatsu} so that the time resolution of the counter is determined by the light yield.
The acrylic cross was cut from an acrylic board having size of 3 mm(W) $\times$ 3 mm(T) $\times$ 150 mm(L)
for one side as shown in Fig.~\ref{fig:Xtype}. 
The opening angle of cross from the beam axis was 45 degrees.
MPPCs were attached to the downstream of both edge of the X-type radiator for measuring $\check{\mathrm{C}}$erenkov photons which reach MPPCs by the shortest path.
The number of detected photoelectrons by using this combination was $\sim$50 p.e. at the center of radiator from the measurement by using the beam of $\beta \sim $1.
For the MPPC signal readout, a shaping amplifier which has a fast operational amplifier(AD8000) was used as ref.~\cite{AMP}.
The shaping amplifier was modified to suppress the overshoot after the signal 
for adding the pole-zero cancellation resistance of 390 $\Omega$ on the circuit for 50-$\mu$m pixel MPPC(S13360-3050PE).
Timing and pulse height information from the counters are measured by the DRS4 module~\cite{DRS4} in which a FPGA based High-resolution TDC (HR-TDC) is implemented.

Discriminator signals measured by HR-TDC were used to measure the time.
The waveform information was only used for measuring signal pulse heights for the pulse-height correction and analyzing pile-up effects in the high-rate conditions.

\begin{figure}[htbp]
\begin{center}
\includegraphics[width=0.48\textwidth]{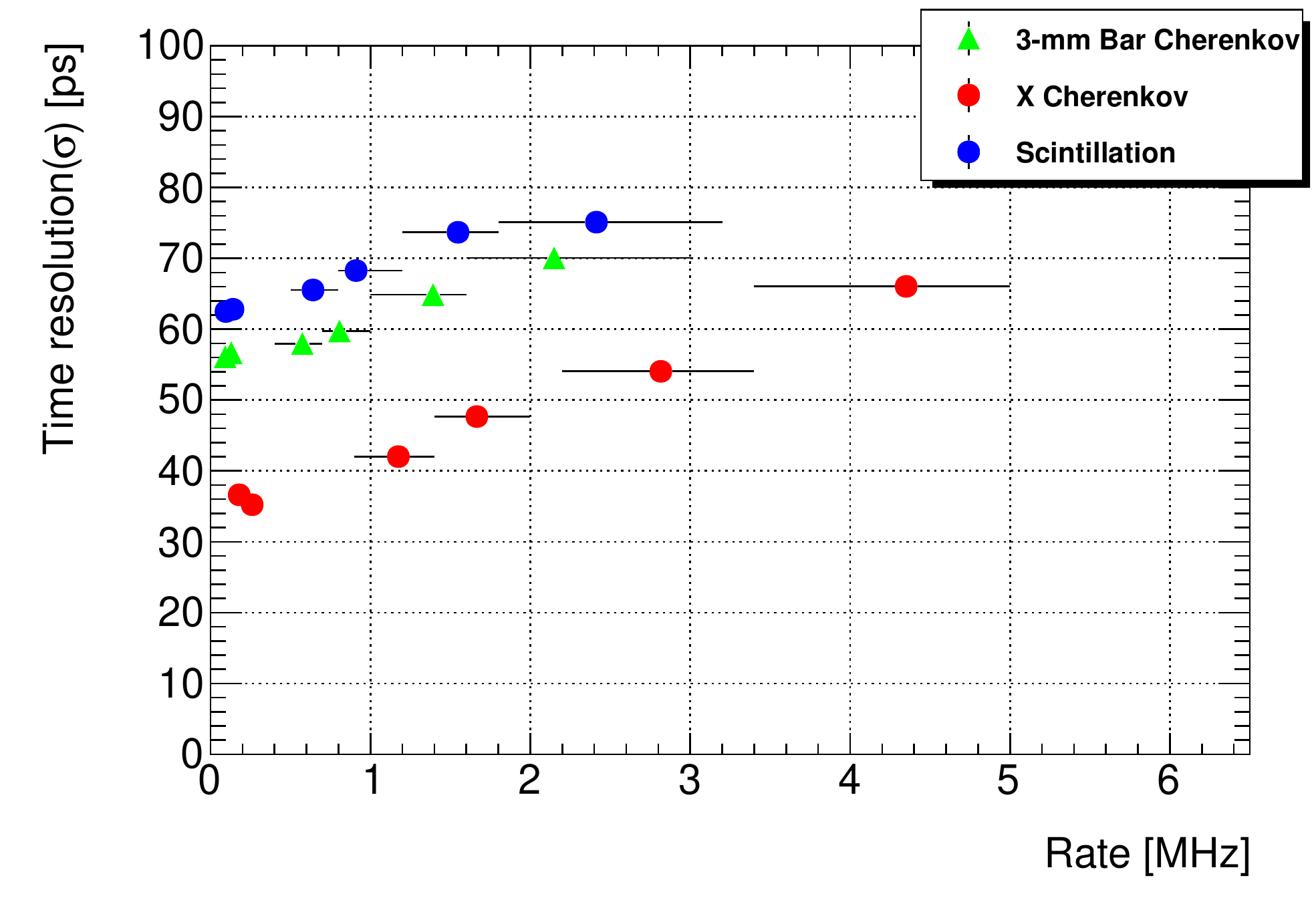}
\caption{Time resolution by changing counting rate conditions. Horizontal error bars show the range of the counting rate by selecting event-by-event scaler counts. }
\label{TimeR}
\end{center}
\end{figure}

Figure~\ref{TimeR} shows the measured time resolution of different detectors as a function of counting rate in the test beam at ELPH.
Horizontal error bars show the range of the counting rate by selecting event-by-event scaler counts. 
The X-type $\check{\mathrm{C}}$erenkov counter showed the best time resolution of $\sim$50 ps(rms) in the range of 3$-$5 MHz. For the EMPHATIC experiment, the time resolution of the counter is expected to be $\sim$30 ps(rms) because of
the lower counting rate of beam ($\mathcal{O}(100)$ kHz).

\begin{figure}[htbp]
\begin{center}
\includegraphics[width=0.48\textwidth]{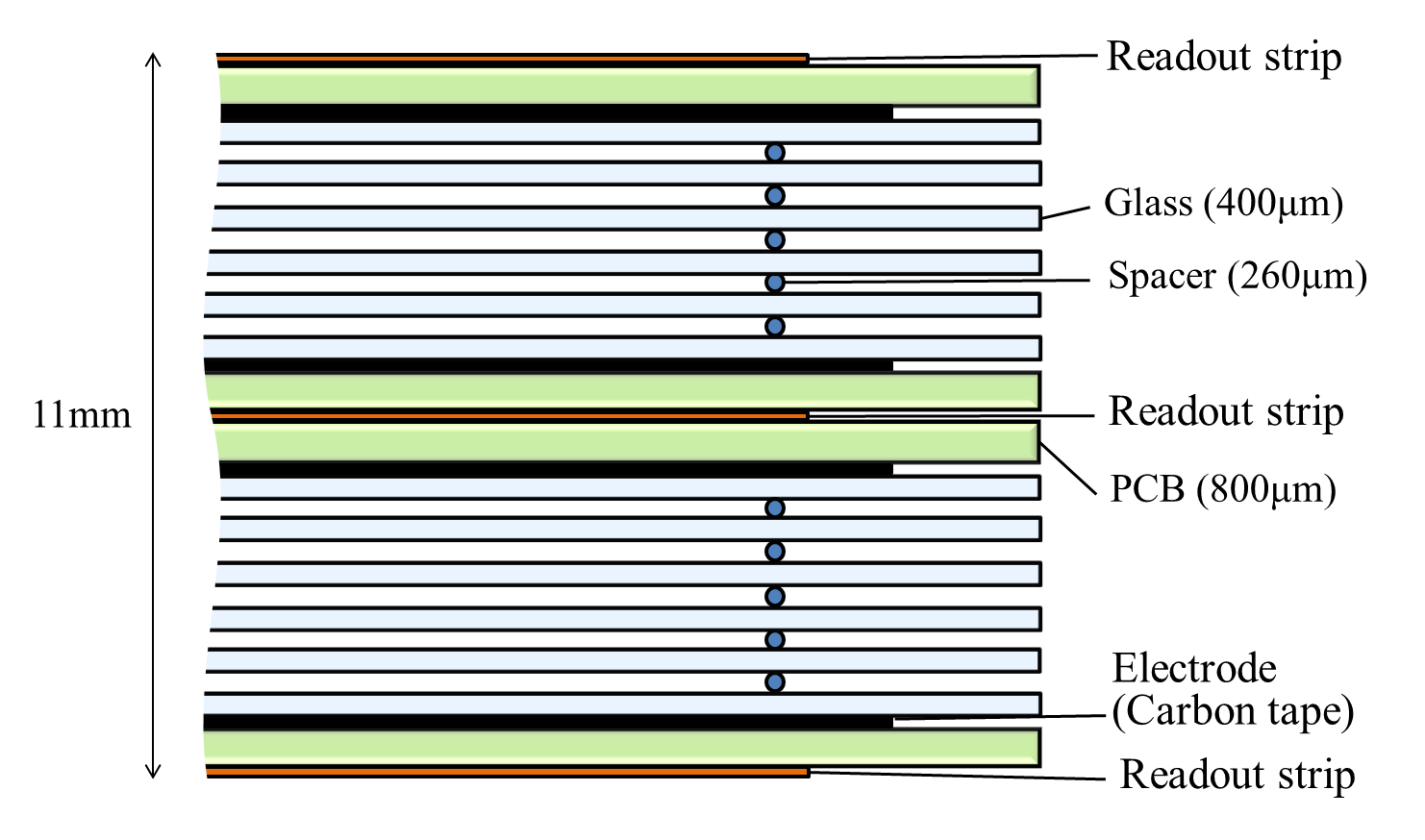}\\
\includegraphics[width=0.48\textwidth]{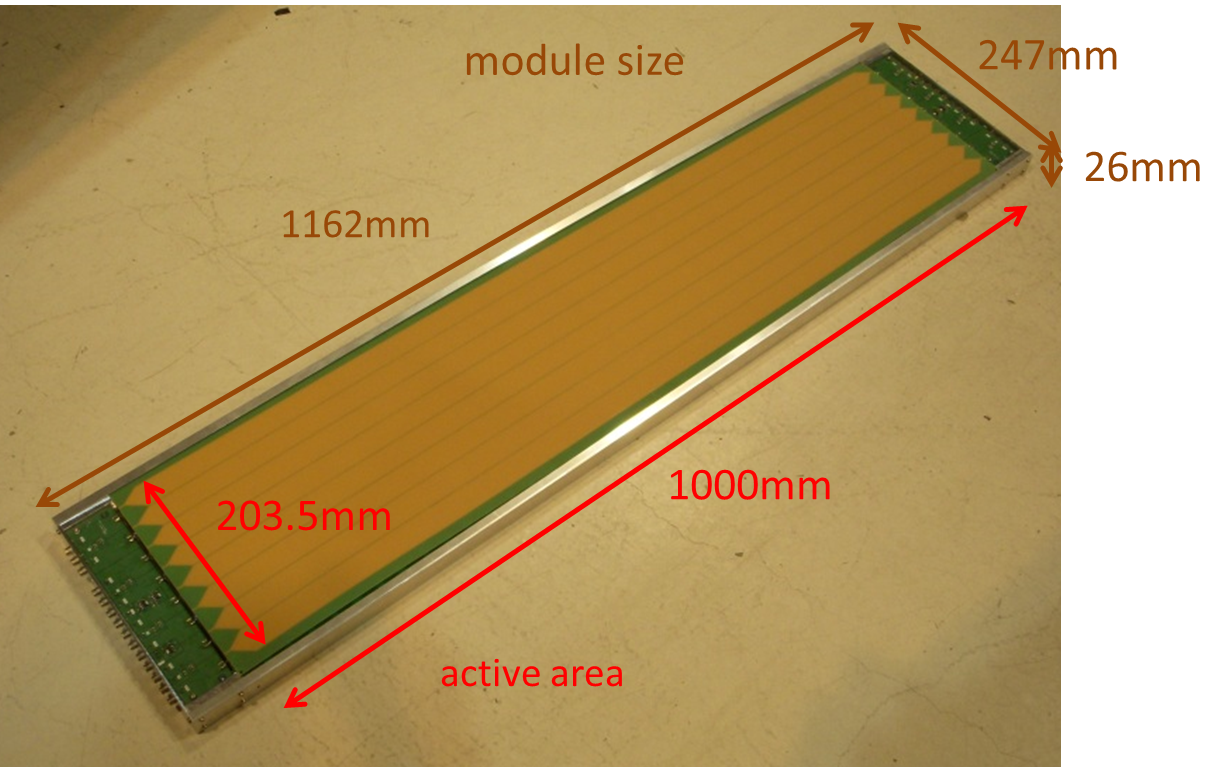} \\
\caption{Top: The cross section of the RPC.  Bottom: Photo of the top view of a RPC module.}
\label{fig:RPC}
\end{center}
\end{figure}

The structure of the resistive plate chamber (RPC) is based on 
the BGOegg-RPC and the LEPS2-RPC used at SPring-8 in Japan \cite{BGOegg,BGOegg2,LEPS2}.
The top of Fig.~\ref{fig:RPC} shows the cross section of the RPC.
It has 5 gaps $\times$ 2 stacks structure.
The gap size is 260 $\mu m$.
The bottom of Fig.~\ref{fig:RPC} is a photo showing the top view of an RPC module.
The readout strip is 25.5 mm wide, 1000 mm long and the gap between strips is 0.5 mm.
There are eight strips per chamber.
In the EMPHATIC experiment, 5 RPC modules will be used to cover 1000 mm $\times$ 1000 mm.

\begin{figure}[htbp]
\begin{center}
\includegraphics[width=0.48\textwidth]{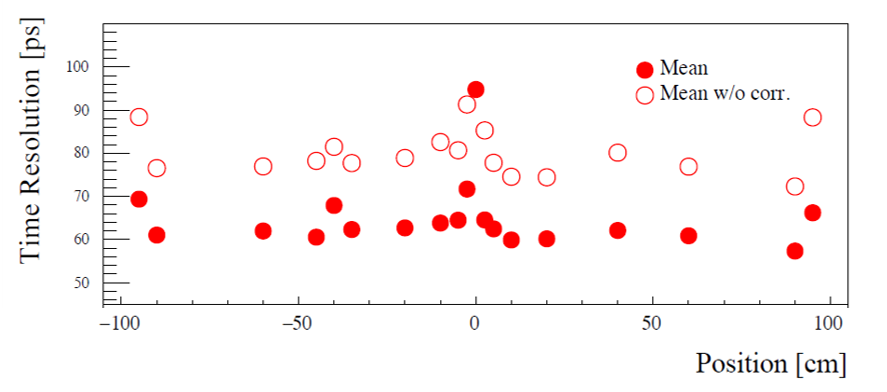}
\caption{The time resolution of the LEPS2-RPC as a function of position across the detector.}
\label{fig:RPCRes}
\end{center}
\end{figure}

The time resolution of the LEPS2-RPC as a function of position across the detector is shown in Fig \ref{fig:RPCRes}.  The average resolution around 60 ps (rms) is obtained.  Combined with the $T0$ detector, the timing resolution of the ToF system is 70 ps (rms).  With a minimum travel distance of 80 cm, this system can provide $>2\sigma$ K-$\pi$ separation up to approximately 1.5 GeV/c.

\subsubsection{Lead Glass Calorimeter }

The lead glass calorimeter at the most downstream location consists of
9$\times$9 lead glass counters to cover a $\pm$400~mrad angle in both transverse directions.
Each lead glass crystal is 12~cm $\times$ 10~cm $\times$
36~cm. Cherevkov light emitted in the crystals is detected by 3 inch
fine-mesh PMTs. These counters are refurbished from K2K long-baseline
neutrino oscillation experiment in Japan.

\subsection{Triggering, Electronics and DAQ }

The overall control of the system will be done by the MIDAS DAQ at the FTBF facility.  Most of the readout electronics will be readout directly by MIDAS, including the beam monitors (currently CAMAC which will need to be be upgraded to VME), ARICH (readout with TRB3 FPGA-TDC), RPC, and 
lead glass calorimeter.  Data from these different detectors will be 
assembled online by the MIDAS event builder using the event fragment 
timestamps.  The silicon strip detectors have its own 
readout DAQ, which will continue to be used.  The data from MIDAS 
and the silicon strip detectors will be merged offline using their 
time stamps. Common reference time and trigger signal will be provided to 
each system for synchronization. 


\section{\label{sec:POPRun}Proof-of-Principle Measurements}

\begin{figure}[htbp]
\begin{center}
\includegraphics[width=0.48\textwidth]{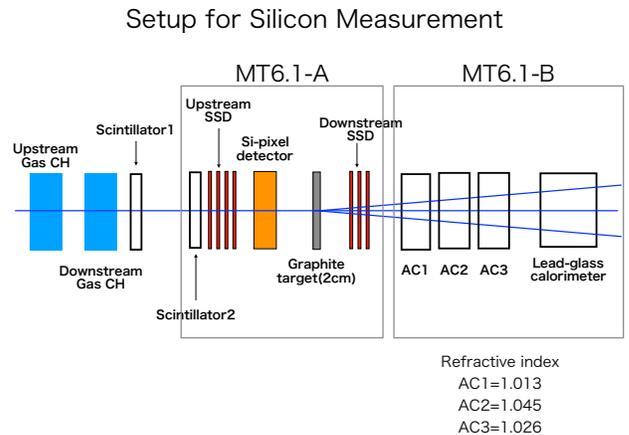}
\caption{Schematic of the 2018 test run.}
\label{fig:2018Setup}
\end{center}
\end{figure}

The feasibility of the measurement of total and quasi-elastic hadron scattering measurements has been demonstrated via the analysis of data collected during a brief run at the FTBF in January 2018.  Fig. ~\ref{fig:2018Setup} shows a schematic of the setup of the experiment.  The SSDs, silicon pixel telescope and target were located in the MT6.1a alcove, and the aerogel Ckov and lead-glass calorimeter were placed in the MT6.1b alcove.  The silicon detectors and lead-glass calorimeter were provided by the FTBF.  No magnet was used in this run.  The silicon pixel telescope was later found to have too low of an efficiency to be used in these measurements.  The response of the aerogel detectors was also found to have a very low efficiency.  Since the 2018 run, however, new aerogel detectors constructed at Chiba university with more photo coverage have been tested and an efficiency $> 97\%$ measured. 

Data were collecting during a two-week period using beam momenta settings of 120, 30, 20, 10 and 2 GeV/c, with carbon, aluminum and iron targets (2-5\% interaction lengths).  Data with no target were also collected for background studies.

\begin{figure}[htpb]
\begin{center}
\includegraphics[width=0.48\textwidth]{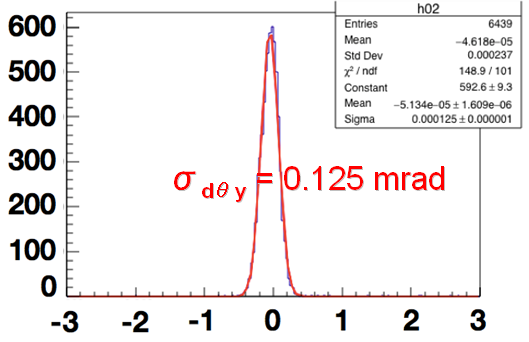}
\caption{Angle difference of tracks connected between two reconstructed tracks by emulsion films before and after the target.  The angular resolution is 0.088 mrad (= 0.125/$\sqrt{2}$).}
\label{fig:angular_resolution}
\end{center} 
\end{figure}

\begin{figure}[htbp]
\begin{center}
\includegraphics[width=0.48\textwidth]{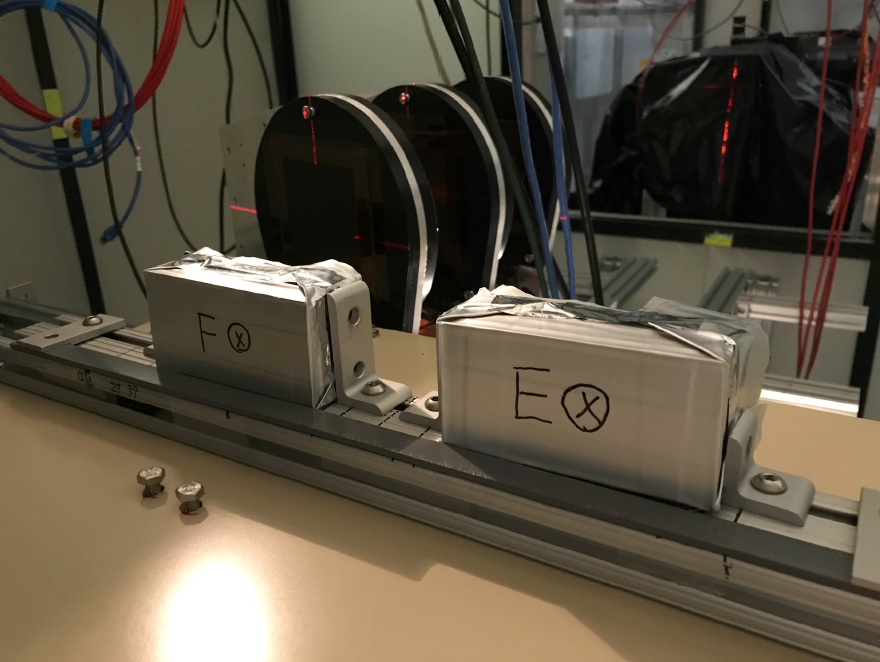}
\caption{Vacuum packed emulsion targets on remotely-controlled sliding support structures.}
\label{fig:ECC2}
\end{center} 
\end{figure}

A fraction of the data were collected using ECCs with graphite targets.  An emulsion handling facility (EHF) was constructed at Fermilab.  The EHF was used to construct ECCs with 2\% interaction length graphite targets, surrounded by 3 layers of emulsion film separated by 2 layers of Rohcell plates, as depicted in Fig.~\ref{fig:ECC1}.  10 ECCs were exposed to 30 GeV/c secondary beam (consisting of a mix of pions, protons and some kaons), and 2 ECCs were exposed to the pure proton 120 GeV/c primary beam.  A motion table moved the ECCs approximately 0.5 cm in the transverse direction to the beam in between beam spills in order to reduce pileup in the emulsion films, resulting in an exposure to approximately 300k beam particles across the face of each ECC.  Fig.~\ref{fig:ECC2} shows a photo of ECCs mounted to the motion table.  After exposure to the beam, the ECCs were disassembled at the EHF and the films developed.  The films were then shipped back to Nagoya University, Japan, where they were digitized.  The digitized emulsion data have been used to determine the internal alignment of the emulsion films and determine the angular resolution ($<$0.1 mrad, shown in Fig.~\ref{fig:angular_resolution}).  Studies of multi-particle production are underway.

\begin{figure}[htbp]
\begin{center}
\includegraphics[width=0.48\textwidth]{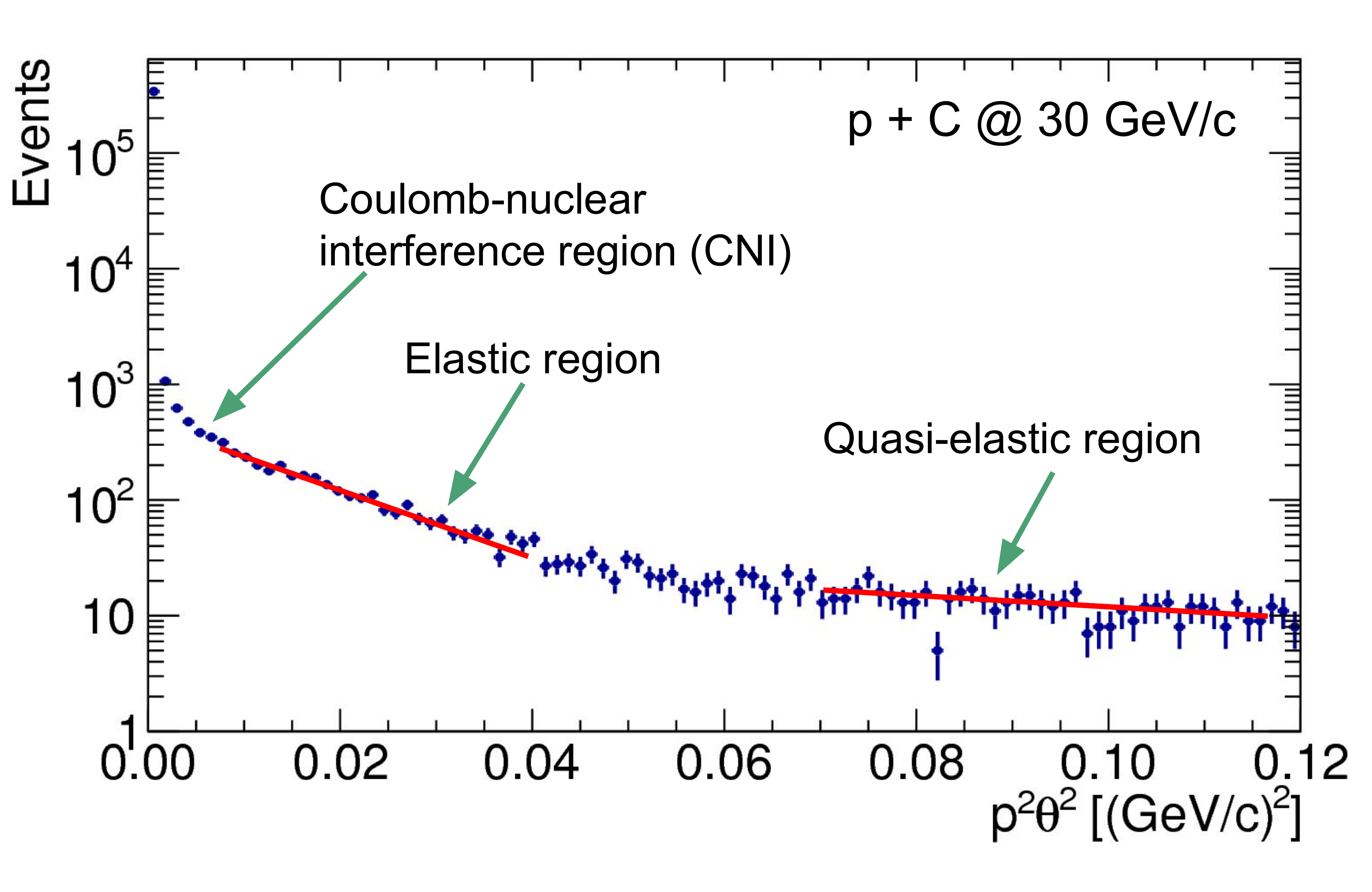}
\caption{Fast-turnaround plot of the distribution of momentum transfer, $t \simeq p^2\theta^2$, for 30 GeV protons scattering off the 2\% interaction length graphite target.}
\label{fig:2018t2Plot}
\end{center}
\end{figure}

\begin{figure*}[htpb]
\begin{center}
\includegraphics[width=0.48\textwidth]{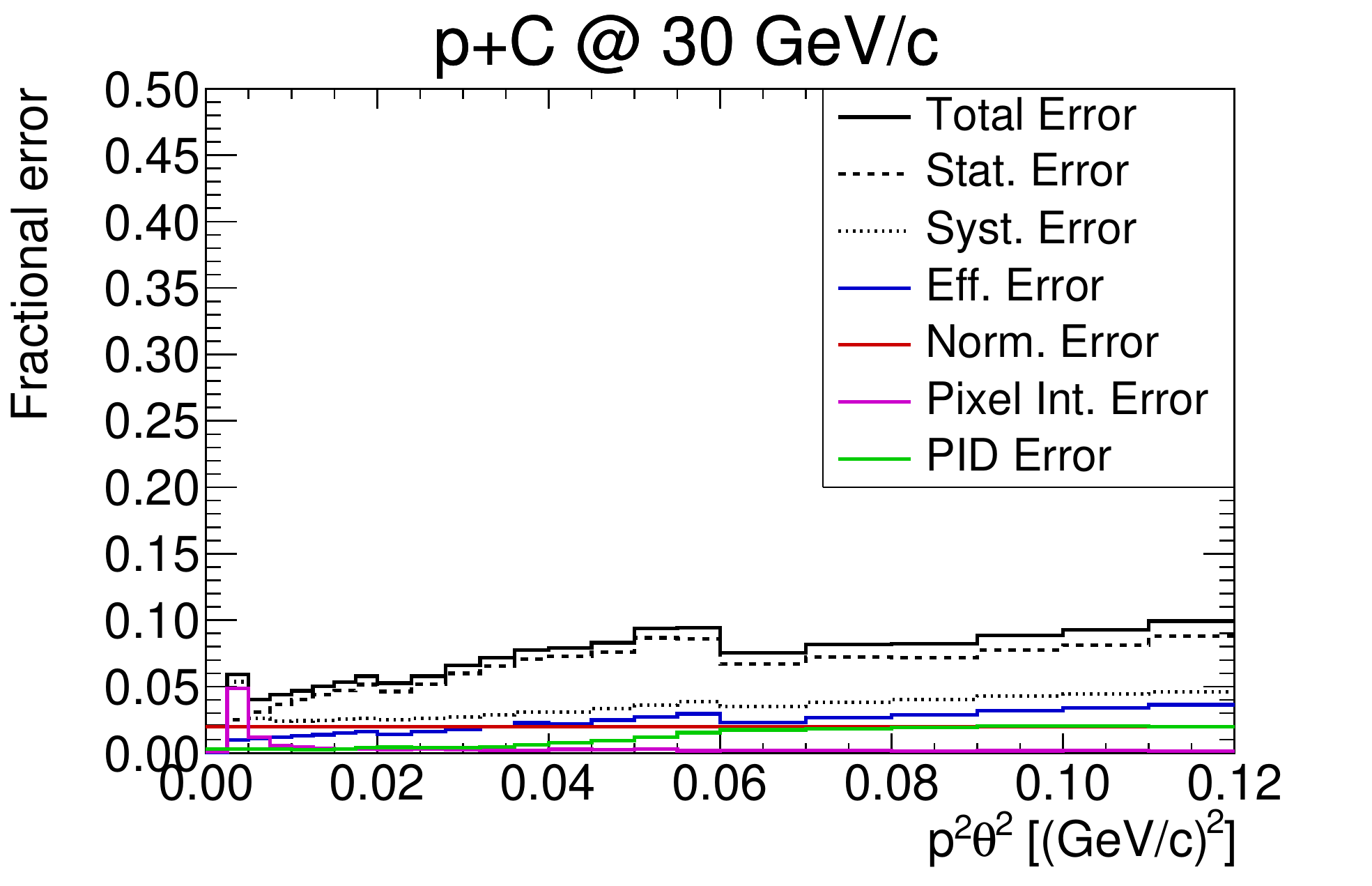}
\includegraphics[width=0.48\textwidth]{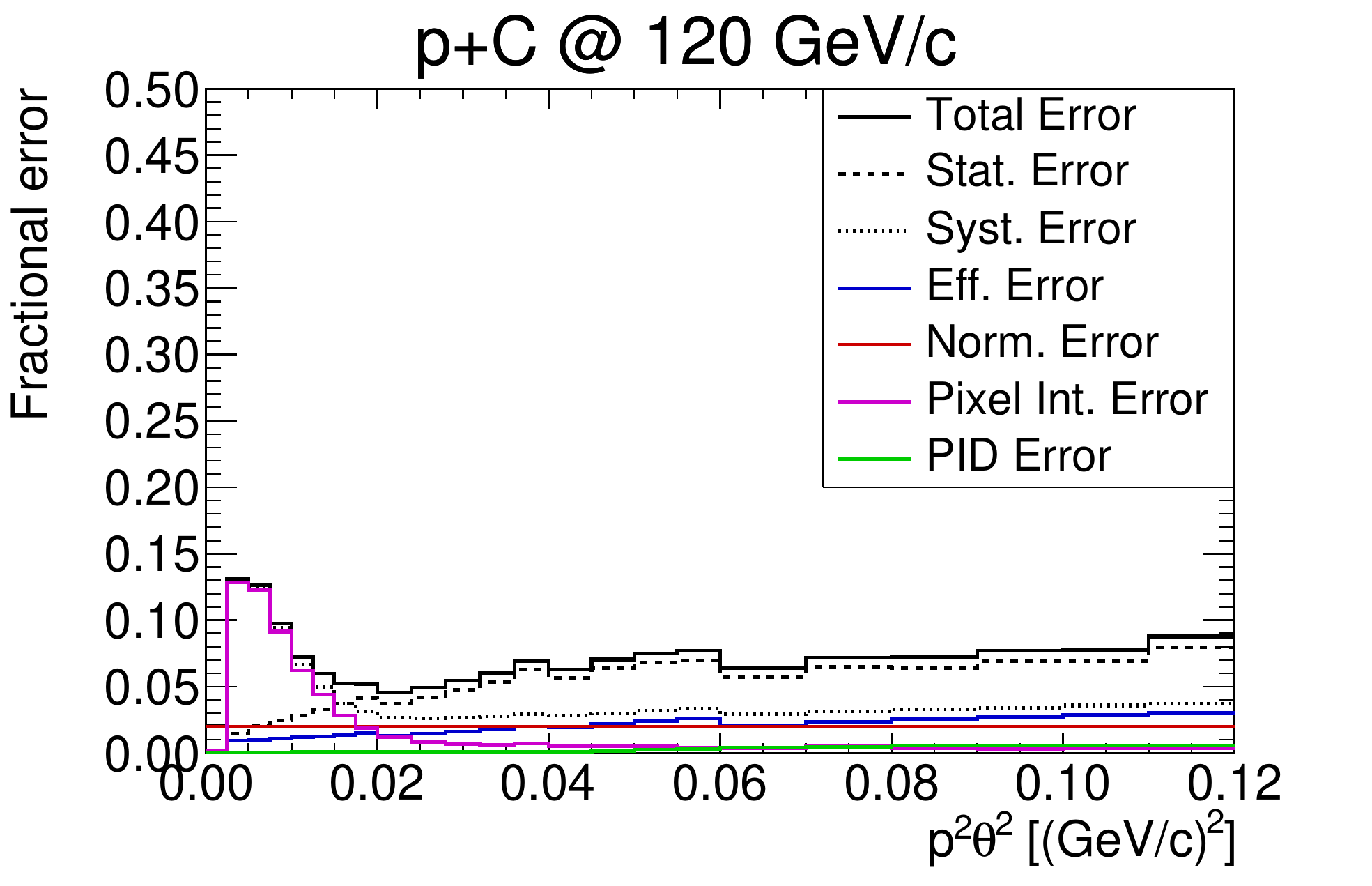}
\caption{Estimated fractional uncertainties on the measurements of the different cross section, $d\sigma/dt$ for the data collected during the 2018 test run, for 30 (left) and 120 (right) GeV/c protons striking the carbon target.}
\label{fig:2018Uncert}
\end{center}
\end{figure*}

An additional $\sim$20M beam triggers were collected in various beam momenta and target configurations, shown in Table \ref{table:2018Data}.  In these configurations, the targets were simply thin slabs of the material.  In the case of the graphite target, the slabs were identical to those used in the T2K neutrino production target, with very well determined density, thickness and purity.  The slabs of aluminum and iron were material found in the FTBF, for which the density and purity will soon be precisely determined.

\begin{table}[t]
\centering
\begin{tabular}{|c|c|c|c|c|} 
 \hline
Beam Momentum & Graphite & Aluminum & Iron & Empty \\  
(GeV/c) & & & & \\
 \hline
120 & 1.63M & 0 & 0 & 1.21M \\
30 & 3.42M & 976k & 1.01M & 2.56M \\ 
-30 & 313k & 308k & 128k & 312k \\
20 & 1.76M & 1.76M & 1.72M & 1.61M \\
10 & 1.18M & 1.11M & 967k & 1.17M \\
2 & 105k & 105k & 183k & 108k \\
 \hline
\end{tabular}
\caption{Number of beam triggers recorded per target material and beam momentum.  Negative momentum reflects the polarity of the beam.}
\label{table:2018Data}
\end{table}

Although the setup of the experiment did not include a spectrometer magnet, one may still measure the differential cross section, $d\sigma/dt$, where $t$ is the 4-momentum difference between the incoming and outgoing particle.  For small scattering angles, the interactions are dominated by elastic and quasi-elastic processes, and $t \simeq p^2\theta^2$.  A measurement of $d\sigma/dt$ for small scattering angles can also be related, with some model-dependence, to the total cross section via the optical theorem.  This method to determine the total cross section was used by \cite{Bellettini} in 1966, and these measurements are still relevant and used to reweight predictions and assess the systematic uncertainties for the NuMI, LBNF, and J-PARC neutrino beams.  Very few of these measurements have been made in the past, and the systematic uncertainties for the existing measurements are very simplistic and not well understood.

The plot in Fig.~\ref{fig:2018t2Plot} shows an example of the statistical precision with which the event rate as a function of $p^2\theta^2$ was measured during the 2018 test run, for 30 GeV/c protons striking the graphite target.  These data are of particular importance for the T2K beam, since this is the energy of the primary proton beam striking the neutrino production target.  However, these data are also very important for the NuMI and LBNF beam, as these cross sections are relevant for the secondary and tertiary protons produced inside the target.  

A mature analysis of the 120, 30 and 20 GeV/c p+C data, presented at a Fermilab Joint Experimental-Theorical Physics Seminar\cite{EMPHATIC_2019JETP} is nearing completion and will be published soon.  At the 30 and 20 GeV/c beam setting, a $>$99\%-pure selection of protons in the beam is achieved via selections based on the response of the upstream gas Ckov detectors provided by the FTBF.  A data-driven procedure uses an independent data set to align all tracking detectors.  Efficiencies have also been derived with a data-driven approach.  The silicon pixel telescope detector represents a small amount of extra material in the path of the particles, and simulations are used to correct for the impact this material has on the reconstructed scattering angle.  Fig.~\ref{fig:2018Uncert} shows the fractional uncertainty on each measurement bin of $p^2\theta^2$ for the 30 and 120 GeV/c data set.  The uncertainties for the 20 GeV/c data set is similar to that of the 30 GeV/c, however with larger statistical uncertainties.  We see that, since the scattering angles of the 120 GeV/c protons are much smaller than an the lower beam momenta, the uncertainty on the pixel telescope material correction is considerably larger at these low angles.  Nevertheless, in general, we see that the systematic uncertainties on these measurements are well below 5\%, and all measurements at $t > 0.02$ GeV$^2$/c$^2$ are limited by statistical uncertainties.

\section{\label{sec:ProposedMeasurements}Proposed Measurements and their Impact on Neutrino Flux Predictions}

\begin{figure*}[th]
\begin{center}
\includegraphics[width=0.49\textwidth]{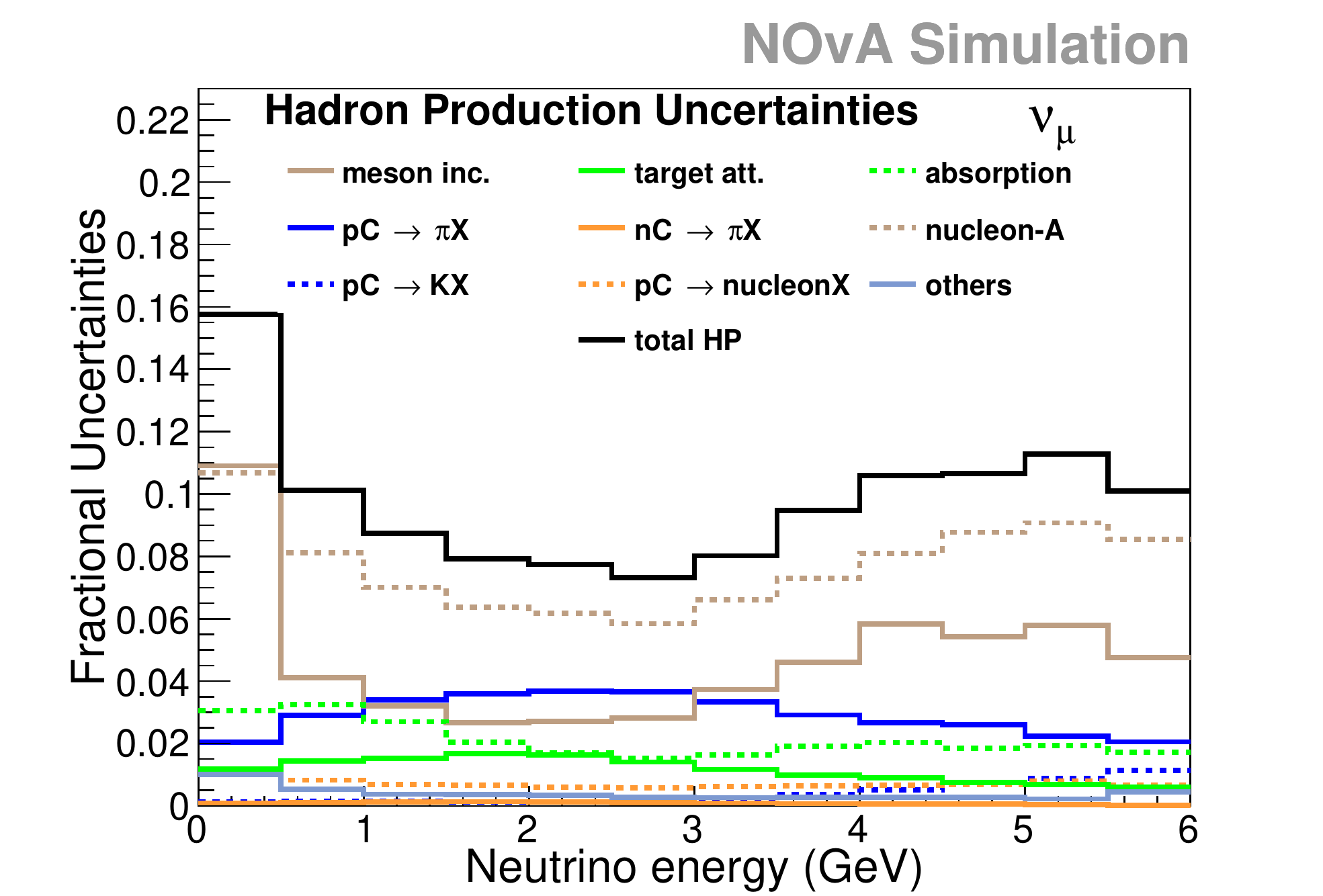}
\includegraphics[width=0.49\textwidth]{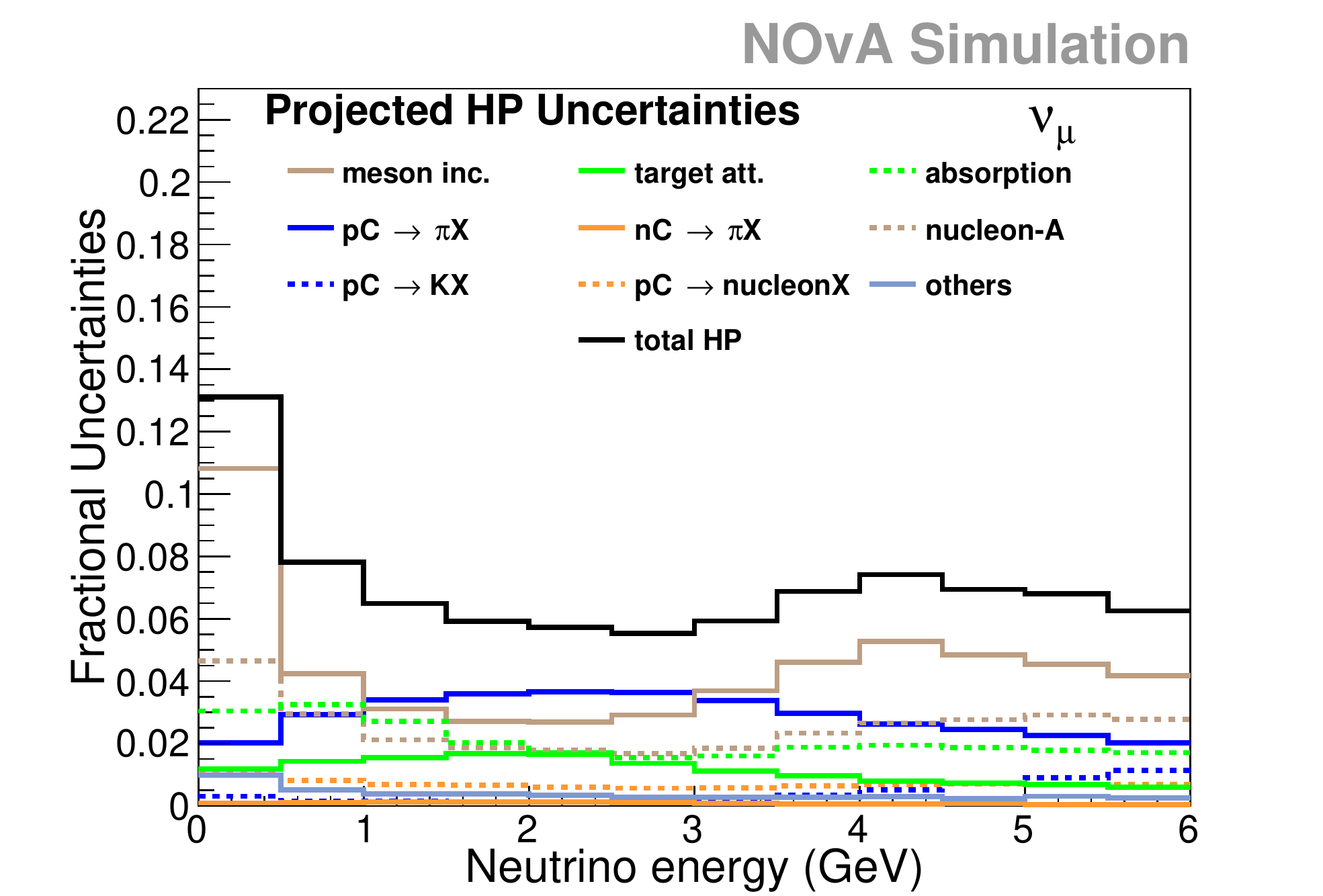}\\
\includegraphics[width=0.49\textwidth]{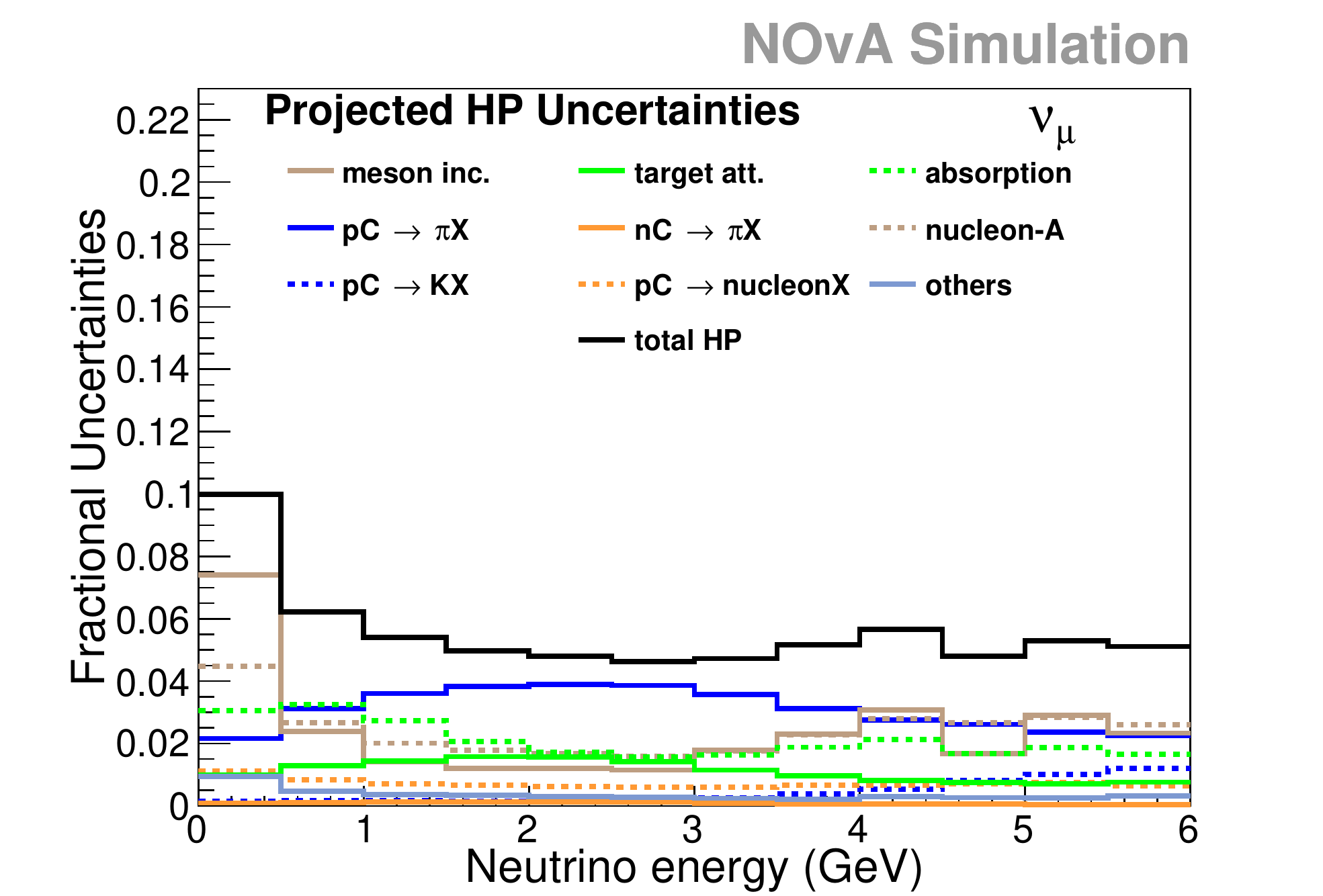}
\caption{Fractional uncertainties of the $\nu_\mu$ flux in the NOvA off-axis beam with current (top left) uncertainties, 10\% uncertainties in the quasi-elastic scattering cross section (top right), and all reduced hadron production uncertainties described in this section (bottom) \cite{NOvAFluxImprovements}.}
\label{fig:NOvAFluxWithEMPHATIC}
\end{center} 
\end{figure*}

We propose a comprehensive measurement of differential pion, kaon and proton interaction and production cross sections at beam momenta between 2-20 GeV, as well as at 30, 60, 80 and 120 GeV/c.  This includes quasi-elastic scattering cross sections with very small scattering angles.  Measurements will be made across a variety of nuclear targets, including at least carbon, aluminum and iron, which are the dominant materials in the production of  accelerator-based neutrino beams.  Other nuclear targets under consideration are Be, H$_2$O, Ca, Hg, and Ti.  Cross sections for oxygen and nitrogen can also be determined from measurements using solid BN, B$_2$O$_3$, and B targets, with the latter cross section subtracted from the first two.  All of these measurements are relevant for atmospheric and accelerator-produced neutrinos, and we expect most of these measurements to have better than 10\% uncertainties.

In the current NuMI and LBNF flux predictions\cite{PPFX}, 40\% uncertainties are assumed for interactions involving hadrons that are not currently covered by existing data.  In the study presented here, we assume that we are able to reduce the uncertainties on quasi-elastic and proton and pion production from 40\% to 10\% (note we have already demonstrated the ability to measure the quasi-elastic scaterring cross section with uncertainties below 5\%).  We also assume that the uncertainty on kaon production is reduced from 40\% to 20\%.  Figure \ref{fig:NOvAFluxWithEMPHATIC} shows the flux uncertainties for the NOvA off-axis experiment with current hadron production uncertainties (top left), with a reduction of only the quasi-elastic scattering cross section uncertainty from 40\% to 10\% (top right), and with all improvements in hadron production uncertainties mentioned above.  Because the current flux uncertainties due to hadron production uncertainties are dominated by uncertainties of the interactions of the lower energy secondary and tertiary interactions, these new measurements reduce the total flux uncertainty by nearly 50\%.  At the time of writing this proposal, similar plots are unavailable from the DUNE collaboration.  However, as shown in Figs.\ref{fig:NOvAFlux} and ~\ref{fig:DUNEFlux}, the level and underlying causes of uncertainties in the DUNE flux are very similar to that of the NuMI flux, so we expect a similar or larger reduction in the DUNE flux uncertainty.

\section{Staged Run Plans \label{sec:RunPlan}}

We propose a staged program of 3 data collection periods, each phase separated by approximately 1 year, and each lasting 3-4 weeks.  The three phases are shown in Table \ref{table:RunPlan}.  The timing and goals of the phases are defined by the subsystems expected to be available at each point in time.  The duration of each run period is limited by available manpower and FTBF schedules.  The upgraded readout electronics will enable the collection data using minimum bias beam triggers at rates around 100 kHz.  Therefore we expect to collect high statistics data for various beam momenta, beam species and nuclear target combinations during these runs.
Use of the beam during these phases will be maximized by pre-testing of all subsystems and developing a detailed installation and commissioning plan.

\begin{table*}[t]
\centering
\begin{tabular}{|c|c|c|c|c|c|} 
 \hline
\textbf{Phase} & \textbf{Date}& \textbf{Subsystems} & \textbf{Momenta} & \textbf{Targets} & \textbf{Goals} \\ 
\hline
 & & Beam Gas Ckov +& & & \\
  & Spring & Beam ACkov + & 4, 8, 12, & & Improved elastic and quasi-elastic\\
 1 & 2020 & FTBF SiStrip Detectors +& 20, 31,& C, Al, Fe& scattering measurements, low-\\
  & & Small-acceptance magnet (borrowed) +&60, 120 & &acceptance hadron production\\
  & & Downstream ACkov & & &measurements\\
  & & Time-of-flight & & & \\
\hline
 & & Beam Gas Ckov + & & & \\
  & & Beam ACkov + & & &\\
  & Spring & FTBF SiStrip Detectors + & 4, 8, 12, &C, Al, Fe, H$_2$O, &Full-acceptance hadron production\\
 2 & 2021 & New Large-area SiStrip Detectors + & 20, 31 &Be, B, BN, B$_2$O$_3$ &with PID up to 8 GeV/c\\
  & & Full-acceptance magnet +& 60, 120 & &\\
  & & Downstream ACkov +& & &\\
  & & Time-of-flight & & &\\
\hline
\multirow{2}{*}{3} & Spring & Same as Phase 2 + &20, 31, 60, & Same as Phase 2 + &Full-acceptance hadron production \\
  &2022 & Extended RICH &80, 120 & Ca, Hg, Ti& with PID up to 15 GeV/c\\
\hline
\end{tabular}
\caption{Details of the proposed 3-phase EMPHATIC run plan.  Each phase will have a duration of 3-4 weeks at the FTBF, not including pre-installation preparations.}
\label{table:RunPlan}
\end{table*}

Phase 1 will focus on hadron scattering and production off carbon, aluminum and iron at beam momenta of 4, 8, 12, 20, 31, 60 and 120 GeV/c.  This is approximately 70 different combinations of beam species, beam momenta and target.  By the start of this phase in Spring 2020, the only subsystems that will likely not be ready to use in the experiment are the new large-area silicon strip detector and the full acceptance magnet (due to funding constraints).  Therefore we will use a small-acceptance magnet borrowed from TRIUMF, and will remount and redistribute the silicon strip detectors available at the FTBF to provide limited angular acceptance.  Such a spectrometer, combined with the PID capabilities of the Ckov detectors, will enable improved background rejection for the elastic and quasi-elastic scattering measurements, as well as low-acceptance hadron production measurements.  Particle production with beams above 31 GeV/c will have limited impact because the secondary PID will be limited to a maximum momentum of 8 GeV/c.  

Phase 2 will incorporate the large-area silicon strip detectors and full acceptance magnet.  This will enable full acceptance hadron production with PID up to 8 GeV/c.  The same targets and beam momenta as in Phase 1 will be remeasured, and new targets will be included.  In Phase 3, the capabilities of the downstream RICH detector will be expanded with the inclusion of a volume with a gas with an index of refraction that enables PID to 15 GeV/c.  Hadron production measurements at higher beam momenta, and an extended list of targets, will be made during this final phase of the experiment.

We estimate that at each beam and target configuration, approximately 100k proton and pion interactions with hadrons produced in the final state can be collected in the matter of a few hours on a 5\% interaction-length target (depending on the beam intensity, which can be only a few kHz at very low momenta).  Kaon rates in the FTBF beam are typically low, and so measurements of kaon interactions will require longer running.  Therefore, in Phase 1, the data collection will take approximately 200 hours.  We expect some extra time will be required to achieve stable operations, but nevertheless, a 3 week run should be enough time to install and commission all detectors and collect all the data.

\section{\label{sec:Conclusion}Conclusion}
The broad-based, international EMPHATIC Collaboration proposes to collect new hadron production measurements with percent-level uncertainties.  EMPHATIC is complementary to existing efforts to collect new hadron production data to reduce neutrino flux uncertainties, since the beam energies at the FTBF are considerably lower than what can easily be achieved elsewhere.  Some data will be collected at higher energies in order to compare to existing measurements.  

EMPHATIC is very low risk for several reasons: very little R\&D is required since the detectors either already exist or the technology is well established, the compact design of the spectrometer results in a low cost of the detectors needed for particle tracking and identification, and the high-rate of data-collection using simple triggers reduces the actual running time of the experiment.  We have already demonstrated the feasibility of high-quality measurements of the quasi-elastic scattering cross section with just a few days of running at the FTBF in 2018.


\bibliography{main}

\begin{thebibliography}{60}%
\makeatletter
\providecommand \@ifxundefined [1]{%
 \@ifx{#1\undefined}
}%
\providecommand \@ifnum [1]{%
 \ifnum #1\expandafter \@firstoftwo
 \else \expandafter \@secondoftwo
 \fi
}%
\providecommand \@ifx [1]{%
 \ifx #1\expandafter \@firstoftwo
 \else \expandafter \@secondoftwo
 \fi
}%
\providecommand \natexlab [1]{#1}%
\providecommand \enquote  [1]{``#1''}%
\providecommand \bibnamefont  [1]{#1}%
\providecommand \bibfnamefont [1]{#1}%
\providecommand \citenamefont [1]{#1}%
\providecommand \href@noop [0]{\@secondoftwo}%
\providecommand \href [0]{\begingroup \@sanitize@url \@href}%
\providecommand \@href[1]{\@@startlink{#1}\@@href}%
\providecommand \@@href[1]{\endgroup#1\@@endlink}%
\providecommand \@sanitize@url [0]{\catcode `\\12\catcode `\$12\catcode
  `\&12\catcode `\#12\catcode `\^12\catcode `\_12\catcode `\%12\relax}%
\providecommand \@@startlink[1]{}%
\providecommand \@@endlink[0]{}%
\providecommand \url  [0]{\begingroup\@sanitize@url \@url }%
\providecommand \@url [1]{\endgroup\@href {#1}{\urlprefix }}%
\providecommand \urlprefix  [0]{URL }%
\providecommand \Eprint [0]{\href }%
\providecommand \doibase [0]{http://dx.doi.org/}%
\providecommand \selectlanguage [0]{\@gobble}%
\providecommand \bibinfo  [0]{\@secondoftwo}%
\providecommand \bibfield  [0]{\@secondoftwo}%
\providecommand \translation [1]{[#1]}%
\providecommand \BibitemOpen [0]{}%
\providecommand \bibitemStop [0]{}%
\providecommand \bibitemNoStop [0]{.\EOS\space}%
\providecommand \EOS [0]{\spacefactor3000\relax}%
\providecommand \BibitemShut  [1]{\csname bibitem#1\endcsname}%
\let\auto@bib@innerbib\@empty
\bibitem [{\citenamefont {Abe}\ \emph {et~al.}(2017)\citenamefont {Abe} \emph
  {et~al.}}]{T2KLatest}%
  \BibitemOpen
  \bibfield  {author} {\bibinfo {author} {\bibfnamefont {K.}~\bibnamefont
  {Abe}} \emph {et~al.} (\bibinfo {collaboration} {The T2K Collaboration}),\
  }\href@noop {} {\bibfield  {journal} {\bibinfo  {journal} {Phys. Rev. D}\
  }\textbf {\bibinfo {volume} {96}},\ \bibinfo {pages} {092006} (\bibinfo
  {year} {2017})}\BibitemShut {NoStop}%
\bibitem [{\citenamefont {Acero}\ \emph {et~al.}(2018)\citenamefont {Acero}
  \emph {et~al.}}]{NOvALatest}%
  \BibitemOpen
  \bibfield  {author} {\bibinfo {author} {\bibfnamefont {M.~A.}\ \bibnamefont
  {Acero}} \emph {et~al.} (\bibinfo {collaboration} {NOvA Collaboration}),\
  }\href@noop {} {\bibfield  {journal} {\bibinfo  {journal} {Phys. Rev. D}\
  }\textbf {\bibinfo {volume} {98}},\ \bibinfo {pages} {032012} (\bibinfo
  {year} {2018})}\BibitemShut {NoStop}%
\bibitem [{\citenamefont {Abe}\ \emph {et~al.}(2018{\natexlab{a}})\citenamefont
  {Abe} \emph {et~al.}}]{H2KProposal}%
  \BibitemOpen
  \bibfield  {author} {\bibinfo {author} {\bibfnamefont {K.}~\bibnamefont
  {Abe}} \emph {et~al.} (\bibinfo {collaboration} {Hyper-Kamiokande}),\
  }\href@noop {} {\enquote {\bibinfo {title} {{Hyper-Kamiokande Design
  Report}},}\ } (\bibinfo {year} {2018}{\natexlab{a}}),\ \Eprint
  {http://arxiv.org/abs/1805.04163} {arXiv:1805.04163} \BibitemShut {NoStop}%
\bibitem [{\citenamefont {Abi}\ \emph {et~al.}(2018)\citenamefont {Abi} \emph
  {et~al.}}]{DUNEIDR}%
  \BibitemOpen
  \bibfield  {author} {\bibinfo {author} {\bibfnamefont {B.}~\bibnamefont
  {Abi}} \emph {et~al.} (\bibinfo {collaboration} {DUNE}),\ }\href@noop {}
  {\bibfield  {journal} {\bibinfo  {journal} {FERMILAB-DESIGN-2018-02}\ }
  (\bibinfo {year} {2018})}\BibitemShut {NoStop}%
\bibitem [{PON(2019)}]{PONDD}%
  \BibitemOpen
  \href {{https://indico.fnal.gov/event/18430/}} {\enquote {\bibinfo {title}
  {Workshop on the {P}hysics {O}pportunities in the {N}ear {DUNE} {D}etector
  {H}all},}\ } (\bibinfo {year} {2019})\BibitemShut {NoStop}%
\bibitem [{\citenamefont {Valencia}\ \emph {et~al.}(2019)\citenamefont
  {Valencia} \emph {et~al.}}]{MINERvA_nu_e}%
  \BibitemOpen
  \bibfield  {author} {\bibinfo {author} {\bibfnamefont {E.}~\bibnamefont
  {Valencia}} \emph {et~al.},\ }\href@noop {} {\enquote {\bibinfo {title}
  {{Constraint of the MINERvA Medium Energy Neutrino Flux using
  Neutrino-Electron Elastic Scattering}},}\ } (\bibinfo {year} {2019}),\
  \Eprint {http://arxiv.org/abs/1906.00111} {arXiv:1906.00111} \BibitemShut
  {NoStop}%
\bibitem [{\citenamefont {{NOvA Collaboration}}(2018)}]{NOvAFluxCurrent}%
  \BibitemOpen
  \bibfield  {author} {\bibinfo {author} {\bibnamefont {{NOvA
  Collaboration}}},\ }\href@noop {} {}\bibinfo {howpublished} {private
  communication} (\bibinfo {year} {2018})\BibitemShut {NoStop}%
\bibitem [{\citenamefont {{DUNE Collaboration}}(2019)}]{DUNEFluxCurrent}%
  \BibitemOpen
  \bibfield  {author} {\bibinfo {author} {\bibnamefont {{DUNE
  Collaboration}}},\ }\href@noop {} {}\bibinfo {howpublished} {private
  communication} (\bibinfo {year} {2019})\BibitemShut {NoStop}%
\bibitem [{\citenamefont {Alt}\ \emph {et~al.}(2007)\citenamefont {Alt} \emph
  {et~al.}}]{NA49}%
  \BibitemOpen
  \bibfield  {author} {\bibinfo {author} {\bibfnamefont {C.}~\bibnamefont
  {Alt}} \emph {et~al.} (\bibinfo {collaboration} {NA49}),\ }\href@noop {}
  {\bibfield  {journal} {\bibinfo  {journal} {Eur. Phys. J.}\ }\textbf
  {\bibinfo {volume} {C49}},\ \bibinfo {pages} {897} (\bibinfo {year}
  {2007})}\BibitemShut {NoStop}%
\bibitem [{\citenamefont {Abgrall}\ \emph {et~al.}(2011)\citenamefont {Abgrall}
  \emph {et~al.}}]{NA61}%
  \BibitemOpen
  \bibfield  {author} {\bibinfo {author} {\bibfnamefont {N.}~\bibnamefont
  {Abgrall}} \emph {et~al.} (\bibinfo {collaboration} {The NA61/SHINE
  Collaboration}),\ }\href@noop {} {\bibfield  {journal} {\bibinfo  {journal}
  {Phys. Rev. C}\ }\textbf {\bibinfo {volume} {84}},\ \bibinfo {pages} {034604}
  (\bibinfo {year} {2011})}\BibitemShut {NoStop}%
\bibitem [{\citenamefont {Aliaga}\ \emph {et~al.}(2016)\citenamefont {Aliaga}
  \emph {et~al.}}]{PPFX}%
  \BibitemOpen
  \bibfield  {author} {\bibinfo {author} {\bibfnamefont {L.}~\bibnamefont
  {Aliaga}} \emph {et~al.} (\bibinfo {collaboration} {MINER\ensuremath{\nu}A
  Collaboration}),\ }\href@noop {} {\bibfield  {journal} {\bibinfo  {journal}
  {Phys. Rev. D}\ }\textbf {\bibinfo {volume} {94}},\ \bibinfo {pages} {092005}
  (\bibinfo {year} {2016})}\BibitemShut {NoStop}%
\bibitem [{\citenamefont {Vladisavljevic}(2018{\natexlab{a}})}]{T2K-2009NA61}%
  \BibitemOpen
  \bibfield  {author} {\bibinfo {author} {\bibfnamefont {T.}~\bibnamefont
  {Vladisavljevic}},\ }\href@noop {} {\enquote {\bibinfo {title} {{Constraining
  the T2K Neutrino Flux Prediction with 2009 NA61/SHINE Replica-Target
  Data}},}\ } (\bibinfo {year} {2018}{\natexlab{a}}),\ \Eprint
  {http://arxiv.org/abs/1804.00272} {arXiv:1804.00272} \BibitemShut {NoStop}%
\bibitem [{\citenamefont {Abgrall}\ \emph
  {et~al.}(2016{\natexlab{a}})\citenamefont {Abgrall} \emph
  {et~al.}}]{Abgrall:2016jif}%
  \BibitemOpen
  \bibfield  {author} {\bibinfo {author} {\bibfnamefont {N.}~\bibnamefont
  {Abgrall}} \emph {et~al.} (\bibinfo {collaboration} {NA61/SHINE
  Collaboration}),\ }\href {\doibase 10.1140/epjc/s10052-016-4440-y} {\bibfield
   {journal} {\bibinfo  {journal} {Eur. Phys. J. C}\ }\textbf {\bibinfo
  {volume} {76}},\ \bibinfo {pages} {617} (\bibinfo {year}
  {2016}{\natexlab{a}})}\BibitemShut {NoStop}%
\bibitem [{\citenamefont {Paley}\ \emph {et~al.}(2014)\citenamefont {Paley}
  \emph {et~al.}}]{MIPP}%
  \BibitemOpen
  \bibfield  {author} {\bibinfo {author} {\bibfnamefont {J.~M.}\ \bibnamefont
  {Paley}} \emph {et~al.} (\bibinfo {collaboration} {MIPP Collaboration}),\
  }\href {\doibase 10.1103/PhysRevD.90.032001} {\bibfield  {journal} {\bibinfo
  {journal} {Phys. Rev. D}\ }\textbf {\bibinfo {volume} {90}},\ \bibinfo
  {pages} {032001} (\bibinfo {year} {2014})}\BibitemShut {NoStop}%
\bibitem [{\citenamefont {Vladisavljevic}(2018{\natexlab{b}})}]{T2KFluxTuning}%
  \BibitemOpen
  \bibfield  {author} {\bibinfo {author} {\bibfnamefont {T.}~\bibnamefont
  {Vladisavljevic}},\ }\href@noop {} {\enquote {\bibinfo {title} {12th
  international workshop on neutrino-nucleus scattering in the few-gev
  region},}\ } (\bibinfo {year} {2018}{\natexlab{b}})\BibitemShut {NoStop}%
\bibitem [{\citenamefont {Abgrall}\ \emph
  {et~al.}(2016{\natexlab{b}})\citenamefont {Abgrall} \emph
  {et~al.}}]{Abgrall:2015hmv}%
  \BibitemOpen
  \bibfield  {author} {\bibinfo {author} {\bibfnamefont {N.}~\bibnamefont
  {Abgrall}} \emph {et~al.},\ }\href {\doibase 10.1140/epjc/s10052-016-3898-y}
  {\bibfield  {journal} {\bibinfo  {journal} {Eur. Phys. J. C}\ }\textbf
  {\bibinfo {volume} {76}},\ \bibinfo {pages} {84} (\bibinfo {year}
  {2016}{\natexlab{b}})}\BibitemShut {NoStop}%
\bibitem [{\citenamefont {Apollonio}\ \emph {et~al.}(2009)\citenamefont
  {Apollonio} \emph {et~al.}}]{Apollonio:2009bu}%
  \BibitemOpen
  \bibfield  {author} {\bibinfo {author} {\bibfnamefont {M.}~\bibnamefont
  {Apollonio}} \emph {et~al.},\ }\href@noop {} {\bibfield  {journal} {\bibinfo
  {journal} {Nucl. Phys. A}\ }\textbf {\bibinfo {volume} {821}},\ \bibinfo
  {pages} {118} (\bibinfo {year} {2009})}\BibitemShut {NoStop}%
\bibitem [{\citenamefont {Bhadra}\ \emph {et~al.}(2015)\citenamefont {Bhadra}
  \emph {et~al.}}]{E61Proposal}%
  \BibitemOpen
  \bibfield  {author} {\bibinfo {author} {\bibfnamefont {S.}~\bibnamefont
  {Bhadra}} \emph {et~al.},\ }\href@noop {} {\bibfield  {journal} {\bibinfo
  {journal} {Proposal to J-PARC PAC}\ } (\bibinfo {year} {2015})}\BibitemShut
  {NoStop}%
\bibitem [{\citenamefont {Aguilar-Arevalo}\ \emph {et~al.}(2009)\citenamefont
  {Aguilar-Arevalo} \emph {et~al.}}]{AguilarArevalo:2008yp}%
  \BibitemOpen
  \bibfield  {author} {\bibinfo {author} {\bibfnamefont {A.~A.}\ \bibnamefont
  {Aguilar-Arevalo}} \emph {et~al.} (\bibinfo {collaboration} {MiniBooNE}),\
  }\href {\doibase 10.1103/PhysRevD.79.072002} {\bibfield  {journal} {\bibinfo
  {journal} {Phys. Rev.}\ }\textbf {\bibinfo {volume} {D79}},\ \bibinfo {pages}
  {072002} (\bibinfo {year} {2009})},\ \Eprint {http://arxiv.org/abs/0806.1449}
  {arXiv:0806.1449 [hep-ex]} \BibitemShut {NoStop}%
\bibitem [{\citenamefont {Catanesi}\ \emph {et~al.}(2007)\citenamefont
  {Catanesi} \emph {et~al.}}]{Catanesi:2007ab}%
  \BibitemOpen
  \bibfield  {author} {\bibinfo {author} {\bibfnamefont {M.~G.}\ \bibnamefont
  {Catanesi}} \emph {et~al.} (\bibinfo {collaboration} {HARP}),\ }\href
  {\doibase 10.1140/epjc/s10052-007-0382-8} {\bibfield  {journal} {\bibinfo
  {journal} {Eur. Phys. J.}\ }\textbf {\bibinfo {volume} {C52}},\ \bibinfo
  {pages} {29} (\bibinfo {year} {2007})},\ \Eprint
  {http://arxiv.org/abs/hep-ex/0702024} {arXiv:hep-ex/0702024 [hep-ex]}
  \BibitemShut {NoStop}%
\bibitem [{\citenamefont {Chemakin}\ \emph {et~al.}(2008)\citenamefont
  {Chemakin} \emph {et~al.}}]{Chemakin:2007aa}%
  \BibitemOpen
  \bibfield  {author} {\bibinfo {author} {\bibfnamefont {I.}~\bibnamefont
  {Chemakin}} \emph {et~al.} (\bibinfo {collaboration} {E910}),\ }\href
  {\doibase 10.1103/PhysRevC.77.049903, 10.1103/PhysRevC.77.015209} {\bibfield
  {journal} {\bibinfo  {journal} {Phys. Rev.}\ }\textbf {\bibinfo {volume}
  {C77}},\ \bibinfo {pages} {015209} (\bibinfo {year} {2008})},\ \bibinfo
  {note} {[Erratum: Phys. Rev.C77,049903(2008)]},\ \Eprint
  {http://arxiv.org/abs/0707.2375} {arXiv:0707.2375 [nucl-ex]} \BibitemShut
  {NoStop}%
\bibitem [{\citenamefont {Aleshin}\ \emph {et~al.}(1977)\citenamefont
  {Aleshin}, \citenamefont {Drabkin},\ and\ \citenamefont
  {Kolesnikov}}]{Aleshin:1977bb}%
  \BibitemOpen
  \bibfield  {author} {\bibinfo {author} {\bibfnamefont {{\relax Yu}.~D.}\
  \bibnamefont {Aleshin}}, \bibinfo {author} {\bibfnamefont {I.~A.}\
  \bibnamefont {Drabkin}}, \ and\ \bibinfo {author} {\bibfnamefont {V.~V.}\
  \bibnamefont {Kolesnikov}},\ }\href@noop {} {\bibfield  {journal} {\bibinfo
  {journal} {ITEP}\ }\textbf {\bibinfo {volume} {80}} (\bibinfo {year}
  {1977})}\BibitemShut {NoStop}%
\bibitem [{\citenamefont {Eichten}\ \emph {et~al.}(1972)\citenamefont {Eichten}
  \emph {et~al.}}]{Eichten:1972nw}%
  \BibitemOpen
  \bibfield  {author} {\bibinfo {author} {\bibfnamefont {T.}~\bibnamefont
  {Eichten}} \emph {et~al.},\ }\href {\doibase 10.1016/0550-3213(72)90120-4}
  {\bibfield  {journal} {\bibinfo  {journal} {Nucl. Phys.}\ }\textbf {\bibinfo
  {volume} {B44}},\ \bibinfo {pages} {333} (\bibinfo {year}
  {1972})}\BibitemShut {NoStop}%
\bibitem [{\citenamefont {Dekkers}\ \emph {et~al.}(1965)\citenamefont
  {Dekkers}, \citenamefont {Geibel}, \citenamefont {Mermod}, \citenamefont
  {Weber}, \citenamefont {Willitts}, \citenamefont {Winter}, \citenamefont
  {Jordan}, \citenamefont {Vivargent}, \citenamefont {King},\ and\
  \citenamefont {Wilson}}]{Dekkers:1965zz}%
  \BibitemOpen
  \bibfield  {author} {\bibinfo {author} {\bibfnamefont {D.}~\bibnamefont
  {Dekkers}}, \bibinfo {author} {\bibfnamefont {J.~A.}\ \bibnamefont {Geibel}},
  \bibinfo {author} {\bibfnamefont {R.}~\bibnamefont {Mermod}}, \bibinfo
  {author} {\bibfnamefont {G.}~\bibnamefont {Weber}}, \bibinfo {author}
  {\bibfnamefont {T.~R.}\ \bibnamefont {Willitts}}, \bibinfo {author}
  {\bibfnamefont {K.}~\bibnamefont {Winter}}, \bibinfo {author} {\bibfnamefont
  {B.}~\bibnamefont {Jordan}}, \bibinfo {author} {\bibfnamefont
  {M.}~\bibnamefont {Vivargent}}, \bibinfo {author} {\bibfnamefont {N.~M.}\
  \bibnamefont {King}}, \ and\ \bibinfo {author} {\bibfnamefont {E.~J.~N.}\
  \bibnamefont {Wilson}},\ }\href {\doibase 10.1103/PhysRev.137.B962}
  {\bibfield  {journal} {\bibinfo  {journal} {Phys. Rev.}\ }\textbf {\bibinfo
  {volume} {137}},\ \bibinfo {pages} {B962} (\bibinfo {year}
  {1965})}\BibitemShut {NoStop}%
\bibitem [{\citenamefont {Abbott}\ \emph {et~al.}(1992)\citenamefont {Abbott}
  \emph {et~al.}}]{Abbott:1991en}%
  \BibitemOpen
  \bibfield  {author} {\bibinfo {author} {\bibfnamefont {T.}~\bibnamefont
  {Abbott}} \emph {et~al.} (\bibinfo {collaboration} {E-802}),\ }\href
  {\doibase 10.1103/PhysRevD.45.3906} {\bibfield  {journal} {\bibinfo
  {journal} {Phys. Rev.}\ }\textbf {\bibinfo {volume} {D45}},\ \bibinfo {pages}
  {3906} (\bibinfo {year} {1992})}\BibitemShut {NoStop}%
\bibitem [{\citenamefont {Marmer}\ and\ \citenamefont
  {Lundquist}(1971)}]{Marmer:1971gd}%
  \BibitemOpen
  \bibfield  {author} {\bibinfo {author} {\bibfnamefont {G.~J.}\ \bibnamefont
  {Marmer}}\ and\ \bibinfo {author} {\bibfnamefont {D.~E.}\ \bibnamefont
  {Lundquist}},\ }\href {\doibase 10.1103/PhysRevD.3.1089} {\bibfield
  {journal} {\bibinfo  {journal} {Phys. Rev.}\ }\textbf {\bibinfo {volume}
  {D3}},\ \bibinfo {pages} {1089} (\bibinfo {year} {1971})}\BibitemShut
  {NoStop}%
\bibitem [{\citenamefont {Vorontsov}\ \emph {et~al.}(1988)\citenamefont
  {Vorontsov}, \citenamefont {Safronov}, \citenamefont {Sibirtsev},
  \citenamefont {Smirnov},\ and\ \citenamefont
  {Trebukhovsky}}]{Vorontsov:1988pu}%
  \BibitemOpen
  \bibfield  {author} {\bibinfo {author} {\bibfnamefont {I.~A.}\ \bibnamefont
  {Vorontsov}}, \bibinfo {author} {\bibfnamefont {G.~A.}\ \bibnamefont
  {Safronov}}, \bibinfo {author} {\bibfnamefont {A.~A.}\ \bibnamefont
  {Sibirtsev}}, \bibinfo {author} {\bibfnamefont {G.~N.}\ \bibnamefont
  {Smirnov}}, \ and\ \bibinfo {author} {\bibfnamefont {{\relax Yu}.~V.}\
  \bibnamefont {Trebukhovsky}},\ }\href@noop {} {\bibfield  {journal} {\bibinfo
   {journal} {ITEP}\ }\textbf {\bibinfo {volume} {11}} (\bibinfo {year}
  {1988})}\BibitemShut {NoStop}%
\bibitem [{\citenamefont {Allaby}\ \emph {et~al.}(1970)\citenamefont {Allaby},
  \citenamefont {Binon}, \citenamefont {Diddens}, \citenamefont {Duteil},
  \citenamefont {Klovning},\ and\ \citenamefont {Meunier}}]{Allaby:1970jt}%
  \BibitemOpen
  \bibfield  {author} {\bibinfo {author} {\bibfnamefont {J.~V.}\ \bibnamefont
  {Allaby}}, \bibinfo {author} {\bibfnamefont {F.~G.}\ \bibnamefont {Binon}},
  \bibinfo {author} {\bibfnamefont {A.~N.}\ \bibnamefont {Diddens}}, \bibinfo
  {author} {\bibfnamefont {P.}~\bibnamefont {Duteil}}, \bibinfo {author}
  {\bibfnamefont {A.}~\bibnamefont {Klovning}}, \ and\ \bibinfo {author}
  {\bibfnamefont {R.}~\bibnamefont {Meunier}},\ }\href@noop {} {\enquote
  {\bibinfo {title} {{High-energy particle spectra from proton interactions at
  19.2-GeV/c}},}\ } (\bibinfo {year} {1970}),\ \Eprint
  {http://arxiv.org/abs/{\bf CERN-70-12}} {{\bf CERN-70-12}} \BibitemShut
  {NoStop}%
\bibitem [{\citenamefont {Abe}\ \emph {et~al.}(2018{\natexlab{b}})\citenamefont
  {Abe} \emph {et~al.}}]{SKLatestAtm}%
  \BibitemOpen
  \bibfield  {author} {\bibinfo {author} {\bibfnamefont {K.}~\bibnamefont
  {Abe}} \emph {et~al.} (\bibinfo {collaboration} {{S}uper-{K}amiokande
  {C}ollaboration}),\ }\href@noop {} {\bibfield  {journal} {\bibinfo  {journal}
  {Phys. Rev. D}\ }\textbf {\bibinfo {volume} {97}},\ \bibinfo {pages} {072001}
  (\bibinfo {year} {2018}{\natexlab{b}})}\BibitemShut {NoStop}%
\bibitem [{\citenamefont {{Fermilab Test Beam Facility}}(2018)}]{FTBF}%
  \BibitemOpen
  \bibfield  {author} {\bibinfo {author} {\bibnamefont {{Fermilab Test Beam
  Facility}}},\ }\href@noop {} {}\bibinfo {howpublished}
  {\url{http://ftbf.fnal.gov}} (\bibinfo {year} {2018})\BibitemShut {NoStop}%
\bibitem [{\citenamefont {Noumi}\ \emph {et~al.}(2012)\citenamefont {Noumi}
  \emph {et~al.}}]{Noumi}%
  \BibitemOpen
  \bibfield  {author} {\bibinfo {author} {\bibfnamefont {H.}~\bibnamefont
  {Noumi}} \emph {et~al.},\ }\href@noop {} {\bibfield  {journal} {\bibinfo
  {journal} {J-PARC E50 Proposal}\ } (\bibinfo {year} {2012})}\BibitemShut
  {NoStop}%
\bibitem [{\citenamefont {Shirotori}\ \emph {et~al.}(2013)\citenamefont
  {Shirotori} \emph {et~al.}}]{CBS}%
  \BibitemOpen
  \bibfield  {author} {\bibinfo {author} {\bibfnamefont {K.}~\bibnamefont
  {Shirotori}} \emph {et~al.},\ }\href@noop {} {\bibfield  {journal} {\bibinfo
  {journal} {PoS (Hadron 2013)}\ } (\bibinfo {year} {2013})}\BibitemShut
  {NoStop}%
\bibitem [{\citenamefont {Kim}\ \emph {et~al.}(2014)\citenamefont {Kim} \emph
  {et~al.}}]{Hosaka}%
  \BibitemOpen
  \bibfield  {author} {\bibinfo {author} {\bibfnamefont {S.~H.}\ \bibnamefont
  {Kim}} \emph {et~al.},\ }\href@noop {} {\bibfield  {journal} {\bibinfo
  {journal} {Prog. Theor. Exp. Phys.}\ }\textbf {\bibinfo {volume} {103D01}}
  (\bibinfo {year} {2014})}\BibitemShut {NoStop}%
\bibitem [{\citenamefont {Nara}\ \emph {et~al.}(2000)\citenamefont {Nara} \emph
  {et~al.}}]{JAM}%
  \BibitemOpen
  \bibfield  {author} {\bibinfo {author} {\bibfnamefont {Y.}~\bibnamefont
  {Nara}} \emph {et~al.},\ }\href@noop {} {\bibfield  {journal} {\bibinfo
  {journal} {Phys. Rev. C.}\ }\textbf {\bibinfo {volume} {61}},\ \bibinfo
  {pages} {024901} (\bibinfo {year} {2000})}\BibitemShut {NoStop}%
\bibitem [{\citenamefont {Sj\"{o}strand}\ \emph {et~al.}(2006)\citenamefont
  {Sj\"{o}strand} \emph {et~al.}}]{PYTHIA}%
  \BibitemOpen
  \bibfield  {author} {\bibinfo {author} {\bibfnamefont {T.}~\bibnamefont
  {Sj\"{o}strand}} \emph {et~al.},\ }\href@noop {} {\bibfield  {journal}
  {\bibinfo  {journal} {JHEP05}\ } (\bibinfo {year} {2006})}\BibitemShut
  {NoStop}%
\bibitem [{\citenamefont {Tabata}\ \emph {et~al.}(2010)\citenamefont {Tabata}
  \emph {et~al.}}]{Aerogel1}%
  \BibitemOpen
  \bibfield  {author} {\bibinfo {author} {\bibfnamefont {M.}~\bibnamefont
  {Tabata}} \emph {et~al.},\ }\href@noop {} {\bibfield  {journal} {\bibinfo
  {journal} {Nucl. Instr. Meth. A}\ }\textbf {\bibinfo {volume} {623}},\
  \bibinfo {pages} {339} (\bibinfo {year} {2010})}\BibitemShut {NoStop}%
\bibitem [{\citenamefont {Tabata}\ \emph {et~al.}(2012)\citenamefont {Tabata}
  \emph {et~al.}}]{Aerogel2}%
  \BibitemOpen
  \bibfield  {author} {\bibinfo {author} {\bibfnamefont {M.}~\bibnamefont
  {Tabata}} \emph {et~al.},\ }\href@noop {} {\bibfield  {journal} {\bibinfo
  {journal} {Nucl. Instr. Meth. A}\ }\textbf {\bibinfo {volume} {668}},\
  \bibinfo {pages} {64} (\bibinfo {year} {2012})}\BibitemShut {NoStop}%
\bibitem [{\citenamefont {{T}he {CMS}~{C}ollaboration}(2014)}]{CMSTracking}%
  \BibitemOpen
  \bibfield  {author} {\bibinfo {author} {\bibnamefont {{T}he
  {CMS}~{C}ollaboration}},\ }\href@noop {} {\bibfield  {journal} {\bibinfo
  {journal} {JINST}\ }\textbf {\bibinfo {volume} {9}},\ \bibinfo {pages}
  {P10009} (\bibinfo {year} {2014})}\BibitemShut {NoStop}%
\bibitem [{\citenamefont {Callier}\ \emph {et~al.}(2011)\citenamefont {Callier}
  \emph {et~al.}}]{SKIROC2}%
  \BibitemOpen
  \bibfield  {author} {\bibinfo {author} {\bibfnamefont {S.}~\bibnamefont
  {Callier}} \emph {et~al.},\ }\href@noop {} {\bibfield  {journal} {\bibinfo
  {journal} {JINST}\ }\textbf {\bibinfo {volume} {6}} (\bibinfo {year}
  {2011})}\BibitemShut {NoStop}%
\bibitem [{\citenamefont {Fukuda}\ \emph {et~al.}(2013)\citenamefont {Fukuda}
  \emph {et~al.}}]{LAtracking1}%
  \BibitemOpen
  \bibfield  {author} {\bibinfo {author} {\bibfnamefont {T.}~\bibnamefont
  {Fukuda}} \emph {et~al.},\ }\href@noop {} {\bibfield  {journal} {\bibinfo
  {journal} {JINST}\ }\textbf {\bibinfo {volume} {8}},\ \bibinfo {pages}
  {P01023} (\bibinfo {year} {2013})}\BibitemShut {NoStop}%
\bibitem [{\citenamefont {Fukuda}\ \emph {et~al.}(2014)\citenamefont {Fukuda}
  \emph {et~al.}}]{LAtracking2}%
  \BibitemOpen
  \bibfield  {author} {\bibinfo {author} {\bibfnamefont {T.}~\bibnamefont
  {Fukuda}} \emph {et~al.},\ }\href@noop {} {\bibfield  {journal} {\bibinfo
  {journal} {JINST}\ }\textbf {\bibinfo {volume} {9}},\ \bibinfo {pages}
  {P12017} (\bibinfo {year} {2014})}\BibitemShut {NoStop}%
\bibitem [{\citenamefont {Ishida}\ \emph {et~al.}(2014)\citenamefont {Ishida}
  \emph {et~al.}}]{HadronAnaly}%
  \BibitemOpen
  \bibfield  {author} {\bibinfo {author} {\bibfnamefont {H.}~\bibnamefont
  {Ishida}} \emph {et~al.},\ }\href@noop {} {\bibfield  {journal} {\bibinfo
  {journal} {PTEP}\ ,\ \bibinfo {pages} {093C01}} (\bibinfo {year}
  {2014})}\BibitemShut {NoStop}%
\bibitem [{\citenamefont {Yoshimoto}\ \emph {et~al.}(2017)\citenamefont
  {Yoshimoto} \emph {et~al.}}]{HTS}%
  \BibitemOpen
  \bibfield  {author} {\bibinfo {author} {\bibfnamefont {M.}~\bibnamefont
  {Yoshimoto}} \emph {et~al.},\ }\href@noop {} {\bibfield  {journal} {\bibinfo
  {journal} {PTEP}\ ,\ \bibinfo {pages} {103H01}} (\bibinfo {year}
  {2017})}\BibitemShut {NoStop}%
\bibitem [{\citenamefont {Halbach}(2008)}]{Halbach}%
  \BibitemOpen
  \bibfield  {author} {\bibinfo {author} {\bibfnamefont {K.}~\bibnamefont
  {Halbach}},\ }\href@noop {} {\bibfield  {journal} {\bibinfo  {journal} {Nucl.
  Instr. Meth. A}\ }\textbf {\bibinfo {volume} {592}},\ \bibinfo {pages} {56}
  (\bibinfo {year} {2008})}\BibitemShut {NoStop}%
\bibitem [{\citenamefont {Gorbunov}\ and\ \citenamefont
  {Kisel}(2006)}]{Kalman1}%
  \BibitemOpen
  \bibfield  {author} {\bibinfo {author} {\bibfnamefont {S.}~\bibnamefont
  {Gorbunov}}\ and\ \bibinfo {author} {\bibfnamefont {I.}~\bibnamefont
  {Kisel}},\ }\href@noop {} {\bibfield  {journal} {\bibinfo  {journal} {Nucl.
  Instr. Meth. A}\ }\textbf {\bibinfo {volume} {559}},\ \bibinfo {pages} {148}
  (\bibinfo {year} {2006})}\BibitemShut {NoStop}%
\bibitem [{\citenamefont {Wolin}\ and\ \citenamefont {Ho}(1993)}]{Kalman2}%
  \BibitemOpen
  \bibfield  {author} {\bibinfo {author} {\bibfnamefont {E.~J.}\ \bibnamefont
  {Wolin}}\ and\ \bibinfo {author} {\bibfnamefont {L.~L.}\ \bibnamefont {Ho}},\
  }\href@noop {} {\bibfield  {journal} {\bibinfo  {journal} {Nucl. Instr. Meth.
  A}\ }\textbf {\bibinfo {volume} {329}},\ \bibinfo {pages} {493} (\bibinfo
  {year} {1993})}\BibitemShut {NoStop}%
\bibitem [{\citenamefont {Fr{\"u}hwirth}(1987)}]{Kalman3}%
  \BibitemOpen
  \bibfield  {author} {\bibinfo {author} {\bibfnamefont {R.}~\bibnamefont
  {Fr{\"u}hwirth}},\ }\href@noop {} {\bibfield  {journal} {\bibinfo  {journal}
  {Nucl. Instr. Meth. A}\ }\textbf {\bibinfo {volume} {262}},\ \bibinfo {pages}
  {444} (\bibinfo {year} {1987})}\BibitemShut {NoStop}%
\bibitem [{\citenamefont {Nishida}\ \emph {et~al.}(2014)\citenamefont {Nishida}
  \emph {et~al.}}]{ARICH1}%
  \BibitemOpen
  \bibfield  {author} {\bibinfo {author} {\bibfnamefont {K.}~\bibnamefont
  {Nishida}} \emph {et~al.},\ }\href@noop {} {\bibfield  {journal} {\bibinfo
  {journal} {Nucl. Instr. Meth. A}\ }\textbf {\bibinfo {volume} {766}},\
  \bibinfo {pages} {28} (\bibinfo {year} {2014})}\BibitemShut {NoStop}%
\bibitem [{\citenamefont {Pestotnik}\ \emph {et~al.}(2017)\citenamefont
  {Pestotnik} \emph {et~al.}}]{ARICH2}%
  \BibitemOpen
  \bibfield  {author} {\bibinfo {author} {\bibfnamefont {R.}~\bibnamefont
  {Pestotnik}} \emph {et~al.},\ }\href@noop {} {\bibfield  {journal} {\bibinfo
  {journal} {Nucl. Instr. Meth. A}\ }\textbf {\bibinfo {volume} {876}},\
  \bibinfo {pages} {265} (\bibinfo {year} {2017})}\BibitemShut {NoStop}%
\bibitem [{\citenamefont {Tabata}\ \emph {et~al.}(2014)\citenamefont {Tabata}
  \emph {et~al.}}]{Aerogel3}%
  \BibitemOpen
  \bibfield  {author} {\bibinfo {author} {\bibfnamefont {M.}~\bibnamefont
  {Tabata}} \emph {et~al.},\ }\href@noop {} {\bibfield  {journal} {\bibinfo
  {journal} {Nucl. Instr. Meth. A}\ }\textbf {\bibinfo {volume} {766}},\
  \bibinfo {pages} {212} (\bibinfo {year} {2014})}\BibitemShut {NoStop}%
\bibitem [{\citenamefont {Yamaga}\ \emph {et~al.}(2014)\citenamefont {Yamaga}
  \emph {et~al.}}]{Yamaga}%
  \BibitemOpen
  \bibfield  {author} {\bibinfo {author} {\bibfnamefont {T.}~\bibnamefont
  {Yamaga}} \emph {et~al.},\ }\href@noop {} {\bibfield  {journal} {\bibinfo
  {journal} {Nucl. Instr. Meth. A}\ }\textbf {\bibinfo {volume} {766}},\
  \bibinfo {pages} {36} (\bibinfo {year} {2014})}\BibitemShut {NoStop}%
\bibitem [{\citenamefont {{Hamamatsu Photonics K.K.}}(2017)}]{Hamamatsu}%
  \BibitemOpen
  \bibfield  {author} {\bibinfo {author} {\bibnamefont {{Hamamatsu Photonics
  K.K.}}},\ }\href@noop {} {\emph {\bibinfo {title} {Opto-Semiconductor
  Handbook}}}\ (\bibinfo  {publisher} {{Hamamatsu Photonics K.K.}},\ \bibinfo
  {year} {2017})\BibitemShut {NoStop}%
\bibitem [{\citenamefont {Nishizawa}\ \emph {et~al.}(2014)\citenamefont
  {Nishizawa} \emph {et~al.}}]{AMP}%
  \BibitemOpen
  \bibfield  {author} {\bibinfo {author} {\bibfnamefont {T.}~\bibnamefont
  {Nishizawa}} \emph {et~al.},\ }\href@noop {} {\bibfield  {journal} {\bibinfo
  {journal} {IEEE TNS}\ }\textbf {\bibinfo {volume} {61}},\ \bibinfo {pages}
  {1278} (\bibinfo {year} {2014})}\BibitemShut {NoStop}%
\bibitem [{\citenamefont {Takahashi}\ \emph {et~al.}(2016)\citenamefont
  {Takahashi} \emph {et~al.}}]{DRS4}%
  \BibitemOpen
  \bibfield  {author} {\bibinfo {author} {\bibfnamefont {T.~N.}\ \bibnamefont
  {Takahashi}} \emph {et~al.},\ }\href
  {http://www.rcnp.osaka-u.ac.jp/~annurep/2016/ResearchReport/3ExperimentalFI/TakahashiTN_3_2016.pdf}
  {\bibfield  {journal} {\bibinfo  {journal} {{RCNP Annual Report}}\ }
  (\bibinfo {year} {2016})}\BibitemShut {NoStop}%
\bibitem [{\citenamefont {Tomida}\ \emph {et~al.}(2014)\citenamefont {Tomida}
  \emph {et~al.}}]{BGOegg}%
  \BibitemOpen
  \bibfield  {author} {\bibinfo {author} {\bibfnamefont {N.}~\bibnamefont
  {Tomida}} \emph {et~al.},\ }\href@noop {} {\bibfield  {journal} {\bibinfo
  {journal} {JINST}\ }\textbf {\bibinfo {volume} {9}},\ \bibinfo {pages}
  {C10008} (\bibinfo {year} {2014})}\BibitemShut {NoStop}%
\bibitem [{\citenamefont {Tomida}\ \emph {et~al.}(2016)\citenamefont {Tomida}
  \emph {et~al.}}]{BGOegg2}%
  \BibitemOpen
  \bibfield  {author} {\bibinfo {author} {\bibfnamefont {N.}~\bibnamefont
  {Tomida}} \emph {et~al.},\ }\href@noop {} {\bibfield  {journal} {\bibinfo
  {journal} {JINST}\ }\textbf {\bibinfo {volume} {11}},\ \bibinfo {pages}
  {C11037} (\bibinfo {year} {2016})}\BibitemShut {NoStop}%
\bibitem [{\citenamefont {Watanabe}\ \emph {et~al.}(2019)\citenamefont
  {Watanabe} \emph {et~al.}}]{LEPS2}%
  \BibitemOpen
  \bibfield  {author} {\bibinfo {author} {\bibfnamefont {K.}~\bibnamefont
  {Watanabe}} \emph {et~al.},\ }\href {\doibase
  https://doi.org/10.1016/j.nima.2019.02.014} {\bibfield  {journal} {\bibinfo
  {journal} {Nucl. Inst. Meth. A}\ }\textbf {\bibinfo {volume} {925}},\
  \bibinfo {pages} {188 } (\bibinfo {year} {2019})}\BibitemShut {NoStop}%
\bibitem [{\citenamefont {Bellettini}\ \emph {et~al.}(1966)\citenamefont
  {Bellettini} \emph {et~al.}}]{Bellettini}%
  \BibitemOpen
  \bibfield  {author} {\bibinfo {author} {\bibfnamefont {G.}~\bibnamefont
  {Bellettini}} \emph {et~al.},\ }\href@noop {} {\bibfield  {journal} {\bibinfo
   {journal} {Nucl. Phys.}\ }\textbf {\bibinfo {volume} {79}} (\bibinfo {year}
  {1966})}\BibitemShut {NoStop}%
\bibitem [{\citenamefont {Pavin}(2019)}]{EMPHATIC_2019JETP}%
  \BibitemOpen
  \bibfield  {author} {\bibinfo {author} {\bibfnamefont {M.}~\bibnamefont
  {Pavin}},\ }\href {https://theory.fnal.gov/events/event/emphatic-results/}
  {\enquote {\bibinfo {title} {{F}ermilab {JETP} {S}eminar, {EMPHATIC}
  {R}esults},}\ } (\bibinfo {year} {2019})\BibitemShut {NoStop}%
\bibitem [{\citenamefont {{NOvA Collaboration}}(2019)}]{NOvAFluxImprovements}%
  \BibitemOpen
  \bibfield  {author} {\bibinfo {author} {\bibnamefont {{NOvA
  Collaboration}}},\ }\href@noop {} {}\bibinfo {howpublished} {private
  communication} (\bibinfo {year} {2019})\BibitemShut {NoStop}%
\end{thebibliography}%


%


\end{document}